\documentclass[AMA,STIX1COL]{WileyNJD-v2}

\usepackage{textcomp}
\usepackage[]{graphicx}
\usepackage{color, soul, threeparttable, multirow, multicol, setspace}
\usepackage{bbm, caption, tikz, booktabs, colortbl, dcolumn, float, enumerate, enumitem, booktabs, makecell}

\usepackage{xr}
\usepackage{xr-hyper}
\usepackage{hyperref}
\usepackage{listings}
\usepackage{xcolor}

\newcommand{\bd}{\boldsymbol}
\newcommand{\Ex}{\mathbb{E}}

\newcommand{\mc}{\mathcal}
\newcommand{\mb}{\mathbf}

\newcommand{\bigCI}{\mathrel{\text{\scalebox{1}{$\perp\mkern-10mu\perp$}}}}

\articletype{Tutorial in Biostatistics}%

\received{26 April 2016}
\revised{6 June 2016}
\accepted{6 June 2016}

\definecolor{Orange}{HTML}{FFA500}      
\definecolor{ForestGreen}{HTML}{228B22} 
\definecolor{Blue1}{HTML}{00B0F6}       
\definecolor{Blue2}{HTML}{00F0F6}       
\definecolor{Pink1}{HTML}{FF61C3}       
\definecolor{Purple1}{HTML}{C77CFF}     

\begin{document}

\title{A tutorial for propensity score weighting methods under violations of the positivity assumption}

\author[1,2]{*Yi Liu}
\author[3]{Yuan Wang}
\author[3]{Ying Gao}
\author[4]{Tonia Poteat}
\author[2,3]{Roland A. Matsouaka}

\authormark{LIU et al.}

\address[1]{\orgdiv{Department of Statistics}, \orgname{North Carolina State University}, \orgaddress{\state{North Carolina}, \country{USA}}}

\address[2]{\orgname{Duke Clinical Research Institute}, \orgaddress{\state{North Carolina}, \country{USA}}}

\address[3]{\orgdiv{Department of Biostatistics and Bioinformatics}, \orgname{Duke University School of Medicine}, \orgaddress{\state{North Carolina}, \country{USA}}}

\address[4]{\orgdiv{School of Nursing}, \orgname{Duke University}, \orgaddress{\state{North Carolina}, \country{USA}}}

\corres{*Yi Liu, Department of Statistics, North Carolina State University, North Carolina, USA. \email{yliu297@ncsu.edu}}


\abstract[Summary]{
Violations of the positivity assumption can render conventional causal estimands unidentifiable, including the average treatment effect (ATE), the average treatment effect on the treated (ATT), and the average treatment effect on the controls (ATC). Shifting the inferential focus to their alternative counterparts---the weighted ATE (WATE), the weighted ATT (WATT), and the weighted ATC (WATC)---offers valuable insights into treatment effects while preserving internal validity. In this tutorial, we provide a comprehensive review of recent advances in propensity score (PS) weighting methods, along with practical guidance on how to select a primary target estimand (while other estimands serve as supplementary analyses), implement the corresponding PS-weighted estimators, and conduct post-weighting diagnostic assessments. The tutorial is accompanied by a user-friendly R package, ChiPS. We demonstrate the pertinence of various estimators through extensive simulation studies. We illustrate the flow of the tutorial on two real-world case studies: (i)  Effect of smoking on blood lead level using data from the 2007–2008 National Health and Nutrition Examination Survey (NHANES); and (ii) Impact of history of sex work on HIV status among transgender women in South Africa.
}

\keywords{Causal inference; Positivity; Target population; Weighted average treatment effect; Overlap weights; Unconfoundedness.}

\maketitle

\section{Introduction}\label{sec:intro}

\subsection{Background}\label{subsec:background}

Suppose we want to compare the effectiveness of a new regimen or exposure (refers to as ``treatment'' and denote $A=1$) against a standard of care, a placebo, or the absence of an exposure (which we refer to as ``control'' and denote $A=0$) on an outcome $Y$, using a sample of $N$ participants from which we have measured a set of baseline covariates $\mb X$. To draw causal inference about the treatment effect by $A\in\{0,1\}$ on outcome $Y$ using the potential outcomes framework,\cite{neyman1990applications, rubin1974estimating} it is essential to clearly define the estimands--the quantities we aim to estimate. Beyond guiding estimation and summarizing causal effects, estimands play a critical role in defining the target population for a given scientific question, offering a structured description the treatment effects a study aims to quantify.\cite{kahan2024estimands} 

The propensity score (PS) is a one-dimensional summary of a participant's characteristics $\mb X$. It is defined as the conditional probability $e(\mb X) = P(A=1\mid\mb X)$ of being assigned to the treatment group, given the measured baseline covariates $\mb X$---is a widely used and arguably the most popular tool in causal inference. \cite{rosenbaum1983central, li2018balancing} When all relevant confounders between treatment and outcome are measured, the PS serves as a balancing score,\cite{rosenbaum1983central} enabling balanced covariate distributions between treatment groups through inverse probability weighting (IPW).  Thus, applying a difference-in-means estimator on the IPW-weighted population provides a valid estimate of the average treatment effect (ATE), for which the target is the whole population of participants. Similarly, depending on the scientific question, one can instead estimate the average treatment effect on the treated (ATT) or the treatment effect on the control (ATC). These are often preferred when studying the effects of policy implementations as well as harmful exposures or behaviors in specific groups of participants.\cite{greifer2021choosing}

At the core of IPW methods lies the positivity assumption, which requires that each participant has a non-zero probability of receiving both treatments ($A = 0, 1$). This assumption can be violated either structurally or randomly.\cite{petersen2012diagnosing, westreich2010invited} Structural violations occur when treatment groups differ systematically in covariate distributions, which prevent proper covariate balance and render conventional estimands---ATE, ATT, and ATC---unidentifiable. Random violations arise even in the absence of structural differences, due to finite-sample imbalances, misspecified PS models, or other sources of randomness. Both types of violations can result in unstable estimates and inefficient IPW estimators.\cite{li2018balancing, liu2024average} The inefficiency arises from the reliance on $e(\mb X)^{-1}$ or $(1 - e(\mb X))^{-1}$ to weight each participant’s contribution to the treatment effect estimation. When $e(\mb X)$ is exactly or nearly 0 or 1, some participants receive extreme PS weights, which can over-amplify their influence in the estimation and inflate variance estimates in finite samples.

To address these challenges, practical methods have been developed to weight participants based on their PS values while mitigating the influence of extreme weights. In recent years, causal inference has undergone a ``revolution'' with the emergence of novel estimands that target treatment effects in specific, PS-defined populations.\cite{li2018balancing, liu2024average, crump2006moving, li2013weighting, mao2019propensity, zhou2020propensity, matsouaka2024overlap, matsouaka2024causal} Beyond conventional estimands---ATE, ATT, and ATC---new approaches have been proposed as generalizations or alternatives. These methods enhance estimation accuracy and efficiency by focusing on subpopulations (or weighted populations) where selected participants align more closely with the target estimand. In the next section, we review several widely used and recently developed PS-based methods. 

The remainder of this paper is organized as follows. The rest of this section reviews recent advances in PS weighting methods and related work in Section \ref{subsec:relate}, and outlines the contributions of our tutorial in Section \ref{subsec:ourcontr}. In Section \ref{sec:notation}, we introduce statistical preliminaries and define the target causal parameters. Section \ref{sec:tutorial} presents the tutorial and implementation details of our user-friendly R package, ``ChiPS'' (\textbf{Ch}oos\textbf{i}ng \textbf{P}ropensity \textbf{S}core Weights). Section \ref{sec:simu} reports results from a Monte Carlo simulation study evaluating various PS-weighted estimators. In Section \ref{sec:data}, we illustrate the methods using two case studies: (i) the effect of smoking on blood lead levels (continuous outcome), and (ii) the effect of sex work history on HIV status among transgender women in South Africa (binary outcome). Finally, Section \ref{sec:conclude} concludes the paper and highlights key takeaways.

\subsection{Related work}\label{subsec:relate}

In the presence of positivity violations, conventional estimands such as the ATE, ATT, and ATC may become non-estimable. To overcome this issue, a growing body of literature has proposed alternative causal estimands that shift focus to subpopulations where the positivity assumption holds. These estimands are often adjacent to the conventional ones and provide meaningful targets for inference in realistic settings.

A seminal example is PS trimming, introduced by Crump et al.,\cite{crump2006moving} which excludes participants with estimated PS values outside a fixed range $[\alpha, 1-\alpha]$ for some $\alpha \in (0, 0.5)$. The resulting trimmed subpopulation is defined as $\{\mb X: \alpha < e(\mb X) < 1 - \alpha\}$ and is equivalent to applying a tilting function $h(\mb X) = \text{I}(\alpha < e(\mb X) < 1 - \alpha)$, where $\text{I}(\cdot)$ is the indicator function.\cite{li2019addressing, li2018balancing, thomas2020overlap} The treatment effect over this subpopulation is referred to as the \textit{weighted average treatment effect (WATE)},\cite{hirano2003efficient, li2018balancing} which reweights the overall population and remains closely related to the ATE.\cite{thomas2020overlap} This approach applies symmetric trimming on both tails of the PS distribution. However, it may be overly restrictive when PS distributions differ {between treatment and control groups}, potentially excluding more participants from one group than the other. To address this, St\"urmer et al.\cite{sturmer2010treatment} proposed an asymmetric trimming: first exclude participants outside the common PS support across treatment groups, then, within each group, exclude participants with PS values below the $q$-th (for treated) or above the $(1 - q)$-th (for controls) quantile of their respective PS distributions, with $q \in (0, 0.5)$. This asymmetric trimming is more flexible and data-adaptive, allowing tailored thresholds for each group’s PS distribution.

Yang and Ding \cite{yang2018asymptotic} proposed a smooth version of PS trimming, which stabilizes the weights near the boundaries $\alpha$ and $1 - \alpha$ by using a smoothing parameter $\varepsilon$ to reduce sensitivity to a possible PS model misspecification. Another related approach is PS truncation, where weights are capped by replacing the PS values that are outside of the interval $[\alpha, 1-\alpha]$ with some given threshold limits.\cite{ju2019adaptive, sturmer2010treatment, cole2008constructing, gruber2022data, austin2015moving, lee2011weight} While trimming, smooth trimming, and truncation require setting specific thresholds $\alpha$, smoothing parameters $\varepsilon$  or quantiles $q$, more strictly data-driven alternatives have also emerged. For example, Ma and Wang,\cite{ma2020robust} Chaudhuri and Hill,\cite{chaudhuri2025heavy} Sasaki and Ura,\cite{sasaki2022estimation} and Ju et al.\cite{ju2019adaptive} proposed methods that ensure that bias from trimming or truncation converges to zero asymptotically. However, these approaches are only effective for random, not structural, violations of the positivity  assumption. Matsouaka and Zhou \cite{matsouaka2024causal} as well as Liu et al.\cite{liu2025assessing} argued that selecting appropriate thresholds can be challenging when prior knowledge of the population characteristics is limited. In many cases, it is difficult to determine an optimal threshold.  

To avoid the need to address the threshold selection challenges, but still circumvent positivity violations, several alternative WATEs have been introduced. Li et al.\cite{li2013weighting} proposed matching weights (MW), 
with the corresponding estimand termed the average treatment effect on the matching population (ATM). Mao et al.\cite{mao2019propensity} extended MW to trapezoidal weights (TW). Li et al.\cite{li2018balancing} developed overlap weights (OW), leading to the average treatment effect on the overlap population (ATO). Zhou et al.\cite{zhou2020propensity} proposed entropy weights (EW) and the corresponding average treatment effect on the entropy population (ATEN), while also evaluating the robustness of IPW, MW, OW, and EW under PS model misspecification and limited overlap. Matsouaka and Zhou\cite{matsouaka2024causal} further generalized OW to beta family weights (BW) and developed criteria for targeting clinical equipoise. Additional insights came from Thomas et al.\cite{thomas2020overlap} and Choi and Lee.\cite{choi2023overlap} Parikh et al.\cite{parikh2025we} proposed a data-driven tilting function, mapping covariates to achieve a more interpretable target population with reduced asymptotic variance.

Recently, Liu et al.\cite{liu2024average} introduced a new class of estimands, the \textit{weighted average treatment effect on the treated (WATT)}, generalizing ATT in the same way WATE extends ATE. Unlike WATE, which applies the tilting function to the overall population of both treatment groups, WATT applies it only to the control participants. Liu et al.\cite{liu2024average} also proposed the overlap weighted ATT (OWATT), which uses OW on the controls, demonstrating greater precision and robustness compared to the conventional ATT. They also briefly discussed the \textit{weighted average treatment effect on the controls (WATC)}, applying the tilting function to the treated group, offering a natural counterpart to WATT.

Finally, several studies have compared differences among estimands within the WATE class.\cite{matsouaka2024overlap, austin2023differences, zhou2020propensity, barnard2024unified} Comparisons within the WATT class have also been explored,\cite{liu2024average, liu2025assessing} though the weighting methods considered---such as overall, trimming, truncation, and OW---do not encompass other approaches in the WATE class, such as MW, EW and BW. 

\subsection{Our contributions}\label{subsec:ourcontr}

While prior research offers thorough investigations of various weighting schemes within the WATE and WATT classes, to our knowledge, no study has integrated these methods---along with those in the WATC class---into a unified tutorial focused on practical implementation. Furthermore, prior literature has not explicitly distinguished between defining ATT under the WATE and WATT frameworks or ATC under the WATE and WATC frameworks; we do believe that such distinctions  are both crucial and meaningful. In addition, no existing work systematically compares these estimands from the following perspectives, particularly under heterogeneous treatment effects, where they are expected to differ: \cite{matsouaka2024overlap} (i) the fundamental differences in the definitions of WATE, WATT, and WATC, and (ii) the implications of using different weights within and across these classes. 

We address these knowledge gaps by providing a principled tutorial that unifies weighting methods across the WATE, WATT, and WATC classes. {First, while the WATC class has been discussed in Liu et al.,}\cite{liu2024average} {it has not been formally introduced in detail.} {We formally introduce the definition of WATC estimands in Section} \ref{subsec:WATT}. {Our tutorial elucidates the theoretical} underpinnings of each method, clarifies their respective target populations and assumptions, and offers practical guidance on implementation using a user-friendly R package. {We also extend all WATE, WATT, and WATC methods to binary outcomes using the risk ratio (RR) and odds ratio (OR) (see Section} \ref{subsec:binout}{), as these measures are the commonly-used treatment effect quantifications for binary outcomes. Finally,  in Section} \ref{sec:tutorial}, {we provide a careful and principled, decision tree-based guideline for selecting an appropriate estimand in the presence of a structural or random violation of positivity, along with detailed instructions for implementing the corresponding estimation procedure.}

Through illustrative applications and simulation studies, we demonstrate how different weighting schemes affect causal effect estimation under violations of the positivity assumption,  thereby equipping practitioners with the tools needed to select and apply appropriate methods in real-world settings.
To facilitate implementation, we provide the ``ChiPS'' R package, available at \url{https://github.com/yiliu1998/ChiPS}. It accommodates a broad range of PS models to meet the needs of different fields. For example, logistic regression is often preferred in epidemiology for its interpretability, while machine learning models such as gradient boosting or random forests are better suited for data-rich settings due to their flexibility in capturing complex relationships. This package enables users to tailor PS models to the specific context and goals of their studies, maximizing both robustness and practical relevance. 

\section{Notation, Assumptions and Estimands}\label{sec:notation}

We consider data from a non-randomized study, where $A = 1$ if a participant received the treatment and $A = 0$ if the participant received the control. Let $Y$ denote the observed outcome, and let $\mb X = (X_1, \ldots, X_p)'$ represent the vector of measured covariates (confounding factors). The observed data $\mc O = \{(A_i, \mb{X}_i, Y_i) : i = 1, \dots, N\}$ is an independently and identically distributed (i.i.d.) sample of size $N$ drawn from a super-population.

Following the potential outcomes framework,\cite{neyman1990applications, rubin1974estimating} we assume that each participant has two potential outcomes $Y(a)$, which correspond to $a\in \{0,1\}$. The potential outcome $Y(a)$ represents the outcome a participant would experience if, possibly contrary to fact, he or she received the treatment $A = a$. We also assume consistency between the potential and observed outcomes, i.e., $Y = AY(1) + (1-A)Y(0)$.

The individual treatment effect is defined as $Y(1) - Y(0)$, though it is not directly observable because for each participant, we can only observe one of the potential outcomes. Thus, we rely on a sample of participants to estimate the causal estimand(s) of interest. To proceed, we assume the stable unit treatment value assumption (SUTVA), i.e., there is only one version of the treatment and the potential outcome $Y(a)$ of one participant is unaffected by the treatment assign to any other participant. \cite{rosenbaum1983central}

We define the PS $e(\mb x)=P(A=1\mid\mb X=\mb x)$, i.e., the participant-specific probability of receiving the treatment, conditional on the covariates $\mb X$. The true PS is unknown in non-randomized studies and must be estimated from the observed data. Thus, one can postulate a parametric model or use a nonparametric machine learning approach (such as the super learner \cite{van2007super}). A commonly used model is the logistic regression $e(\mb X) = e(\mb X;\bd\beta)=\{1+\exp(-\mb X'\bd\beta)\}^{-1}$. 

\subsection{{Assumptions and conventional estimands}}\label{subsec:estimands}

The following structural assumptions are made throughout the paper:\cite{rosenbaum1983central}

\begin{assumption}[Unconfoundedness]\label{assp:unconfound}
    $(Y(0), Y(1))\bigCI A\mid \mb X$, where ``$\bigCI$'' stands for independence. 
\end{assumption}

\begin{assumption}[Positivity]\label{assp:overlap}
    There exist two positive constants $c_0$ and $c_1 \in (0,1)$, such that $c_0\leq e(\mb X)\leq 1-c_1$ with probability 1.
\end{assumption}

Assumption \ref{assp:unconfound} implies that all possible confounders of the relation between the treatment $A$ and the outcome $Y$ have been  measured in the observed data. Assumption \ref{assp:overlap} requires that each participant has a certain non-zero probability of being assigned to either treatment group for ATE estimation. In that case, it is required that both $c_0$ and $c_1$ to be positive. Nevertheless, to estimate ATT (resp. ATC), we only need $e(\mb X)\geq c_0>0$ (resp. $1- e(\mb X)\geq c_1>0$) in the control (resp. treated) group.\cite{liu2024average, abadie2005semiparametric, heckman1998matching} 

The three conventional causal estimands (ATE, ATT and ATC) are defined, respectively, by
\begin{align}\label{eq:convEstimands}
    \text{ATE: } & \tau^{\text{ate}} = \Ex\{Y(1)-Y(0)\} = \Ex\{\tau(\mb X)\},\nonumber \\
    \text{ATT: } & \tau^{\text{att}} = \Ex\{Y(1)-Y(0)\mid A=1\}= \Ex\{Y\mid A=1\}-\Ex\{Y(0)\mid A=1\}, \text{ and }\nonumber\\
     \text{ATC: } & \tau^{\text{atc}} = \Ex\{Y(1)-Y(0)\mid A=0\}= \Ex\{Y(1)\mid A=0\}-\Ex\{Y\mid A=0\},
\end{align}
 where $\tau(\mb X)=\Ex\{Y(1)-Y(0)\mid\mb X\}$ is the conditional average treatment effect (CATE). Based on these definitions, the ATE targets the overall population from which the data were sampled, while the ATT (resp. ATC) targets the population of participants with $A=1$ (resp. $A=0$). 

It can be shown that, under Assumptions \ref{assp:unconfound} and \ref{assp:overlap}, the above estimands can be written as the following functions of the PS\cite{matsouaka2024causal,liu2024average, matsouaka2024overlap} 
\begin{align}\label{eq:convEstimandswithPS}
    \text{ATE: } & \tau^{\text{ate}} = \Ex\left\{\frac{(A-e(\mb X))Y}{e(\mb X)(1-e(\mb X))}\right\},\nonumber\\
    \text{ATT: } &\tau^{\text{att}} = P(A=1)^{-1} \Ex\left\{\frac{(A-e(\mb X))Y}{1-e(\mb X)}\right\}, \text{ and }\nonumber\\
     \text{ATC: } & \tau^{\text{atc}} = P(A=0)^{-1}\Ex\left\{\frac{(A-e(\mb X))Y}{e(\mb X)}\right\}.
\end{align}
These functions are also known as the identification formulas of ATE, ATT and ATC. 
Without restrictions in Assumption \ref{assp:overlap}, there might be regions of the distributions of the covariates $\mb X$ where the ratios $e(\mb X)^{-1}(1-e(\mb X))^{-1}$, $e(\mb X)(1-e(\mb X))^{-1}$, or $e(\mb X)^{-1}(1-e(\mb X))$ becomes arbitrarily large, which render the point and variance estimations particularly unstable and difficult. 
Thus, the point and variance estimations of ATE, ATT and ATC are possible only under Assumption \ref{assp:overlap} for some $c_0$ and $c_1 \in (0, 1)$. 

In practice, both Assumptions \ref{assp:unconfound} and \ref{assp:overlap} are not directly testable. However, we can evaluate whether they are {plausible} to the best of our knowledge based on the domain expertise and the data at hand. The plausibility of Assumption \ref{assp:unconfound} may be assessed with the aid of  clinical expertise and the use of directed acyclic graphs (DAGs).\cite{tennant2021use} 
To evaluate Assumption \ref{assp:overlap}, we often examine the estimated PS values and visualize the distributions (or densities) of the estimated PS  for each treatment group.

For this paper, we assume Assumption \ref{assp:unconfound} always holds, i.e., the data  contain all possible relevant confounders of the treatment and the outcome, as it often the case in the literature.\cite{li2018balancing, liu2024average, rosenbaum1983central, yang2018asymptotic, li2013weighting} Our main focus is Assumption \ref{assp:overlap}, which might be violated regardless whether Assumption \ref{assp:unconfound} is satisfied. 

\subsection{The weighted average treatment effect}\label{subsec:WATE}

When the positivity assumption (Assumption \ref{assp:overlap}) is violated, the use of ATE, ATT, and ATC can become problematic. If the violation occurs by chance---e.g., random errors in a specific finite sample---estimates may be unstable due to the influence of excessively large weights assigned to a few participants. A more severe and pernicious issue arises when there is a structural violation of positivity,\cite{petersen2012diagnosing} i.e., the two treatment groups differ systematically due to inherent characteristics of the population. In such cases, ATE, ATT, and ATC may not be identifiable. For a more detailed discussion on the violations of the positivity assumption, see Remark \ref{rmk:posivity} after we introduce different PS weights and their corresponding target populations. To address violations of the  positivity assumption, much of the literature proposes shifting the focus from the overall population to a subpopulation (or weighted population) to gain insights into the treatment effect of interest.\cite{hirano2003efficient,yang2018asymptotic,li2013weighting,matsouaka2024causal,matsouaka2024overlap} This motivates the definition of a general class of estimands, known as the weighted average treatment effect (WATE),\cite{li2018balancing} which generalizes the ATE.

The WATE is defined by 
\begin{align*}
\tau^{\text{wate}} & = \frac{\Ex\{h(\mb X)\tau(\mb X)\}}{\Ex\{h(\mb X)\}}, 
\end{align*} 
where $\tau(\mb X) = \Ex\{Y(1) - Y(0) \mid \mb X\}$ is the CATE, and $h(\mb X)$ is the tilting function. {Furthermore, under Assumption \ref{assp:unconfound}, $\tau^{\text{wate}}$ can be re-written as (see Li et al.\cite{li2018balancing}):
\begin{align}\label{eq:WATEform}
    \tau^{\text{wate}} = \frac{1}{\Ex\{h(\mb X)\}}\Ex\left[h(\mb X)\left\{\frac{A}{e(\mb X)}-\frac{1-A}{1-e(\mb X)}\right\}Y\right]. 
\end{align}
}

Table \ref{tab:WATEs} provides a comprehensive review of popular tilting functions and their WATEs, as well as their target populations, names of PS weights, and summaries of their connections.

\begin{table}[ht]
\caption{Summary of PS weights, related literature, and their tilting functions when targeting WATE}\label{tab:WATEs}
\centering
\begin{tabular}{ccccccccccccccccccccccccccccccc}
    \toprule 
    PS weights & Target population & Related literature  & Tilting function $h(\mb x)$   \\
    \midrule 
    IPW & Overall &  Rosenbaum and Rubin\cite{rosenbaum1983central} & $1$ \\
    IPW treated & Treated & Hirano et al.\cite{hirano2003efficient} & $e(\mb x)$  \\
    IPW controls & Controls & Tao and Fu\cite{tao2019doubly} & $1-e(\mb x)$  \\
    IPW trimming & Trimmed &  Crump et al.\cite{crump2006moving} & $\text{I}(\alpha<e(\mb x)<1-\alpha)$ \\
    IPW smooth trimming & Trimmed & Yang and Ding\cite{yang2018asymptotic} & $\Phi_{\epsilon}(e(\mb x)-\alpha)\Phi_{\epsilon}(1-\alpha-e(\mb x))$ \\
    \addlinespace
    IPW truncation & Overall & \makecell[c]{
    Cole and Hernan,\cite{cole2008constructing}\\
    St\"urmer et al.\cite{sturmer2010treatment}} & \makecell[l]{$\text{I}(\alpha<e(\mb x)<1-\alpha) + $  $\text{I}(e(\mb x)\leq\alpha)\dfrac{e(\mb x)}{\alpha}$\\
    $ +$ $ \text{I}(e(\mb x)\geq 1-\alpha)\dfrac{1-e(\mb x)}{1-\alpha}$} \\ 
    \addlinespace
    Matching weights (MW) & Equipoise & Li and Greene\cite{li2013weighting} & $\min\{e(\mb x),1-e(\mb x)\}$ \\
    Trapezoidal weights (TW)& Equipoise & Mao et al.\cite{mao2019propensity} & $\min\{1,\min\{e(\mb x),1-e(\mb x)\}K], K>1$ \\
    Overlap weights (OW)& Equipoise & Li et al.\cite{li2018balancing} & $e(\mb x)(1-e(\mb x))$ \\
    Entropy weights (EW)& Equipoise & Zhou et al.\cite{zhou2020propensity} & $ - e(\mb x)\log\left\{\dfrac{e(\mb x)}{1-e(\mb x)}\right\}-\log\{1-e(\mb x)\}$ \\
    Beta family weights (BW)& Equipoise & Matsouaka and Zhou\cite{matsouaka2024causal} & $e(\mb x)^{\nu_{1}-1}\{1-e(\mb x)\}^{\nu_{2}-1}$ $(\nu_{1}, \nu_{2}\geq2)$ \\
    \bottomrule 
\end{tabular}
\begin{tablenotes}\scriptsize
    \item $\Phi_\epsilon(\cdot)$ is the cumulative distribution function of \text{N}$(0, \epsilon^2), \epsilon~>0$; $\Phi_{\epsilon}(t-\alpha)\Phi_{\epsilon}(1-\alpha-t)\to \text{I}(\alpha<t<1-\alpha)$ for every $t\in[0,1]$ as $\epsilon\to 0$. OW is the special case of BW when $\nu_1=\nu_2=2$. IPW, IPW treated and IPW controls are special cases of BW, respectively, when $(\nu_1, \nu_2)=(1,1), (2,1)$ and $(1,2)$.
\end{tablenotes}
\end{table}

The three conventional estimands, ATE, ATT and ATC,  fall within the class of WATE.\cite{li2019propensity} Specifically, when $h(\mb X) = 1$, $e(\mb X)$, and $1 - e(\mb X)$, respectively, the WATE becomes ATE, ATT, and ATC. We should note that within the WATE class, ATT and ATC have different interpretations compared to their original definitions, although they target the same numerical values. For ATT (resp. ATC), with the original definition, the target population consists only of the treated (resp. control) participants. Under the WATE definition, however, it represents a weighted average of the treatment effect on the whole population of participants, weighted by the tilting function $e(\mb X)$ (resp. $1 - e(\mb X)$). Hence, participants (from treatment and control groups) with higher PSs will have larger (resp. smaller) contribution for ATT (resp. ATC) estimation. This indicates that the estimands in the WATE class can all be viewed as targeting the overall population where the weights are defined in part by the tilting functions. 

Table \ref{tab:WATEs} also summarizes the different WATEs used in the literature, together with their respective names (``PS weights'' column) and the corresponding tilting functions. 
Beyond ATE, ATT, and ATC, another notable member is the ATO, defined by overlap weights (OW) in the table. The mathematical formulation of the average treatment effect on the overlap population (ATO) was first introduced in Hirano et al.\cite{hirano2003efficient} and given the moniker ``overlap weights'' by Li et al.\cite{li2018balancing}  The intuition behind OW is straightforward: since its tilting function $h(\mb X) = e(\mb X)(1-e(\mb X))$ reaches its maximum at $e(\mb X) = 0.5$; it assigns higher weights to those with high certainty to be assigned to either treatment group, which is also referred to as ``targeting the clinical equipoise.''\cite{matsouaka2024causal, thomas2020overlap, li2019addressing} In addition, OW has the following advantageous statistical properties: 
\begin{enumerate} 
\item When the PSs are estimated through logistic regression, the corresponding OW-weighted covariate distributions between the two treatment groups achieve exact balance; 
\item The PSW estimator based on OW has the smallest asymptotic variance among the PSW estimators of estimands in the WATE class. 
\end{enumerate}

\begin{figure}[ht]
    \centering \includegraphics[width=0.6\textwidth]{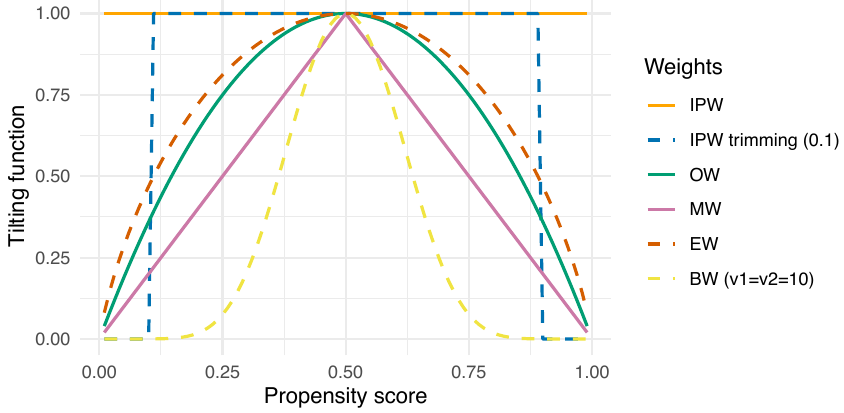}
    \caption{Examples of the tilting function $h(\mb X)$ for WATE.}
    \label{fig:tiltWATE_PS}
\end{figure}

Figure \ref{fig:tiltWATE_PS} illustrates examples of tilting functions $h(\mb X)$ for the WATE class, as defined in Table \ref{tab:WATEs} with $\alpha=0.1$ and $\nu_1=\nu_2=10$. The tilting function of IPW is a constant line, $h(\mb X)=1$, which assigns equal value (or importance) to  all the weights $\displaystyle {A}e(\mb X)^{-1}$ and $\displaystyle {(1-A)}(1-e(\mb X))^{-1}$ of the participants, $A \in \{0, 1\}$. For IPW trimming, its function $h(\mb X)$ is also equal to 1 inside the interval $[\alpha,1-\alpha]$. It drops to 0 for $e(\mb X)$ that are outside of the interval $[\alpha,1-\alpha]$, with $\alpha\in (0, 0.5]$, effectively excluding participants with extreme $\displaystyle {A}e(\mb X)^{-1}$ and $\displaystyle {(1- A)}(1-e(\mb X))^{-1}$ weights. The (scaled) tilting functions of OW, MW, EW, and BW (with $\nu_1=\nu_2=10$) exhibit similar symmetric patterns: they peak at $e(\mb X)=0.5$ and gradually decrease toward 0 as $e(\mb X)$ approaches 0 or 1 (at different rates). {Although, in theory, OW assigns zero weight when $e(\mb X) = 0$ or $1$, estimated PSs are rarely exactly 0 or 1 in practice. Thus, participants with extreme estimated PSs receive weights that are very close to zero rather than exactly zero. } 

The above weighting schemes define different target populations, a critical information needed by investigators when selecting an appropriate approach. IPW targets the original, overall population, while IPW trimming restricts inference to participants with less extreme PSs, i.e., those whose PS are within the interval $[\alpha,1-\alpha]$. OW, MW, and EW shift the target population toward participants with greater clinical equipoise.\cite{thomas2020overlap} Unlike IPW trimming, which depends on the threshold $\alpha\in (0, 0.5]$ and only keeps PS weights $\displaystyle {A}e(\mb X)^{-1}$ and $\displaystyle {(1- A)}(1-e(\mb X))^{-1}$  whose $e(\mb X)$  are  within $[\alpha,1-\alpha]$,  OW, MW and EW assign strictly higher weights to participants whose PS values are in the vicinity of  0.5. Viewed from this perspective,  OW, MW and EW act like the ultimate inherent data-driven pruning tools (i.e., that do not depend on any $\alpha\in (0, 0.5]$). Additional properties of  OW, MW and EW as well as their similarities are provided by Matsouaka and Zhou.\cite{matsouaka2024causal}

BW extends OW by introducing tuning parameters $\nu_1$ and $\nu_2$. When $\nu_1 = \nu_2 = 2$, BW reduces to OW; as these parameters increase, the tilting function becomes more concentrated around $e(\mb X) = 0.5$, as shown in Figure \ref{fig:tiltWATE_PS}. Investigators may consider BW when seeking a balance between a refined focus on participants with PSs near 0.5 and the stability of OW. While increasing $(\nu_1, \nu_2)$ enhances emphasis on clinical equipoise, it also over-weights a small subset of participants, making the estimation more sensitive to PS misspecification and local weight variations around $e(\mb X) \approx 0.5$. Empirical results from our simulations in Section \ref{sec:simu} and previous studies \cite{matsouaka2024causal} confirm this phenomenon.

\begin{remark}\label{rmk:truncation}
    Truncation stabilizes the weights $\displaystyle {A}e(\mb X)^{-1}$ and $\displaystyle {(1- A)}(1-e(\mb X))^{-1}$ by capping their values outside of the interval $[\alpha,1-\alpha]$ to $\displaystyle {A}\alpha^{-1}$ and $\displaystyle {(1- A)}(1-\alpha)^{-1}$, respectively. Thus, truncation improves efficiency (i.e., reduces variance) but may introduce bias when estimating the treatment effect for the overall population of participants it is supposed to target. In contrast, trimming removes from the data analysis participants whose PSs are  outside of $[\alpha,1-\alpha]$. By removing those participants, trimming shifts the target population and yields an estimate for a totally different estimand. This distinction between trimming and truncation is extremely important: truncation controls instability while preserving the original target population, whereas trimming changes the interpretation of the causal effect altogether. Investigators should carefully align their choice with their research objective. If the goal is to estimate the effect in the full population but extreme weights introduce finite-sample bias, truncation can be a practical adjustment. However, if restricting inference to a subpopulation with more balanced PS values is acceptable, IPW trimming provides a more interpretable and potentially unbiased alternative.
\end{remark}

\begin{remark}\label{rmk:posivity}
    As discussed earlier, violations of the positivity assumption can be random or structural, influencing the choice of target populations. When such a violation is random (e.g., due to chance in a finite sample or a misspecification of the PS model), ATE, ATT, and ATC remain estimable, but researchers should be aware that extreme weights may lead to high variability and imprecision. When the violation is structural---due to fundamental, inherent differences between treatment groups---these estimands are no longer well-defined or identifiable.\cite{matsouaka2024causal} In such a case, it is necessary to explicitly redefine the target population or use alternative weighting methods to implicitly redefine the target population and ensure valid inference.\cite{mao2019propensity,matsouaka2024causal} OW and its close alternatives, MW and EW, assign  weights that {are nearly 0 to participants who strongly violate the positivity assumption (those with estimated PSs close to 0 or 1)}. These methods gradually increase weights for participants whose PS moves away from 0 or 1, emphasizing those in regions of better overlap. BW follows a similar trend but may also assign relatively lower weights to other participants with PS values that are not always near 0 and 1, especially when the parameters $\nu_1$ and $\nu_2$ become larger. 
    
    A key point is that all of the weights discussed in this paper require some degree of overlap between the two treatment groups, even if limited. Structural violations of positivity assumption may still arise from certain subsets of participants in both groups (see examples in Matsouaka and Zhou\cite{matsouaka2024causal}). In the extreme case of no overlap---though rare in practice---all estimands become unidentifiable. Therefore, our framework fundamentally relies on a meaningfully defined target population where at least some weighted effect can be assessed.
\end{remark}

\begin{remark}\label{rmk:bias-var}
    {Another perspective on how these weighting techniques address violations of positivity is through the lens of the bias-variance trade-off. Compared to methods that focus primarily on bias reduction,}
    \cite{ma2020robust, sasaki2022estimation} {shifting the target population toward regions that satisfy the positivity assumption can reduce estimation variance (and thereby enhance efficiency) by uniformly avoiding extreme weights. This, in turn, improves the internal validity of the treatment effect estimation within the defined target population. }
\end{remark}

\subsection{Weighted average treatment effect on the treated and on the controls}\label{subsec:WATT}

Although the WATE class includes ATT and ATC as special cases, it also provides alternative interpretations for these estimands, as previously discussed. In our view, the central role of the ATE is paralleled by the importance of ATT and ATC, particularly in policy-making contexts where these estimands are often the primary focus.\cite{heckman1997matching, dynarski2003does, jin2025policy, matsouaka2023variance} However, the WATE class is a generalization centered around ATE; there was no direct extension centered around ATT and ATC before Liu et al.\cite{liu2024average} Thus, while there are many alternatives to estimate ATE in the WATE class, there are no corresponding measures that are (and adjacent to) ATT- and ATC-centered treatment effects. 

Liu et al.\cite{liu2024average} fill this gap by proposing the weighted ATT (WATT) class of estimands as a generalization of ATT. The generalization of ATC to WATC follows a similar argument as it only requires switching the role of treated and control groups. Furthermore, unlike the tilting functions for the WATE class, which apply to the entire population of participants, the tilting functions for the WATT and WATC classes only apply to one of the treatment groups. The motivation for weighting only one treatment group stems from how the positivity assumption is defined for ATT and ATC and how its  violations occur. As shown by equations \eqref{eq:convEstimandswithPS}, violations of the positivity assumption in ATT (resp. ATC) arise when the PS of the control group (resp. treated group) is too close to 1 (resp. 0). To address this issue, we first consider the weighting formula for ATT from \eqref{eq:convEstimandswithPS}, i.e.,
\begin{align*}
    \tau^{\text{att}} = \frac{\Ex(AY)}{\Ex(A)} - \frac{\Ex\{(1-A)w_0(\mb X)Y\}}{\Ex\{(1-A)w_0(\mb X)\}}, \text{ where } w_0(\mb X) = e(\mb X)/\{1- e(\mb X)\}.
\end{align*}
The ATT is the difference between the mean outcomes of the treated participants $(A=1)$ and the mean weighted outcomes of the control participants using the weights $w_0(\mb X)$. These weights can be understood as a covariate density ratio of the treated to the control group, which creates a pseudo-population among the original control participants such that it mimics the traits of the treated participants. 

Unfortunately, the weight $w_0(\mb X)$ can become extremely large when $e(\mb X) \approx 1$. To avoid such extreme values, we often use specific tilting functions  on the control participants (similar to those used in the  WATE class) and define the following WATT class of estimands:
\begin{align}\label{eq:WATTform}
    \tau^{\text{watt}} = \frac{\Ex(AY)}{\Ex(A)} - \frac{\Ex\{(1-A)\omega_{0g}(\mb X)Y\}}{\Ex\{(1-A)\omega_{0g}(\mb X)\}}, \text{ where } \omega_{0g}(\mb X) =  g(\mb X)e(\mb X)/\{1 - e(\mb X)\}.
\end{align}
We distinguish the tilting function $g(\mb X)$ from $h(\mb X)$ in the WATE estimands since some tilting functions with the same name may have different definitions. For example, for trimming, we only need $g(\mb X) = \text{I}(e(\mb X) < 1 - \alpha)$, for some threshold $\alpha \in (0, 0.5)$, to ensure the positivity needed to identify WATT based on \eqref{eq:WATTform}.

{The tuning parameter $\alpha$ plays a central role in the trimming and truncation tilting functions $g(\mb X)$, as it governs the extent to which control participants with extreme estimated PS values are down-weighted (with truncation) or excluded (for trimming). Conceptually, $\alpha$ defines the empirical boundary of overlap between the treated and control groups. For instance, when $\alpha = 0.1$ in WATT trimming, the effective analysis population excludes (or, in the case of truncation, heavily de-emphasizes) control participants whose PSs exceed $0.9$. The resulting pseudo-population retains only control participants whose covariate profiles resemble those of the treated group, while preserving overlap and internal validity. Smaller values of $\alpha$ (e.g., $0.05$) correspond to broader but potentially less stable populations, while larger values (e.g., $0.2$) yield narrower but more robust inference domains with stronger overlap. In practice, $\alpha$ can be pre-specified based on subject-matter knowledge or chosen using diagnostic plots of the estimated PS distribution. Following the recommendations by Crump et al.,}\cite{crump2009dealing} {we generally consider $\alpha \in [0.05, 0.2]$, i.e., those that often better satisfies bias-variance trade-off.}

When we consider shifting the target from ATT to some other WATTs, we implicitly make a trade-off: we improve efficiency and mitigate the impact of the violations of the positivity assumption, but deviate the target estimand away from ATT.\cite{jin2025policy,liu2025assessing} Thus, an important goal under the WATT framework is to find candidate functions $g(\mb X)$ that better make this trade-off. 
We can extend the above WATT framework to ATC to derive the WATC class of estimands as follows:
\begin{align} \label{eq:WATCform}
    \tau^{\text{watc}} = \frac{\Ex\{A\omega_{1g}(\mb X)Y\}}{\Ex\{A\omega_{1g}(\mb X)\}} - \frac{\Ex\{(1-A)Y\}}{\Ex\{1-A\}}, ~\text{ where }~ \omega_{1g}(\mb X) = g(\mb X)\{1-e(\mb X)\}/e(\mb X).
\end{align}
 In Table \ref{tab:TiltWATT}, we provide examples of tilting functions $g(\mb X)$ for WATT and WATC. {The interpretation of $\alpha$ in WATC is analogous to that in WATT, except that the tilting function is applied to the group of treated participants to down-weight or annihilate the influence of those with low treatment propensity, thereby stabilizing estimation and preserving overlap. }

\begin{table}[ht]
\caption{Tilting functions $g(\mb x)$ in for WATT and WATC estimands}\label{tab:TiltWATT}
\centering
\begin{tabular}{ccccccccccccc}
    \toprule 
    PS weights & Estimand name & $g(\mb x)$ for WATT & $g(\mb x)$ for WATC   \\
    \midrule
    Conventional & ATT / ATC & 1 & 1 \\
    Trimming & ATT / ATC trimming & $\text{I}(e(\mb x)<1-\alpha)$ & $\text{I}(e(\mb x)>\alpha)$ \\
    Smooth trimming & Smooth ATT / ATC trimming & $\Phi_\varepsilon(1-\alpha-e(\mb x))$ & $\Phi_\varepsilon(e(\mb x)-\alpha)$ \\
    \addlinespace\addlinespace
     \multirow{2}{*}{Truncation} & \multirow{2}{*}{ATT / ATC truncation} & $\text{I}(e(\mb x)\geq 1-\alpha)\dfrac{(1-e(\mb x))\alpha}{(1-\alpha)e(\mb x)}$ & $\text{I}(e(\mb x)>\alpha)$\\
       &  & + ~$\text{I}(e(\mb x)<1-\alpha)$ &  + $~ \text{I}(e(\mb x)\leq\alpha)\dfrac{e(\mb x)(1-\alpha)}{\alpha(1-e(\mb x))}$ \\
\addlinespace\addlinespace
    Matching weights (MW) & MWATT / MWATC & \multicolumn{2}{c}{$\min\{e(\mb x),1-e(\mb x)\}$} \\
    Overlap weights (OW) & OWATT / OWATC & \multicolumn{2}{c}{$e(\mb x)(1-e(\mb x))$} \\
    Entropy weights (EW) & EWATT / EWATC & \multicolumn{2}{c}{$-e(\mb x)\log\left\{\dfrac{e(\mb x)}{1-e(\mb x)}\right\}-\log\{1-e(\mb x)\}$} \\
    Beta family weights (BW) & BWATT / BWATC & \multicolumn{2}{c}{$e(\mb x)^{\nu_{1}-1}\{1-e(\mb x)\}^{\nu_{2}-1}$ $(\nu_{1}, \nu_{2}\geq2)$} \\
    \bottomrule 
\end{tabular}
\begin{tablenotes}\scriptsize
    \item $\Phi_\epsilon(\cdot)$ is the cumulative distribution function of normal with mean 0 and standard error $\epsilon>0$; $\Phi_{\epsilon}(t-\alpha)\to \text{I}(t>\alpha)$ and $\Phi_{\epsilon}(1-\alpha-t)\to \text{I}(t<1-\alpha)$ for every $t\in[0,1]$ as $\epsilon\to 0$. OW is the special case of BW when $\nu_1=\nu_2=2$. 
\end{tablenotes}
\end{table}

\begin{figure}
    \centering 
    \includegraphics[width=\textwidth]{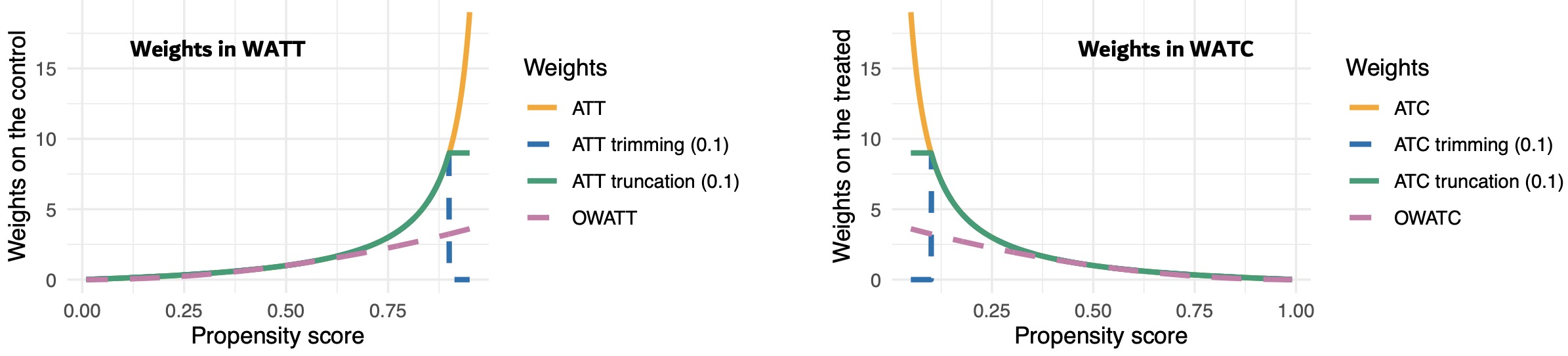}
    \caption{The PS weights used in WATT, i.e., $\omega_{0g}(\mb x)$ (the left panel) and WATC, i.e., $\omega_{1g}(\mb x)$ (the right panel). {\scriptsize [For OWATT and OWATC, we actually plot $4\cdot \omega_{0g}(\mb x)$ and $4\cdot\omega_{1g}(\mb x)$ to compare the curves{; this scaling does not affect the estimand, as any constant factor cancels out in the weighting definitions given in} \eqref{eq:WATTform} and \eqref{eq:WATCform}.]} }
    \label{fig:tiltPS}
\end{figure}

We compare some of the PS weights  $\omega_{0g}(\mb x)$ and $\omega_{1g}(\mb x)$, in Figure \ref{fig:tiltPS}, with $\alpha = 0.10$ for trimming and truncation. Such a comparison is important as it (i) helps illustrate how different tilting functions influence the relative weighting of control (or treated) participants in WATT (or WATC) estimands---thereby affecting the interpretation of these estimands; (ii) provides insights into the stability and variability of the estimation (more details in Section \ref{sec:tutorial}), and (iii) highlights differences in efficiency between estimands, guiding the choice of appropriate weighting schemes in practice. 

For $\omega_{0g}(\mb x)$ in WATT, the ATT weights increase to infinity as the PS approaches 1, ATT trimming (resp. truncation) sets the weights to 0 (resp. a non-zero constant) for participants with PS $\geq 1-\alpha = 0.9$. Only OWATT assigns different non-zero, monotonically increasing, and bounded weights as the PSs increases. Similar trends can be seen with $\omega_{1g}(\mb x)$ as the PS decreases toward 0. Unlike trimming and truncation, both OWATT and OWATC do not require the selection of any specific threshold parameter This property simplifies their implementation and make them better, data-driven pruning tools. Moreover, when the positivity is satisfactory (e.g., $e(\mb x)<0.8$ for WATT or $e(\mb x)>0.2$ for WATC), OWATT (resp. OWATC) produces almost the same weight curve as that of ATT (resp. ATC). In this case,  ATT and WATT as well as ATC and WATC are expected to be similar. On the other hand, when the positivity assumption is violated, we know that conventional ATT and ATC can be inefficient. However, due to the boundedness of their weights, both OWATT and OWATC are shown to be efficient with internal validity.\cite{liu2024average}

We also consider the titlting functions of MW, EW and BW in the context of WATT and WATC, as shown in Table \ref{tab:TiltWATT}. We  expect them to produce similar weight curves as those with OWATT and OWATC, respectively, since their tilting functions  $g(\mb X)$   share similar features as shown in Figure \ref{fig:tiltWATE_PS} and demonstrated by Matsouaka and Zhou.\cite{matsouaka2024causal}

\subsection{Binary outcomes}\label{subsec:binout}

Many real-world applications involve binary outcomes. For example, studies that focus on the social and medical determinants of people living with HIV,\cite{poteat2020prep, poteat2024transgender, poteat2022stigma, poteat2025social} often consider HIV status as binary response (with 1 if HIV-negative and 0 otherwise). 
 {However, extensions of the WATE, WATT, and WATC frameworks to relative effect measures (beyond the weighted difference) for binary outcomes are limited. For example, methodological developments that focus on the risk ratio (RR) or odds ratio (OR) are sparse, yet the RR and OR often offer additional interpretability and clinical relevance for binary outcomes.}\cite{austin2017estimating, austin2018assessing, boughdiri2025unified} 
In this section, we extend these estimands to binary outcomes and define the weighted RR and OR by leveraging such versatile frameworks. For the WATE, we have
\begin{align*}
    \tau^{\text{wate}} & = \frac{\Ex\{h(\mb X)\tau(\mb X)\}}{\Ex\{h(\mb X)\}} = \frac{\Ex\{h(\mb X)\tau_1(\mb X)\}}{\Ex\{h(\mb X)\}} - \frac{\Ex\{h(\mb X)\tau_0(\mb X)\}}{\Ex\{h(\mb X)\}} := p_1^h-p_0^h,
\end{align*}
where $\tau_a(\mb X) = \Ex\{Y(a) \mid \mb X\}$ is the conditional mean potential outcome under treatment $A = a$. \\
Since $Y(a) \in \{0, 1\}$,  $\tau_a(\mb X) \in [0, 1]$, thus $p_a^h \in [0, 1]$ is a weighted average of $\tau_a(\mb X)$, with weight $h(\mb X)/\Ex\{h(\mb X)\}$ for $a = 0, 1$. Therefore, both $p_0^h$ and $p_1^h$ are proportions in a weighted population. 

Similarly, using $p_a^h$ and $p_a$ , we can define the WATT and WATC estimands as 
\begin{align*}
    \tau^{\text{watt}} & = \frac{\Ex(AY)}{\Ex(A)} - \frac{\Ex\{(1-A)\omega_{0g}(\mb X)Y\}}{\Ex\{(1-A)\omega_{0g}(\mb X)\}} := p_1-p_0^g,\\
    \tau^{\text{watc}} & = \frac{\Ex\{A\omega_{1g}(\mb X)Y\}}{\Ex\{A\omega_{1g}(\mb X)\}} - \frac{\Ex\{(1-A)Y\}}{\Ex\{1-A\}} := p_1^g-p_0. 
\end{align*}
We can also verify that all $p_0^g, p_1^g, p_0, p_1\in[0,1]$ are weighted proportions. 
Thus, we derive the weighted RR and weighted OR  for the WATE, WATT, and WATC classes defined in Table \ref{tab:weightedRR}.  

\begin{table}[!htbp]
    \centering
    \begin{tabular}{rlllll}
    \toprule
       & WATE & WATT & WATC \\
       \midrule
       Weighted RR & $p_1^h/p_0^h$ & $p_1/p_0^g$ & $p_1^g/p_0$  \\
       Weighted OR & $p_1^h(1-p_0^h)/\{p_0^h(1-p_1^h)\}$ & $p_1(1-p_0^g)/\{p_0^g(1-p_1)\}$ & $p_1^g(1-p_0)/\{p_0(1-p_1^g)\}$  \\
       \bottomrule
    \end{tabular}
    \caption{Definition of weighted RR and weighted OR}
    \label{tab:weightedRR}
\end{table}

\section{Tutorial for Data Analysis}\label{sec:tutorial}

Building on the preliminary exposition from the previous sections, we present a practical tutorial for PS analysis, tailored to a chosen estimand from the WATE, WATT, or WATC class and guided by the researcher's scientific questions. We thus recommend that the choice of the class of estimands (WATE, WATT, or WATC) be made in advance. This can be informed by the scientific questions, {the assessment of possible violation of positivity (and its type), and subject-domain knowledge.} The refinements related to each specific estimand one can choose within a given class should be dictated by the data at hand. 

\begin{figure}
    \centering
    \includegraphics[width=0.65\linewidth]{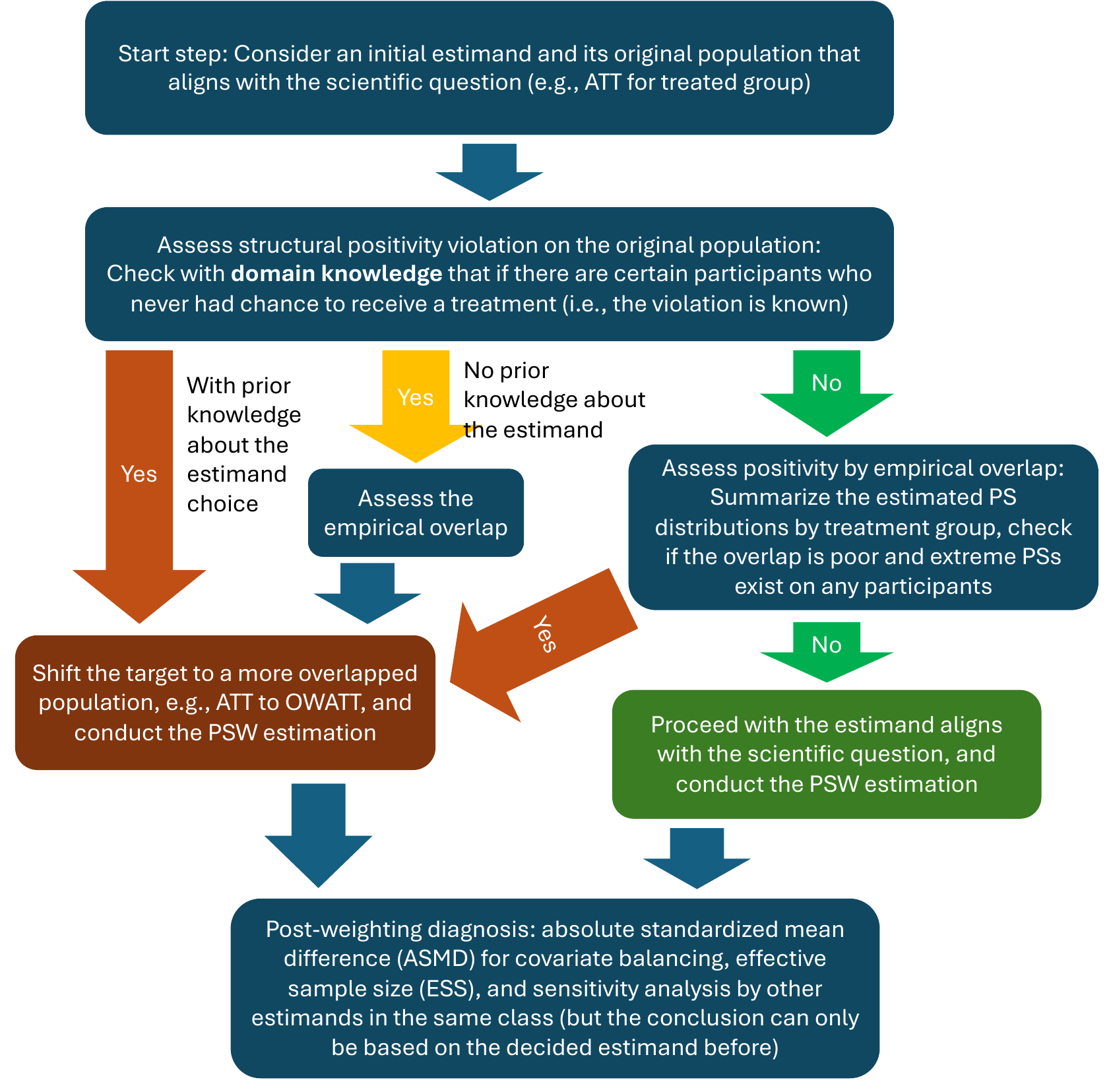}
    \caption{{Decision-tree flow of the tutorial; it help decide which estimand to choose based on the scientific question(s) of interest and the assessments of the positivity assumption.} }
    \label{fig:flow}
\end{figure}

{An important point to consider is \textit{how to select an estimand that best aligns with the underlying scientific question while ensuring interval validity and robustness of estimation.} In Figure}~\ref{fig:flow}{, we present a step-by-step decision tree as a practical guideline to conduct a comprehensive analysis that integrates estimand selection along with the assessment of the positivity assumption. The process consists of three key steps: (i) define the estimand that conceptually matches the scientific question—for example, ATT for the treated population or ATE trimming for a restricted subgroup of the overall population; (ii) evaluate whether the target should be shifted due to potential violations of positivity, prioritizing structural violations identified through domain knowledge, followed by empirical assessment of overlap; and (iii) after the target and estimand are determined, perform post-weighting diagnostics and sensitivity analyses. The final inference on the treatment effect should be based solely on the chosen estimand, while alternative estimands should be interpreted only as supplementary sensitivity analyses rather than competing primary results.} 

\begin{remark}\label{rmk:flow}
{In the step highlighted by the dark-red box in Figure} \ref{fig:flow}{, further consideration is warranted when deciding which estimand to select, after we have determined that a target shift is necessary. When overlap is severely limited---such that many participants have propensity scores near 0 or 1---the OW-type estimands (ATO, OWATT, or OWATC) serve as suitable alternatives to their counterparts (ATE, ATT, or ATC) due to their desirable statistical properties discussed earlier. However, as noted by Austin et al.,}\cite{austin2023differences} {in this context, these OW-type estimands may yield values that differ substantively from the true ATE, ATT, or ATC. Hence, we should not expect or consider that the OW-type estimands will be closer to either of them. In contrast, when the lack of overlap is modest---e.g., only a small fraction of participants has extreme weights, likely driven by random rather than structural violations of the positivity assumption---trimming or truncation with a small $\alpha$ parameter may be preferable, as it improves estimation efficiency while maintaining closer alignment with the original estimand. In practice, the choice of $\alpha$ should balance the sample size after trimming or truncating as well as the exclusion of participants with extreme weights. We must acknowledge that the literature lacks clear quantitative criteria for making an optimal decision when it comes to selecting $\alpha$. The use of effective sample size is a step in that direction.}\cite{matsouaka2024causal,zhang2024three} {Developing principled, data-driven approaches to guide the selection of estimands and target populations under varying degrees of positivity violations represents an important direction for future research.}
\end{remark}

{In addition, for estimation and inference on the chosen estimand, we} consider the following PS-weighted (PSW) estimators derived based on equations \eqref{eq:WATEform}, \eqref{eq:WATTform}, and \eqref{eq:WATCform}. 
\allowdisplaybreaks\begin{align}\label{eq:pswEst}
    \widehat\tau^{\text{wate}} & = \frac{\displaystyle\sum_{i=1}^N A_i\widehat\omega_{1h}(\mb X_i)Y_i}{\displaystyle\sum_{i=1}^N A_i\widehat\omega_{1h}(\mb X_i)}-\frac{\displaystyle\sum_{i=1}^N (1-A_i)\widehat\omega_{0h}(\mb X_i)Y_i}{\displaystyle\sum_{i=1}^N (1-A_i)\widehat\omega_{0h}(\mb X_i)}, ~~\text{ with }~~\widehat\omega_{1h}(\mb X_i) = \frac{A_i\widehat h(\mb X_i)}{\widehat e(\mb X_i)}  ~~\text{ and }~~\widehat\omega_{0h}(\mb X_i)= \frac{(1-A_i)\widehat h(\mb X_i)}{(1-\widehat e(\mb X_i))},\nonumber\\
    \widehat\tau^{\text{watt}} & = \frac{\displaystyle\sum_{i=1}^N A_iY_i}{\displaystyle\sum_{i=1}^NA_i} - \frac{\displaystyle\sum_{i=1}^N (1-A_i)\widehat\omega_{0g}(\mb X_i)Y_i}{\displaystyle\sum_{i=1}^N(1-A_i)\widehat\omega_{0g}(\mb X_i)}, ~~~ \text{ and }~~~ \widehat\tau^{\text{watc}}  =  \frac{\displaystyle\sum_{i=1}^N A_i\widehat\omega_{1g}(\mb X_i)Y_i}{\displaystyle\sum_{i=1}^NA_i\widehat\omega_{1g}(\mb X_i)}-\frac{\displaystyle\sum_{i=1}^N (1-A_i)Y_i}{\displaystyle\sum_{i=1}^N(1-A_i)},\\
   \text{ where }~~& \widehat\omega_{0g}(\mb X_i)  = \frac{\widehat g(\mb X_i)\widehat e(\mb X_i)}{1-\widehat e(\mb X_i)},~~ \text{ and } ~~ \widehat\omega_{1g}(\mb X_i) = \frac{\widehat g(\mb X_i)(1-\widehat e(\mb X_i))}{\widehat e(\mb X_i)}. \nonumber
\end{align}
where $\widehat e(\mb X_i)$,  $\widehat h(\mb X_i)$, and $\widehat g(\mb X_i)$ are respectively estimators of  $e(\mb X_i)$, $h(\mb X_i),$ and $g(\mb X_i),$ for  $i=1, \dots N.$ 

The resulting PSW estimators are then interpreted within the context of the specified tilting function, $h(\mb X)$ or $g(\mb X)$, and the target estimand (WATE, WATT, or WATC).
{We summarize the technical procedure as follows:} 
\allowdisplaybreaks\begin{itemize} 
\item \textbf{Step 1:} Estimate the PS using the entire dataset $\mc O =  \{(A_i, \mb{X}_i, Y_i) : i = 1, \dots, N\}$. The chosen PS model  can a logistic regression model or algorithms from \texttt{SuperLearner}.\cite{van2007super}
\item \textbf{Step 2:} {Decide on the causal estimand of interest, based on the decision-tree in Figure }\ref{fig:flow}. 
\item \textbf{Step 3:} 
Compute the corresponding PSW estimate, using the estimators in \eqref{eq:pswEst}, with $\widehat e(\mb X)$ the estimated PS from Step 1.  
\end{itemize}
{Both point and variance estimations can be conveniently implemented using our accompanying R package, \texttt{ChiPS}. }

\begin{remark}\label{rmk:inference}  The uncertainty associated with the PSW estimators \eqref{eq:pswEst} arises from two specific sources, which pertain to how we (i) estimate the PS and (ii) evaluate the estimand itself. There are two primary methods to estimate the variance corresponding to the point estimate that takes into account those two sources of uncertainty: the nonparametric bootstrap method \cite{efron1994introduction} and the empirical sandwich variance estimator method. \cite{lunceford2004stratification, matsouaka2024overlap} In this tutorial, we adopt the bootstrap method because it is model-agnostic, accommodating any PS models that users may choose in Step 1 of the framework. In contrast, the sandwich variance estimator's formula varies with changes in the PS model, making it less adaptable to different models. In addition, its exact form is unclear when we deal with machine learning algorithms. 

To simplify, let $\widehat\tau$ denote a generic  estimated WATE, WATT or WATC. With $B$ bootstrap replicates sampled with repetition from data $\mc O$, we  calculate the bootstrap estimates $\mc B =\{\widehat\tau_1,\dots,\widehat\tau_B\}$ and derive the standard bootstrap variance as 
\begin{align*}
V^{\text{boot}}(\widehat\tau) = \frac1B\sum_{b=1}^B(\widehat\tau_b-\bar{\widehat\tau}_B)^2, ~~\text{where} ~~\bar{\widehat\tau}_B=\frac1B\displaystyle\sum_{b=1}^B\widehat\tau_b,
\end{align*} 
from which we construct the 95\% confidence interval (CI) as 
$ \widehat\tau\pm 1.96\cdot\sqrt{\widehat V^{\text{boot}}(\widehat\tau)}.$
\end{remark}
\begin{remark}\label{rmk:boot}
The rationale for such variance estimation follows from semi-parametric theory. Under some regularity conditions (see van der Vaart\cite{van2000asymptotic}), the PSW estimator \eqref{eq:pswEst} is regular and asymptotically linear (RAL), which implies it is also asymptotically normally distributed. Thus, the bootstrap variance of the PSW estimator approximates the asymptotic variance  consistently.\cite{shao2012jackknife} 
\end{remark}

The same arguments also applies to the weighted RR and weighted OR estimations for binary outcomes. However, due to the skewed nature of their distributions, we suggest two possible alternative bootstrap procedures we can consider: 
\begin{enumerate}
    \item Use the quantile method to construct the 95\% CI, in lieu of the standard bootstrap variance estimation. With $\mc B = \{\widehat\tau_1,\dots,\widehat\tau_B\}$, the 95\% quantile bootstrap CI is  [$\mc B_{0.025},\mc B_{0.975}$], for $\mc B_q$ the $q$th-quantile of set $\mc B$, where $q\in \{0.025,0.975\}$. 
    \item Apply the log transformation to the weighted RR or OR, followed by the standard bootstrap variance estimation, as the log transformation often improves normal approximation.\cite{haldane1956estimation, west2022best} Let $\log\widehat\tau$ denote the estimated log weighted RR or OR. We first calculate the bootstrap estimates $\{\log\widehat\tau_1,\dots,\log\widehat\tau_B\}$. Then, we derive the bootstrap variance $$\widehat V^{\text{boot}}\left(\log\widehat\tau\right) = B^{-1}\displaystyle\sum_{b=1}^B\left(\log\widehat\tau_b-\overline{\log\widehat\tau}_B\right)^2,~~ \text{ where } ~~\overline{\log\widehat\tau}_B=B^{-1}\displaystyle\sum_{b=1}^B\log\widehat\tau_b.$$
\end{enumerate}
The 95\% CI for $\tau$ is given by
$\widehat\tau\exp\left\{\pm 1.96\cdot\sqrt{\widehat V^{\text{boot}}(\log \widehat\tau)}\right\}$. 

\subsection{R code example}\label{subsec:R}

In this section, we demonstrate how to use the proposed tutorial. To facilitate its rapid adoption and use, we have developed an R package, \texttt{ChiPS} (\url{https://github.com/yiliu1998/ChiPS}) that specifically helps implement the proposed methods. The R package can be installed using the following lines of code:
\begin{lstlisting}
if (!require("devtools")) install.packages("devtools")
devtools::install_github("yiliu1998/ChiPS")
\end{lstlisting}
The core functions of the package are \texttt{ChiPS()} for continuous outcomes and \texttt{ChiPS\_bin()} for binary outcomes. 

Below, we show how to use the package with an example call to \texttt{ChiPS()}. We first generate two independent covariates $X_1$ and $X_2$ that follow the normal distributions $\mc N(0,1)$. We then generate the treatment $A$ using a logistic regression model and the potential outcomes $(Y(0), Y(1))$ using linear models. The observed outcome is generated as $Y=AY(1)+(1-A)Y(0)$.  

\begin{lstlisting}
### Generate the observational data
n <- 1000 # sample size
X1 <- X2 <- rnorm(n) # two covariates
X <- cbind(X1, X2)
expit <- function(x) 1/(1+exp(-x))
A <- rbinom(n, 1, prob=expit(-1+0.2*X1+0.3*X2)) # treatment
Y0 <- 10 + 0.3*X1^2 + 0.8*X2^3 + rnorm(n)
Y1 <- (4+X1^2+X2^2)*A + Y0
Y <- A*Y1 + (1-A)*Y0 # observed outcome
\end{lstlisting}

Next, we run the analysis using the following code:

\begin{lstlisting}
library(ChiPS)
result <- ChiPS(A, Y, X,
                ps.library="SL.glm",
                beta=TRUE, v=c(2,4),
                trim=TRUE, trim.alpha=c(.05,.1),
                trun=TRUE, trun.alpha=c(.05,.1),
                boot=TRUE, n.boot=200,
                seed=4399,
                conf.level=.05)
\end{lstlisting}

In the argument specification of the \texttt{ChiPS()} function, \texttt{A} denotes the binary treatment vector, \texttt{Y} the observed outcome vector, and \texttt{X} the covariate matrix. The \texttt{ps.library} argument is to specify the PS model. To see the full list of available model options, use \texttt{SuperLearner::listWrappers()} in R. Here, we choose the option \texttt{"SL.glm"}, which corresponds to the logistic regression model with \texttt{X} as covariate matrix. This argument could also be a vector, such as \texttt{c("SL.glm", "SL.randomForest")}, meaning that an ensemble learning by logistic regression and random forest will be used to estimate the PSs. 

The function \texttt{ChiPS()} returns the results for multiple WATE, WATT, and WATC estimands. By default, the WATE class includes ATE, ATT, ATC, ATO, ATM, and ATEN; the WATT and WATC classes include ATT, OWATT, MWATT, EWATT and ATC, OWATC, MWATC, EWATC, respectively. To obtain additional estimates based on BW, trimming, or truncation, users must specify \texttt{beta=TRUE}, \texttt{trim=TRUE}, or \texttt{trun=TRUE}, respectively. In this example, we illustrate the results for BW weights with $\nu = 2, 4$ via \texttt{v=c(2,4)}, for trimming with $\alpha = 0.05, 0.1$ via \texttt{trim.alpha=c(.05,.1)}, and for truncation with $\alpha = 0.05, 0.1$ via \texttt{trun.alpha=c(.05,.1)}. The argument \texttt{boot} controls whether bootstrap-based variance and interval estimates are returned. If \texttt{boot=FALSE}, only point estimates are provided. The argument \texttt{n.boot} specifies the number of bootstrap replicates; in this example, we set \texttt{n.boot=200} for illustration. 

The function \texttt{ChiPS()} also returns a list that comprises three data frames, each containing the PSW estimation and inference results for the WATE, WATT, and WATC estimands, respectively. For brevity, we only present here the results for the WATE estimands; the outputs for WATT and WATC follow the same format.

\begin{lstlisting}
> result$df.WATE
                             Est   Std.Err      Upr      Lwr
overall                 6.132390 0.2050156 6.534213 5.730566
treated                 6.226569 0.3561650 6.924640 5.528499
control                 6.096214 0.1659719 6.421513 5.770915
overlap                 6.016689 0.2430086 6.492978 5.540401
matching                6.128316 0.2857209 6.688318 5.568313
entropy                 6.030264 0.2320513 6.485076 5.575451
beta (v=2)              6.016689 0.2430086 6.492978 5.540401
beta (v=4)              6.110358 0.3350225 6.766990 5.453726
trimming (alpha=0.05)   6.132390 0.2087713 6.541574 5.723205
trimming (alpha=0.1)    6.049817 0.2274571 6.495625 5.604010
truncation (alpha=0.05) 6.132390 0.2032605 6.530773 5.734006
truncation (alpha=0.1)  6.200870 0.2147957 6.621862 5.779878
\end{lstlisting}

For complete details on how to use the R package, users can enter \texttt{?ChiPS()} or \texttt{?ChiPS\_bin()} in the R console (after installing the package) to view the full documentation.

\section{Simulation Study}\label{sec:simu}

In this section, we use a Monte Carlo simulation study to compare the performances of the PSW estimator for different estimands. For reproducibility purpose, we first detail our data generating processes (DGPs) in Table \ref{tab:DGPsum}. We began by generating the baseline covariates $X_1$--$X_7$, where $X_1$--$X_2$ are binary and $X_3$--$X_4$ are correlated continuous covariates. Next, we generated $X_5$--$X_7$ as higher-order terms from $X_1$--$X_4$, which we later incorporated into the true PS model. In practice, when modeling the PS, data analysts may include only the main terms while omitting higher-order terms. To examine the impact of such misspecification, we consider a scenario where the PS model is intentionally misspecified, by excluding these higher-order terms, and evaluate how different estimators perform. 

\begin{table}[h]
    \centering
    \begin{tabular}{rllllllll}
    \toprule
      Covariates $\mb X$   &  \makecell[l]{${X_1}\sim \text{Bern}(0.5)$,\\  
      ${X_2}\sim \text{Bern}(0.4+0.2{X_1})$, \\ 
      ${(X_3,X_4)}\sim\mc N_2(\bd\mu,\bd\Sigma)$, a bivariate normal where \\    
    $
    \bd\mu = \begin{pmatrix}
    X_1-0.25X_2+X_1X_2 \\ -0.25X_1+X_2+X_1X_2 
    \end{pmatrix}$ and 
    $\bd\Sigma = X_2\begin{pmatrix}
    1&0.5\\0.5&1
    \end{pmatrix} + 
    (1-X_2)\begin{pmatrix}
    2&0.25\\0.25&2
    \end{pmatrix}
    $, \\  $X_5=X_3^2$, $X_6=X_3X_4,$ and $X_7=X_4^2.$ }\\
    \midrule
     PS and $A$  & \makecell[l]{$e(\mb X) = \text{expit}[{(-0.4X_1-0.4X_2-0.4X_3-0.4X_4-0.1X_5+0.1X_6+0.1X_7)}\cdot\gamma-\alpha_{0}]$,$^\dagger$ \\ where expit$(u)=\{1+\exp(-u)\}^{-1}$, with {$(\gamma,\alpha_0)$} $\in\{(0.5,0.407), (2.5,2.074)\}$, \\ leading to 2 different PS models, corresponding to good and poor PS overlap, respectively. \\
     {The true PS densities by treatment group are presented in Figure} \ref{fig:simPS}. \\
     Then, $A\sim \text{Bern}(e(\mb X))$. } \\
    \midrule
    $Y(0)$ & $ = {0.5-1.2X_1+2.2X_2+X_3+0.6X_4+X_5+2X_6+X_7}+\varepsilon$, with $\varepsilon\sim\mc N(0,2^2)$, and \\ 
    $Y(1)$ & $ = Y(0)+{4+X_2X_3+3X_5+6X_6+3X_7}$.
    \\
    \midrule
    $Y$ & $=Y(A)=AY(1)+(1-A)Y(0).$ \\
    \midrule
    & Correctly specified model: includes all $X_1$--$X_7$ in a logistic regression model \\
    \makecell[r]{PS model \\ specification} & \makecell[l]{Misspecified model: includes only $X_1$--$X_4$ in a logistic regression model \\ $\qquad\qquad\qquad\qquad$ (omitting the higher-order terms $X_5$--$X_7$)} \\
     & Methods ensemble by super learner: \texttt{SL.glm}, \texttt{SL.glm.interaction}, \texttt{SL.glmnet}, and \texttt{SL.gam} \\
    \bottomrule
    \end{tabular}
    \begin{tablenotes}\footnotesize
        \item {$\dagger$: For the PS model, $\alpha_0$ denotes the intercept of the logistic regression model, and $\gamma$ is a scaling parameter for the slope coefficients. A larger $\gamma$ reduces PS overlap, while $\alpha_0$ controls the proportion of treated participants under a given overlap scenario. }
    \end{tablenotes}
    \caption{Data generating processes (DGPs) and PS model specifications in simulation}
    \label{tab:DGPsum}
\end{table}

\begin{figure}[h]
    \centering
    \includegraphics[width=0.75\linewidth]{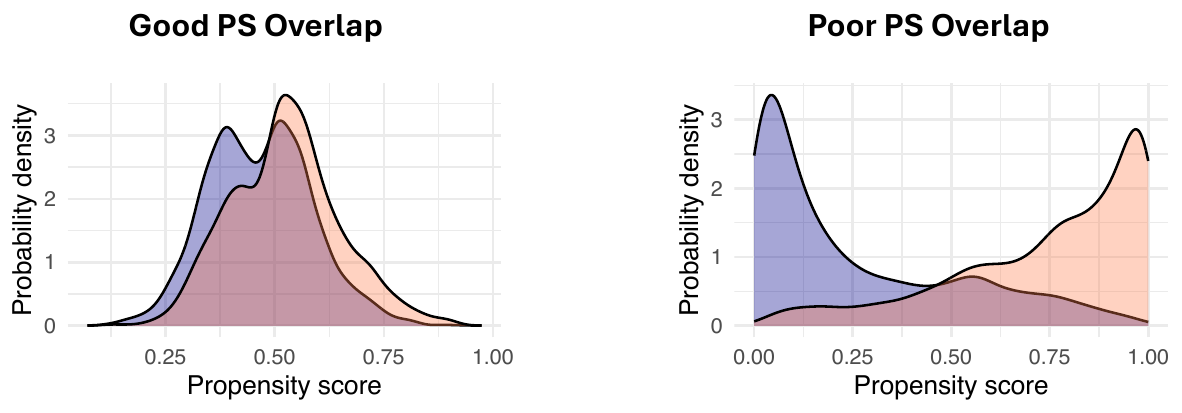}
    \caption{{Example of a density plot of propensity score by treatment group, based on the simulation data generating process in Table} \ref{tab:DGPsum}. {Left panel: $(\gamma,\alpha_0) = (0.5,0.407)$; Right panel: $(\gamma,\alpha_0) = (2.5,2.074)$.}}
    \label{fig:simPS}
\end{figure}

Under each DGP, we generate two sets of Monte Carlo data. The first set used as super population (to reduce sampling variability) based on $N = 10^7$ independent participants, where $Y(1)$, $Y(0)$, and the true PSs are generated for each participant. By replacing ``$\Ex$'' in the WATE, WATT and WATC estimands by the ``empirical average'' over the $10^7$ participants, we can view the calculated values as the true values. The values of estimands from the super population serve as benchmarks; we use them to compare and contrast the results we obtain from the second set of data. For the second set of data, we generated in $M=1000$ independent Monte Carlo simulation replicates. Within each replicate, we generated observed data of size $N=2000$, where only $(A,Y,\mb X)$ are provided for each participant. We estimated the causal estimands and their variances on these $M$ replicates using the corresponding PSW estimators in \eqref{eq:pswEst} and $B=200$ bootstrap replicates. 

\subsection{Weights, competing methods and performance metrics}\label{subsec:performance}

We consider the following estimands and compare their corresponding estimators: 
\begin{itemize}
    \item WATE class: ATE, ATT, ATC, ATO, ATEN, ATM, ATB, ATE trimming, ATE truncation
    \item WATT class: ATT, OWATT, EWATT, MWATT, BWATT, ATT trimming, ATT truncation 
    \item WATC class: ATC, OWATC, EWATC, MWATC, BWATC, ATC trimming, ATC truncation 
\end{itemize}
For beta weights, we consider $\nu_1=\nu_2=\nu\in\{3,10\}$, and for all trimming and truncation, we consider $\alpha\in\{0.05,0.1,0.15\}$. 
The different estimators are compared based on the following measures and metrics (based on $M=1000$ simulated data sets): 
\begin{enumerate}[label=(\arabic*)]
\item The relative percent biases RBias\%$ =\vert 100\%*(\widehat\tau-\tau)/\tau\vert$,  $\tau$ is the true value of an estimand and $\widehat\tau$ is its PSW estimator; \label{sim:item2}
\item The coverage probability (CP), i.e., the proportion of the number of times $\tau$ falls within the 95\% CI. A CP is considered significantly deviating from the 95\% nominal level if it falls outside of $0.95\pm1.96\sqrt{0.95\times0.05/1000}=[0.937, 0.964]$. \label{sim:item3}
\end{enumerate}

\subsection{Results}\label{subsec:simuResu}

\begin{figure}[ht]
    \centering
    \includegraphics[width=0.9\linewidth]{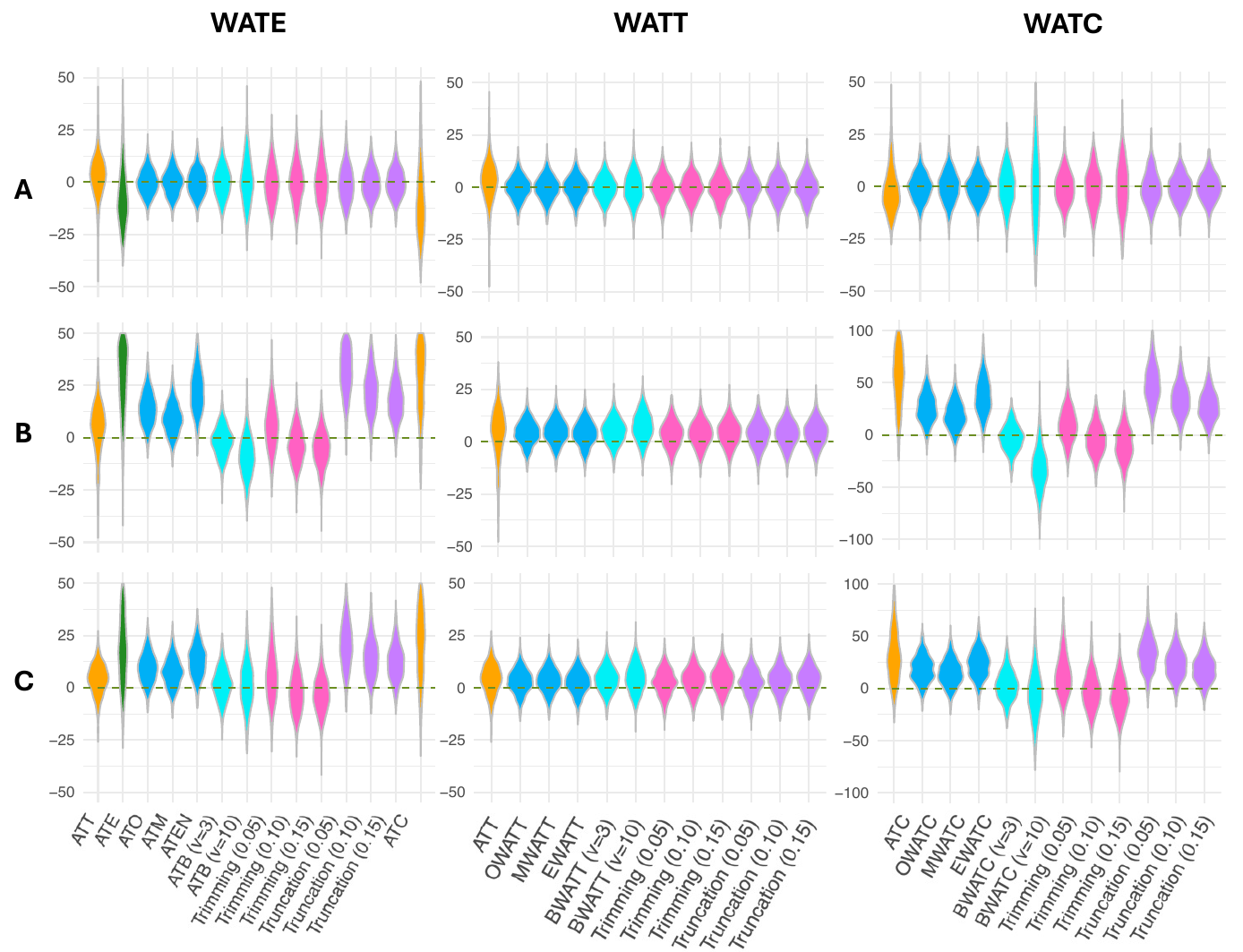}
    \caption{{Simulation results for estimation bias by violin plots under poor overlap.} }
    \label{fig:simBiaspoor}
\end{figure}

\begin{figure}[ht]
    \centering
    \includegraphics[width=0.75\linewidth]{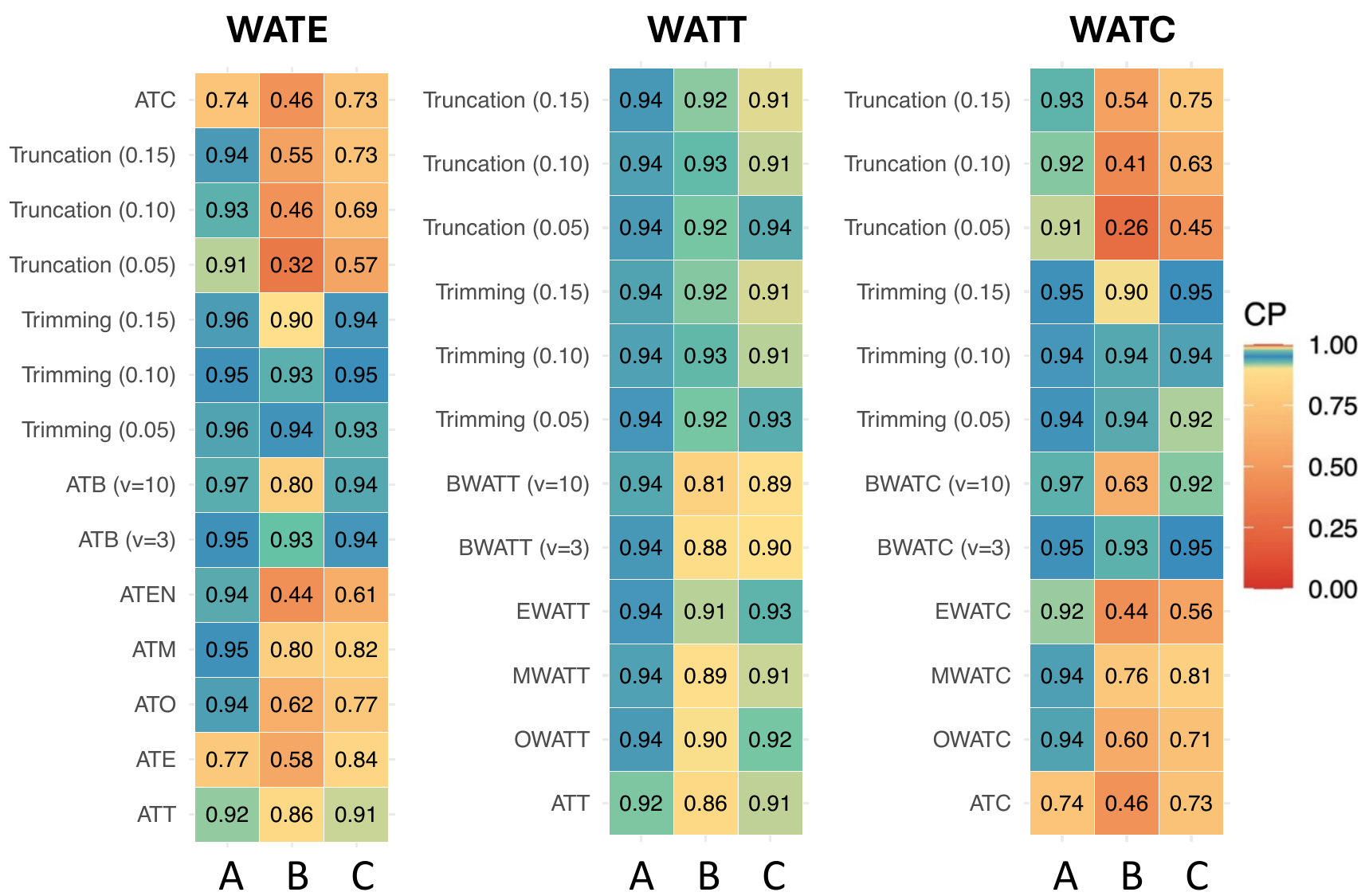}
    \caption{{Simulation results for coverage probability under poor overlap.}}
    \label{fig:simCPpoor}
\end{figure}

{For brievity, in this section, we only present the simulation results under poor overlap; we defer the results under good overlap to the Online Supplemental Material. Both Figures} \ref{fig:simBiaspoor} and \ref{fig:simCPpoor} {illustrate the performance of the PSW estimators for different estimands under poor overlap.} We use the following color panel to distinguish different types of estimands: (i) \textcolor{Orange}{orange} for the \textcolor{Orange}{ATT and ATC}; (ii)  \textcolor{ForestGreen}{green} for the \textcolor{ForestGreen}{ATE} (so only the WATE figures have this color); (ii) \textcolor{Blue1}{blue} for the estimands defined by \textcolor{Blue1}{OW, MW and EW} (since they have similar tilting functions as shown in Figure \ref{fig:tiltPS} and are often considered as ``close friends''\cite{zhou2020propensity, matsouaka2024overlap}); (iii) \textcolor{Blue2}{cyan} for the estimands defined by the two chosen \textcolor{Blue2}{BWs}; (iv) \textcolor{Pink1}{pink} for the three \textcolor{Pink1}{trimmings}; and finally (v) \textcolor{Purple1}{purple} for the three \textcolor{Purple1}{truncations}. The use of these colors helps better distinguish visually estimands that have similar interpretation and traits on their tilting functions, even where within each type, from the other estimands that are completely different. 

{Figure} \ref{fig:simBiaspoor} {shows} the violin plots of RBias\% for the three model specifications indicated in Table \ref{tab:DGPsum} {under poor overlap}. Across all estimands {(or weighting methods)}, those defined by OW, MW, and EW (e.g., ATO, OWATT, OWATC) are more stable with lower RBias\%. Notably, even with a correctly specified PS model (Case A), ATE, ATT, ATC, ATB, BWATC (especially with $\nu=10$), ATE trimming, and ATC trimming (at $\alpha=0.05, 0.1, 0.15$) exhibit inefficiencies, likely due to extreme PS weights induced by the poor PS overlap. 

{Figure} \ref{fig:simCPpoor} displays the CP heat maps {under poor overlap}. For the correctly specified PS model (Case A), all estimands---except ATE, ATT, and ATC---achieve valid CPs near 0.95. This indicates that, even when point estimates are consistent, inference for conventional estimands can be unreliable due to extreme weights. When the PS model is misspecified or estimated via ensemble learning (Cases B and C), most CPs deviate from 0.95, primarily due to estimation bias by model misspecification. 

Comparing model specifications, only a correctly specified PS model ensures consistent estimates. While the super learner ensemble (Case C) can improve some of the RBias\%'s and CPs relative to a misspecified model (Case B), e.g., the CPs of {ATE, ATT, ATC,} truncations in WATE and WATC ($\alpha=0.05$, 0.1 and 0.15), {OWATT, MWATT, EWATT, OWATC, MWATC and EWATC}, it cannot fully restore consistency or valid coverage, even though it includes the correctly specified PS model (\texttt{SL.glm.interaction}) in the ensemble learning. 

{Finally, we briefly comment on the results for good overlap in the Online Supplemental Material. Overall, the findings under model misspecifications are similar to those of poor overlap. However, because of higher overlap and less violations of positivity, the performances of all weighting methods perform better (except for beta weights when $\nu=10$, ATC and ATC trimmings under PS model misspecification). }

\section{Data Applications}\label{sec:data}

To illustrate our tutorial to real-world studies, we considered two data applications. The first data, from the 2007--2008 U.S. National Health and Nutrition Examination Survey (NHANES),\cite{hsu2013calibrating, yang2018asymptotic} provide health-related information on $N=3,340$ participants. We use the data to evaluate the effects of smoking on blood lead level. The second data are survey data from an HIV/AIDS prevention program. The original study assessed awareness and engagement with Pre-Exposure Prophylaxis (PrEP) among transgender women in South Africa.\cite{poteat2020prep, poteat2025social} In this paper, we investigate the effect of sex work on HIV status among $N=196$ transgender women. {To illustrate our tutorial and reflect its intended purpose, we follow strictly the decision flow outlined in Figure}~\ref{fig:flow} {and the estimation procedure described in Section}~\ref{sec:tutorial} {for all analyses presented in this section.} 

{For both studies, we first specify one primary estimand and present other estimands within the same class as supplementary results. For example, if ATT is chosen as the primary estimand, other estimands in the WATT class are provided as supplements. Nevertheless, for completeness, all results under the WATE, WATT, and WATC frameworks are included in the Online Supplemental Material. }

{To assess the quality of covariate balance and the precision of the weighted estimators, we report the effective sample size (ESS) and absolute standardized mean difference (ASMD) for each weighting method. The ESS quantifies the amount of information retained after weighting and is defined as}\cite{matsouaka2024overlap, li2018balancing}
$$
\text{ESS} = \frac{\left(\sum_i w_i\right)^2}{\sum_i w_i^2},
$$
{where $w_i$ denotes the PS weight for participant $i$. A smaller ESS indicates higher variability in the weights and potential loss of precision.} 

{To evaluate covariate balance, the ASMD for covariate $X_j$ is calculated as}\cite{austin2009balance}
$$
\text{ASMD}_j = 
\frac{\big|\bar{X}_{1j}^{(w)} - \bar{X}_{0j}^{(w)}\big|}
{\sqrt{(s_{1j}^2 + s_{0j}^2)/2}},
$$
{where $\bar{X}_{aj}^{(w)}$ is the weighted mean of covariate $X_j$ in treatment group $A=a$ ($a=0,1$), and $s_{aj}^2$ is the corresponding sample variance. Smaller ASMD values indicate improved covariate balance, with values below an empirically recommended threshold 0.1}\cite{austin2009balance} {generally considered acceptable.}

\subsection{Study I: Effects of smoking on blood lead level}\label{subsec:studyI}

{The NHANES 2007–2008 data were collected using a complex, multistage probability sampling design to ensure national representativeness of the U.S. civilian, noninstitutionalized population. Each participant was assigned a sample weight reflecting their selection probability to population totals. In this analysis, we included the 2-year Mobile Examination Center weight (\texttt{WTMEC2YR}) as a survey-weight covariate in the PS model to account for unequal sampling probabilities when estimating the probability of smoking. The normalized survey weight was also used to construct a survey-weighted outcome (weighted blood lead level, in $\mu$g/dl) when conducting PS weighting analyses, ensuring that both treatment assignment and outcome components reflect the NHANES survey design and population representation. This approach follows that of Liu et al.},\cite{liu2025assessing} {who conducted similar PS analyses using NHANES data.}

The data consist of 3,340 participants, with 679 (20.33\%) smokers (exposed group, $A=1$). {In addition to the survey-weight variable \texttt{WTMEC2YR}, we} included various demographic and health-related covariates in the PS model, such as age, income-to-poverty ratio, gender, education level, and race to estimate the PSs. We use a super learner ensemble, including GLM (\texttt{SL.glm} and \texttt{SL.glm.interaction}), LASSO (\texttt{SL.glmnet}) and extreme gradient boosting (\texttt{SL.xgboost}) to model and predict the PSs. 

{Our scientific question of interest is: ``\textit{What is the impact of smoking on blood lead levels?}'' Thus, we focus on the overall population and target ATE or a quantity close to it. In this example, both smokers and non-smokers may have had a chance to smoke, so it is unclear whether there are structural violations of positivity. We then proceed to assess the empirical PS overlap. }

\begin{figure}[ht]
    \centering
    \includegraphics[width=0.65\linewidth]{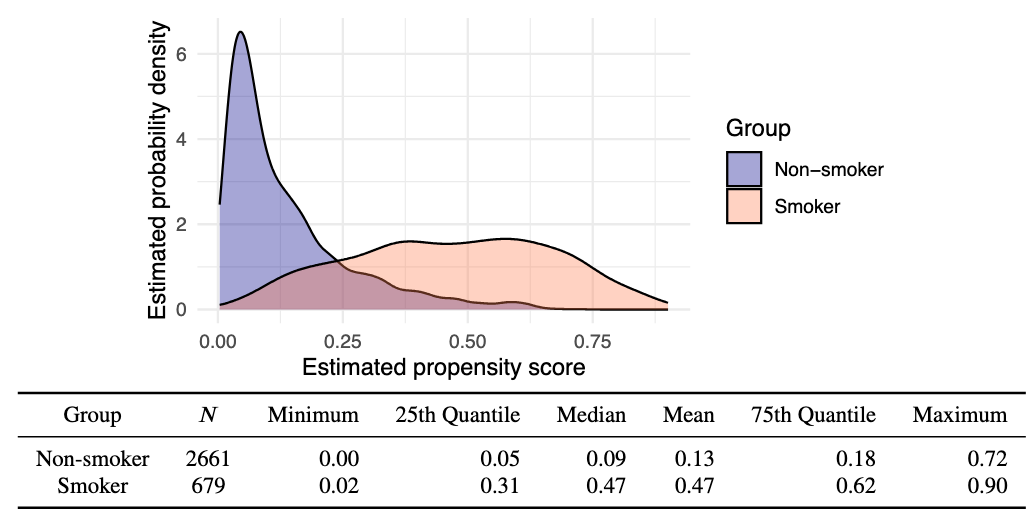}
    \caption{Density plots and summary statistics of the estimated propensity scores of smoking. }
    \label{fig:ps-blood}
\end{figure}

The density plots of estimated PSs (along with a summary table by smoking status) is presented in Figure \ref{fig:ps-blood}, showing a poor overlap. Some smokers have their PS close to 0. Thus, we expect to see some smokers with extreme weights {when considering ATE or ATC as the primary estimand.}

{Based on the exploratory analysis of the PS distributions, we decide to shift the target within the WATE class where ATE belongs to. Since the overlap is poor, we choose the estimand as ATO. We obtain an estimate of 0.84 (95\% CI: [0.67, 1.01]). This indicates that,  the average blood lead level would increase by approximately 0.84$\mu$g/dl higher among smokers over the overlap population, which suggests a statistically significant increase in blood lead concentration due to smoking. }

\begin{figure}[ht]
    \centering
    \includegraphics[width=\linewidth]{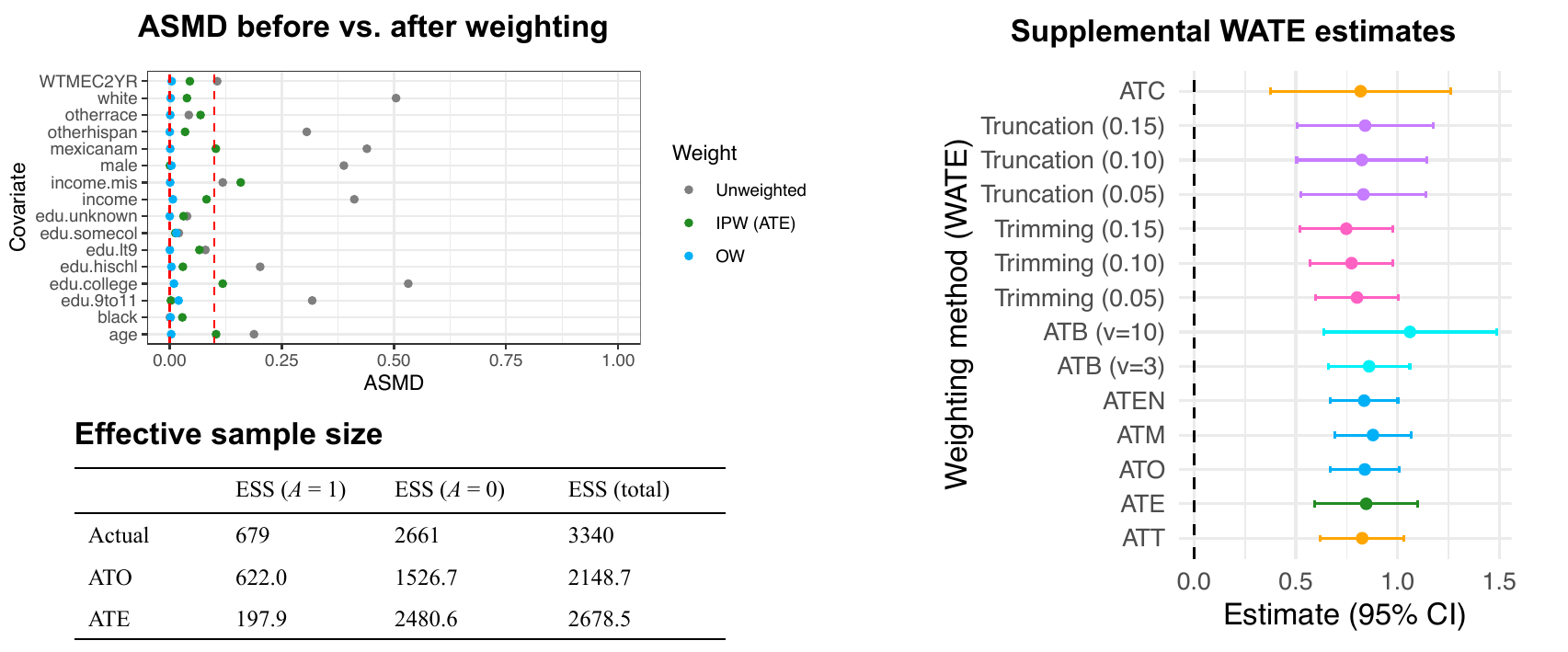}
    \caption{{Supplemental results for the primary ATO analysis: (i) ASMD for covariate balancing and ESS; the two red dashed lines in the love plot indicate 0 and 0.1 ASMDs; the ATE result in ESS is solely for reference; and (ii) supplemental analyses (WATE estimands) for effects of smoking on blood lead level.}}\label{fig:blood-supp}
\end{figure}

{Furthermore, we conducted supplemental analyses using other WATE estimands to evaluate the robustness of our findings (right forest plot panel in Figure}~\ref{fig:blood-supp}{). The ATO estimand produced point estimates very similar to those from ATE, ATM, ATEN, and ATB ($\nu=3$), with ATE showing a slightly wider confidence interval that still excluded zero. In contrast, ATB ($\nu=10$), ATC, and ATE truncation estimands ($\alpha = 0.05, 0.1,$ and $0.15$) yielded noticeably different point estimates and wider confidence intervals, suggesting potential instability. 

The same figure also displays the ASMDs before and after weighting, showing that ATO (OW) weighting achieved excellent covariate balance, with all ASMD values near zero, whereas ATE (IPW) weighting left a few covariates with ASMDs above 0.1, likely due to the influence of extreme weights. The ESS results further support these observations: the ESS for the $A=1$ (smoker) group under ATO was substantially improved compared to ATE and approached the actual sample size. Although the ESS for the $A=0$ (non-smoker) group was smaller than both the actual sample size and the ESS under ATE (1526.7 vs. 2661 and 2408.6, respectively), this reduction is expected because OW assigns near-zero weights to non-smokers with estimated PS values close to 0. Overall, these results indicate that the ATO estimand provides stable and robust estimates.} 

{Finally, for completeness, we refer readers to the Online Supplemental Material B.1 for the full set of results under all WATE, WATT, and WATC estimands for this dataset. These include (i) all ESS and ASMD summaries across weighting methods and (ii) numerical tables and forest plots of the estimated treatment effects, with the corresponding 95\% CIs. 

In this example, all treatment effect estimates are statistically significant at the 0.05 level. Among all classes, the BW method with $\nu=10$ produces relatively unstable results, consistent with our simulation findings that larger $\nu$ values can increase variability. The results with the WATT class are particularly more similar since there are no extreme WATT weights in this example. We emphasize that although the results for all estimands are presented for completeness and illustration, our intention is not to encourage practitioners to perform multiple analyses and select one post hoc. Rather, we caution against such practices as they are not recommended by the more of statistics.}

\subsection{Study II: Effects of sex work history on HIV status among transgender women in South Africa}\label{subsec:studyII}

The second study examines the effects of sex work history on HIV-positive status {using data from a cross-sectional, community-engaged mixed methods study conducted among transgender women in South Africa.}\cite{poteat2020prep} {The study was designed specifically to assess HIV vulnerability and PrEP engagement among transgender women. These women were recruited between June and November 2018 in three urban areas---Cape Town, East London, and Johannesburg---where dedicated transgender health clinics were planned. The dataset represents a non-probability community sample rather than a population-based survey like NHANES; hence, no design or sampling weights were applied in the original study. All analyses in our paper therefore treat this as an observational convenience sample for causal inference.}

Of the 196 participants, 128 (65.30\%) had a history of sex work ($A=1$), while 68 did not ($A=0$). With only 196 participants, we used a multivariable logistic regression model to estimate the PS {by several social variables: threaten violence history, substance use, sexual violence history, physical violence history, homeless history and food insecurity}. We chose a simpler logistic regression model instead of complex (yet probably more predictive) machine learning models or super learner ensembles because they often have more model parameters and they typically require a larger sample size to converge adequately. The logistic regression provides straightforward, robust parameter estimates that are less prone to overfitting in this example. 

{A meaningful scientific question in this example is, ``\textit{What is the average effect of having a history of sex work on HIV status among transgender women who have not engaged in sex work?}'' This aligns with ATC as an initial estimand consideration, which captures the potential increase in HIV risk if these participants were to engage in sex work and provides valuable insights for developing preventive strategies to reduce HIV acquisition risk in this population. Conceptually, this population has potentially structural violation of positivity due to the presence of subgroups of transgender women with characteristics (e.g., specific sociodemographic or behavioral factors) that make engagement in sex work nearly impossible or highly unlikely. At the same time, no prior knowledge exists regarding which estimand is best to target as alternative to ATC, so we proceed to assess the empirical overlap. }

\begin{figure}[ht]
    \centering
    \includegraphics[width=0.65\linewidth]{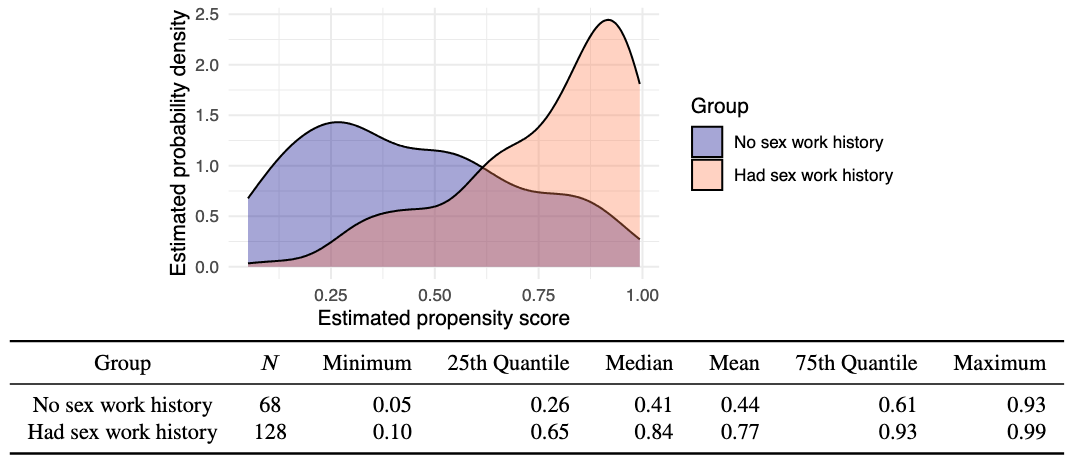}
    \caption{Estimated density and summary statistics of the estimated propensity score of the transgender women. }
    \label{fig:ps-hiv}
\end{figure}

The density and descriptive statistics of the estimated PSs are presented in Figure  \ref{fig:ps-hiv}, which shows a moderate overlap of the distributions of the PSs. {Note that the ATC weight on the treated group (had sex work history) is given by $\{1-\widehat e(\mb X)\}/\widehat e(\mb X)$, but we did not observe extremely small $\widehat e(\mb X)$ in this example or a severely poor overlap. Because of the conceptually violation of positivity, we consider ATC trimming with $\alpha=0.05$ to exclude participants with extremely low risk of engaging in sex work. } 

\begin{table}[ht]
    \centering
    \begin{tabular}{rcccc}
    \toprule
        Measure for binary outcome & Estimate & 95\% CI \\
    \midrule
        Risk difference (RD) & 0.19 & (0.04, 0.42) \\
        Risk ratio (RR) & 2.74 & (1.11, 7.94) \\
        Odds ratio (OR) & 3.42 & (1.41, 12.21) \\
    \bottomrule
    \end{tabular}
    \caption{{Results of primary analysis (ATC trimming with $\alpha=0.05$) for effects of sex work history on HIV status among participants without sex work history.} }
    \label{tab:hiv-main-res}
\end{table}

\begin{figure}
    \centering
    \includegraphics[width=\linewidth]{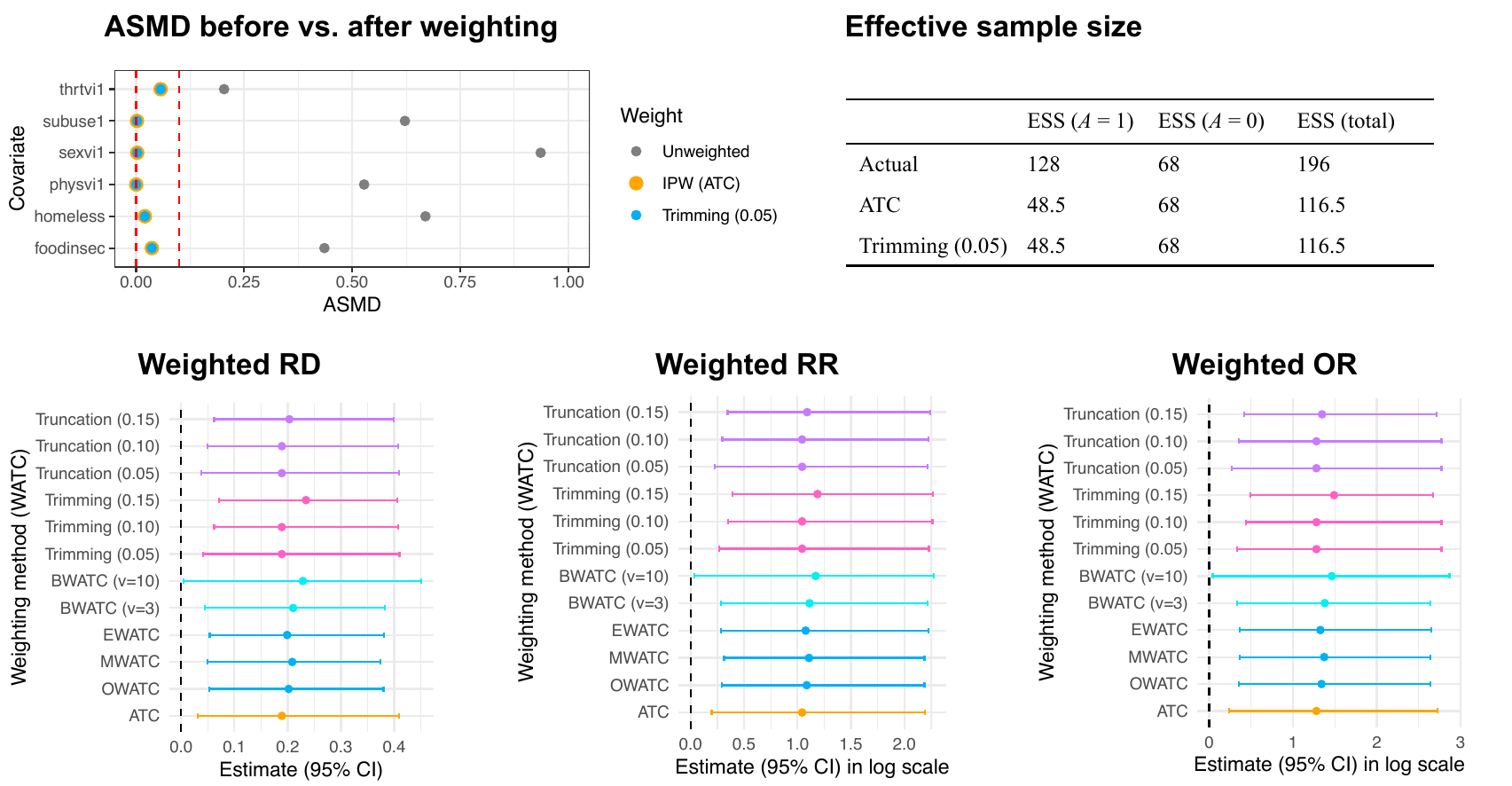}
    \caption{{Supplemental results for the primary ATC trimming $(\alpha=0.05)$ analysis: (i) ASMD for covariate balancing and ESS; the two red dashed lines in the love plot indicate 0 and 0.1 ASMDs; the ATC (without trimming) result in ESS is solely for a reference; and (ii) supplemental analyses (WATC estimands in weighted RD, RR and OR forms) for effects of sex work history on HIV status among participants without sex work history.}}\label{fig:hiv-supp}
\end{figure}

{Table} \ref{tab:hiv-main-res} {shows the point and 95\% CI estimates under RD, RR and OR scales for the primary ATC trimming analysis. The results} can be interpreted as follows: on average, among transgender women in South Africa without a history of sex work (controls) who also have certain risk of engaging in sex work (a PS of greater than 0.05), the probability of being HIV-positive increases {by 0.19 with a 95\% CI of (0.04, 0.42)}, if they engage in sex work. The estimated RR is 2.74 with a 95\% CI of (1.11, 7.94), and the estimated OR is 3.42 with a 95\% CI of (1.41, 12.21). 

{Supplemental sensitivity analyses by other estimands (lower panels in Figure }\ref{fig:hiv-supp}) {in the WATC class further support these findings, as all methods produced similar estimates, 95\% CIs and significant treatment effects (except for BWATC with $\nu=10$ who has notably wider CIs). In addition, from the ASMD results in upper panels of Figure }\ref{fig:hiv-supp}, {we find that the ATC trimming ($\alpha=0.05$) weights balance all covariates well with ASMD close to 0. Furthermore, although the ESS for the $A=1$ group (those with a history of sex work) is smaller than the actual sample size (48.5 vs. 128), the pattern is consistent across weighting methods, including the OWATC and similar alternatives as shown in the Online Supplemental Material.} 

{Finally, for completeness, we refer readers to the Online Supplemental Material B.2 for the full set of results under all WATE, WATT, and WATC estimands for this dataset. These include (i) all ESS and ASMD results across weighting methods, and (ii) numerical tables and forest plots presenting the estimated treatment effects and the corresponding 95\% CIs for the weighted RD, RR, and OR measures. In the context of this study, if the target estimand is the overall (ATE) or the population of participants with a history of sex work (ATT), the estimates will be unstable due to large weights assigned to a few participants. The supplemental results under the WATE and WATT classes corroborate this, showing that ATE and ATT yield non-significant results, whereas other WATE and WATT estimands suggest the opposite conclusions. Again, we caution against any post hoc analysis as we commented in Section} \ref{subsec:studyI}. 

\section{Concluding Remarks}\label{sec:conclude}

\subsection{Summary}

To improve causal inference from observational data, it is recommended to define the protocol of a hypothetical randomized pragmatic trial---often referred to as the ``target trial''---that best addresses the causal question of interest, and then emulate this trial using observational data.\cite{hernan2022target, hernan2025target, Hansforde074626} This framework aligns well with the ICH E9 (R1) guidance, which outlines key principles for constructing estimands in clinical trials to enhance planning, design, analysis, and interpretation.\cite{kahan2024estimands} By explicitly defining the target population, treatment strategies, outcomes, potential intercurrent events, and population-level summaries, ICH E9 (R1) encourages a clear articulation of what is being estimated and for whom.\cite{mayo2023can, rubin1974estimating}

The estimand framework---used alongside the potential outcomes framework---has been foundational in causal inference long before the release of ICH E9 (R1).\cite{rubin1974estimating} Estimating and drawing inference about causal estimands requires assumptions, most of which are untestable. It is the investigator’s responsibility to justify their plausibility based on prior literature and domain expertise. Key assumptions such as unconfoundedness and positivity are assessed differently: the former often through expert knowledge and DAGs,\cite{tennant2021use} and the latter by distinguishing structural from random violations of positivity,\cite{matsouaka2024causal, petersen2012diagnosing} as well as by examining the distribution of PSs across treatment groups.

In this tutorial, we summarized and presented three specific classes of estimands that can be used to responds to wide range of scientific questions: the WATE class of estimands, which centers around and includes the conventional and widely used ATE estimand; the WATT class of estimands, which is adjacent to and generalizes the ATT estimand; and the WATC class of estimands that is centered around and expand the on the ATC estimand. These classes of estimands have some common features: they can be defined in terms of PS, and their estimators are rooted in and defined with estimated weights that are also functions of the PS. The related estimands are useful to account for both random and structural violations of the positivity assumption.\cite{matsouaka2024causal} We have also demonstrated that they include the underlying description of the population of participants targeted, the population-level measures, the possible causal contrasts, and the attributes inherent to the way they handle violations of the positivity assumptions, and clear interpretations of the study findings. 

The WATE targets the whole population of participants or its subsets thanks to the tilting functions of Table \ref{tab:WATEs}, while the WATT (resp. WATC) is estimated for the population of treated (resp. control) participants or its subsets defined by specific tilting functions (provided in Table \ref{tab:TiltWATT}).  While often implicit, we have made it clear that all trimming and truncation methods that are based on IPW also target subpopulation of participants than those they are often attributed to in practice. In a recently published paper, Liu et al. applied this framework to assess racial disparities in healthcare expenditure across different racial and ethnic groups.\cite{liu2025assessing} In such a context were violations of the positivity assumption is expected and the distribution of participants across racial and ethnic is disproportionate, the paper provide an appropriate and well-throughout illustration of these methods.

When should we keep our analyses simple and straightforward, instead of exploring the options provided in this paper? When the unconfoundedness is assumed and there is no significant violations of the positivity assumption. In this case, estimates of the WATE  will be similar to ATE estimate, those from the WATT (resp. WATC) will mimic the ATT (resp. ATC) estimate. When violations of the positivity assumption occur, the results are expected to be drastically different from estimands within the same class of estimands. We have provided guidelines on specific choices to make, to what extent and what to expect. We must recognize that there are three schools of thoughts when it comes to methods that mitigate the lack of adequate positivity. While the classes presented in this paper are purposefully selected to handle both structural and random violations of the positivity assumption, several other methods have been proposed in the literature, including those that either focus directly in estimating weights that lead to covariate balance or ``fix'' the PS estimation as to obtain PS estimates that are exempt from any violations of the positivity assumption.\cite{matsouaka2024causal} Unfortunately, these methods can still lead to unstable estimates of treatment effects with larger variances, especially in the presence of structural violations of the positivity assumption. \cite{matsouaka2024causal,zhao2019covariate} When such a violation of positivity assumption is expected, it often requires a priori subject-matter expertise to incorporate such information into the causal analysis, without the need to fix the lack of positivity by brute force. As argued by Matsouaka and Zhou,\cite{matsouaka2024causal} in practice, the reliance on one statistical prowess to solve the issue of positivity can be detrimental; sometimes, it is better to recognize the inherent limitations of the data at hand and find acceptable alternative. 

\subsection{Discussion}

While the framework we have presented in this paper has a number of strengths, it is not without limitations. {A key limitation of our tutorial is that we have not developed formal statistical tools to test or identify the type of violations of the positivity assumption; to the best of our knowledge, we have not also seen this specified anywhere in the literature as well. Therefore, we recommend that practitioners rely primarily on (i) domain knowledge and study design considerations to assess the plausibility of the positivity assumption, and (ii) graphical diagnostics such as the overlap in the distributions of estimated PSs across treatment groups to evaluate empirical support. Future work is needed to develop principled testing procedures to detect and differentiate these two types of violations of the positivity assumption. }

{In addition, it} is well known that when the PS model is misspecified, the resulting PSW estimate may be biased.\cite{zhou2020propensity} This raises a practical question: can the PSW estimate still meaningfully interpret the targeted estimand? Our simulation results demonstrate that such misspecifications can indeed lead to biased estimations. Even when using a super learner ensemble that includes a correctly specified model, the estimation can still be affected. Nonetheless, we argue that the PSW estimate remains interpretable based on the estimand the user selects after eliciting subject-matter expert input. The tilting functions $h(\cdot)$ or $g(\cdot)$ in WATE, WATT, or WATC estimands are known to the user, which ensure that the interpretation aligns with the intended target population.\cite{thomas2020overlap} Although bias may occur due to PS model misspecifications, such bias is quantifiable asymptotically if the user has prior knowledge of potential bias in the PS model. For example, Zhou et al.\cite{zhou2020propensity} provide asymptotic bias quantification for WATE, and Liu et al.\cite{liu2024average} extend this to WATT. Moreover, both our simulation results and previous studies\cite{liu2024average, zhou2020propensity, gao2024does} indicate that estimands defined by OW, EW, and MW are more robust and less sensitive to PS model misspecifications. As a result, these estimands exhibit higher internal validity and are recommended for consideration in a user's study plan.

From this work, we can pursue several other potential avenues. First, this paper assumes the unconfoundedness throughout and discusses the lack of positivity without considering their connection; we can explore the potential relationships between these two assumptions and provide way to analyze the data when both are violated. In fact, the presence of extreme PS estimate distributions may sometimes signal unmeasured confounding.\cite{sturmer2010treatment, kang2007demystifying} Intuitively, if no participants in certain regions of the covariate space can receive a particular treatment level, this may indicate the presence of an unobserved factor that deterministically influences treatment assignment. Therefore, incorporating methods to address violations of unconfoundedness, such as through sensitivity analyses,\cite{imai2010identification, vanderweele2017sensitivity} would be a valuable addition to our proposed framework. 

{Second, we did not include the normalized-weighted (H\'ajek-type) estimators for the estimands considered in this paper. The H\'ajek-type estimator differs from the PSW estimators discussed in Section} \ref{sec:tutorial} {primarily in the implementation details rather than in the underlying methodology. Moreover, prior studies have shown that the ability of the H\'ajek-type estimators to handle structural violations of positivity is limited.}\cite{matsouaka2024overlap} {Therefore, to maintain the conceptual clarity and practical consistency of our framework and R package---and to avoid potential confusion regarding the choice of estimators---we did not incorporate the H\'ajek-type version into the current implementation. }

{Third}, our current tutorial does not consider augmented estimators, such as doubly robust estimators for the ATE, which incorporate additional outcome models to enhance the (semiparametric) efficiency.\cite{robins1995analysis, ogburn2015doubly, wang2025rate} This omission is intentional, as the primary focus of this paper is on PS weights. In addition, augmented estimators can underperform when both the outcome and PS models are misspecified,\cite{kang2007demystifying, li2019propensity} which complicates their application. For this reason, we have refrained ourselves from including augmented estimators until there is more insight into their bias quantification--an aspect that, to the best of our knowledge, remains unaddressed in the literature. 

{Furthermore, we did not discuss the connection between our framework and the covariate balancing propensity score (CBPS) method.}\cite{imai2014covariate, yang2018propensity} {The CBPS differs, in principle, from the methods covered in our tutorial. Specifically, CBPS formulates the PS estimation as a data-driven optimization problem that directly enforces covariate balance through moment conditions or calibration constraints. From this perspective, CBPS aims to obtain nonparametric estimates of PSs---rather than model-based ones---to better achieve covariate balance in finite samples, but it does not explicitly account for shifts in the target population. As a future direction, we plan to investigate how CBPS-based PS estimates may affect (and potentially improve) the estimation precision of WATE, WATT, and WATC estimands. It also remains an open question whether CBPS methods can adequately handle structural violations of the positivity assumption, making this an interesting avenue for future research}

{Moreover, while we included the survey-weight variable to account for the contribution of the complex survey design in our Study I (Section} \ref{subsec:studyI}{), we have not yet incorporated other design features such as strata and primary sampling units (PSUs). To our knowledge, methodological developments that integrate these factors with PS weighting analyses remain limited in the current causal inference literature and are particularly unexplored for the general classes of WATE, WATT, and WATC estimands we considered. This therefore represents an important avenue for future research.} Finally, we are interested in extending the tutorial to more complex settings, such as multi-valued treatments,\cite{li2019propensity, yang2016propensity, liu2025assessing} survival outcomes,\cite{cheng2022addressing, cao2024using} multi-source data,\cite{zhuang2024assessment, liu2025targeted, wang2025integrating} high-dimensional covariate,\cite{liu2025coadvise, belloni2014inference} missing data,\cite{zhao2024covariate} and conformal predictions.\cite{liu2024multi}

Last but not least, while our tutorial offers a number of PS weighting methods, we strongly discourage data-driven selection of estimands after exploring multiple options. Researchers should pre-specify their causal estimand(s) in collaboration with clinical experts to ensure principled inference and to avoid issues such as p-hacking and multiplicity bias.\cite{simmons2011false} As with most clinical research, we recommend that they elaborate a statistical analysis plan in advance (that is agreed upon by all the stakeholders) to promote transparency and reproducibility of study findings. 

\section*{Acknowledgment}

{The authors are truly thankful to the Associate Editor and the two anonymous reviewers for their insightful comments and constructive suggestions, which have greatly improved the clarity, rigor, and overall contribution of this manuscript.}

\section*{Data Availability Statement}

The data used in Section \ref{subsec:studyI} are available at \url{https://onlinelibrary.wiley.com/action/downloadSupplement?doi=10.1111%2Fbiom.12101&file=biom12101-sm-0002-SupExampleData.txt}, {and the survey-weight variable (WTMEC2YR) can be found at }\url{https://wwwn.cdc.gov/nchs/nhanes/search/variablelist.aspx?Component=Demographics&Cycle=2007-2008}.
The data used in Section \ref{subsec:studyII} are not publicly available due to their sensitive nature and the vulnerability of the study population; however, they can be made available upon reasonable request to the fourth author, Poteat (tonia.poteat@duke.edu).

\section*{Software}

The user-friendly software ``\textit{ChiPS}'' implementing the framework in this tutorial is available at the author's Github page: \url{https://github.com/yiliu1998/ChiPS}. 

\section*{Financial Disclosure}

Research reported in this manuscript was supported by the National Heart, Lung, and Blood Institute (NHLBI) of the National Institutes of Health (NIH) under Award Number T32HL079896. The content is solely the responsibility of the authors and does not necessarily represent the official views of the NIH. 

\bibliography{_refs}

@article{liu2024multi,
  title={Multi-source conformal inference under distribution shift},
  author={Liu, Yi and Levis, Alexander W and Normand, Sharon-Lise and Han, Larry},
  journal={Proceedings of the 41st International Conference on Machine Learning},
  volume={235},
  pages={31344--31382},
  year={2024}
}

@article{yang2018propensity,
  title={Propensity score weighting for causal inference with clustered data},
  author={Yang, Shu},
  journal={Journal of Causal Inference},
  volume={6},
  number={2},
  pages={20170027},
  year={2018},
  doi={10.1515/jci-2017-0027},
  publisher={De Gruyter}
}

@article{imai2014covariate,
  title={Covariate balancing propensity score},
  author={Imai, Kosuke and Ratkovic, Marc},
  journal={Journal of the Royal Statistical Society Series B: Statistical Methodology},
  volume={76},
  number={1},
  pages={243--263},
  year={2014},
  doi={10.1111/rssb.12027},
  publisher={Oxford University Press}
}

@article{westreich2010invited,
  doi={10.1093/aje/kwp436},
  title={Invited commentary: positivity in practice},
  author={Westreich, Daniel and Cole, Stephen R},
  journal={American Journal of Epidemiology},
  volume={171},
  number={6},
  pages={674--677},
  year={2010},
  publisher={Oxford University Press}
}

@article{hsu2013calibrating,
  doi={10.1111/BIOM.12101},
  title={Calibrating sensitivity analyses to observed covariates in observational studies},
  author={Hsu, Jesse Y and Small, Dylan S},
  journal={Biometrics},
  volume={69},
  number={4},
  pages={803--811},
  year={2013},
  publisher={Wiley Online Library}
}

@article{tennant2021use,
  title={Use of directed acyclic graphs (DAGs) to identify confounders in applied health research: review and recommendations},
  author={Tennant, Peter WG and Murray, Eleanor J and Arnold, Kellyn F and Berrie, Laurie and Fox, Matthew P and Gadd, Sarah C and Harrison, Wendy J and Keeble, Claire and Ranker, Lynsie R and Textor, Johannes and others},
  journal={International Journal of Epidemiology},
  volume={50},
  number={2},
  pages={620--632},
  year={2021},
  publisher={Oxford University Press}
}

@article{austin2015moving,
  doi={10.1002/sim.6607},
  title={Moving towards best practice when using inverse probability of treatment weighting (IPTW) using the propensity score to estimate causal treatment effects in observational studies},
  author={Austin, Peter C and Stuart, Elizabeth A},
  journal={Statistics in Medicine},
  volume={34},
  number={28},
  pages={3661--3679},
  year={2015},
  publisher={Wiley Online Library}
}

@article{ju2019adaptive,
  doi={10.1177/0962280218774817},
  title={On adaptive propensity score truncation in causal inference},
  author={Ju, Cheng and Schwab, Joshua and van der Laan, Mark J},
  journal={Statistical Methods in Medical Research},
  volume={28},
  number={6},
  pages={1741--1760},
  year={2019},
  publisher={SAGE Publications Sage UK: London, England}
}

@article{matsouaka2024causal,
doi={10.1002/bimj.202300156},
  title={Causal inference in the absence of positivity: The role of overlap weights},
  author={Matsouaka, Roland A and Zhou, Yunji},
  journal={Biometrical Journal},
  volume={66},
  number={4},
  pages={2300156},
  year={2024},
  publisher={Wiley Online Library}
}

@article{vanderweele2017sensitivity,
   doi={10.7326/M16-2607},
title={Sensitivity analysis in observational research: introducing the E-value},
  author={VanderWeele, Tyler J and Ding, Peng},
  journal={Annals of Internal Medicine},
  volume={167},
  number={4},
  pages={268--274},
  year={2017},
  publisher={American College of Physicians}
}

@article{sturmer2010treatment,
  doi={10.1093/aje/kwq198},
  title={Treatment effects in the presence of unmeasured confounding: dealing with observations in the tails of the propensity score distribution—a simulation study},
  author={St{\"u}rmer, Til and Rothman, Kenneth J and Avorn, Jerry and Glynn, Robert J},
  journal={American Journal of Epidemiology},
  volume={172},
  number={7},
  pages={843--854},
  year={2010},
  publisher={Oxford University Press}
}

@article{gruber2022data,
  doi={10.1093/aje/kwac087},
  title={Data-adaptive selection of the propensity score truncation level for inverse-probability--weighted and targeted maximum likelihood estimators of marginal point treatment effects},
  author={Gruber, Susan and Phillips, Rachael V and Lee, Hana and van der Laan, Mark J},
  journal={American Journal of Epidemiology},
  volume={191},
  number={9},
  pages={1640--1651},
  year={2022},
  publisher={Oxford University Press}
}

@article{yang2016propensity,
  doi={10.1111/biom.12505},
  title={Propensity score matching and subclassification in observational studies with multi-level treatments},
  author={Yang, Shu and Imbens, Guido W and Cui, Zhanglin and Faries, Douglas E and Kadziola, Zbigniew},
  journal={Biometrics},
  volume={72},
  number={4},
  pages={1055--1065},
  year={2016},
  publisher={Wiley Online Library}
}

@article{kang2007demystifying,
  doi={10.1214/07-STS227},
  title={Demystifying double robustness: A comparison of alternative strategies for estimating a population mean from incomplete data},
  author={Kang, Joseph DY and Schafer, Joseph L},
  journal={Statistical Science},
  year={2007}
}

@article{van2007super,
  doi={10.2202/1544-6115.1309},
  title={Super learner},
  author={Van der Laan, Mark J and Polley, Eric C and Hubbard, Alan E},
  journal={Statistical Applications in Genetics and Molecular Biology},
  volume={6},
  number={1},
  year={2007},
  publisher={De Gruyter}
}

@article{wang2025rate,
  title={Rate doubly robust estimation for weighted average treatment effects},
  author={Wang, Yiming and Liu, Yi and Yang, Shu},
  journal={arXiv preprint arXiv:2509.14502},
  year={2025}
}

@article{lunceford2004stratification,
  doi={10.1002/sim.1903},
  title={Stratification and weighting via the propensity score in estimation of causal treatment effects: a comparative study},
  author={Lunceford, Jared K and Davidian, Marie},
  journal={Statistics in Medicine},
  volume={23},
  number={19},
  pages={2937--2960},
  year={2004},
  publisher={Wiley Online Library}
}

@article{austin2023differences,
  doi={10.1002/pds.5639},
  title={Differences in target estimands between different propensity score-based weights},
  author={Austin, Peter C},
  journal={Pharmacoepidemiology and Drug Safety},
  volume={32},
  number={10},
  pages={1103--1112},
  year={2023},
  publisher={Wiley Online Library}
}

@article{parikh2025we,
  title={Who Are We Missing?: A Principled Approach to Characterizing the Underrepresented Population},
  author={Parikh, Harsh and Ross, Rachael K and Stuart, Elizabeth and Rudolph, Kara E},
  journal={Journal of the American Statistical Association},
  volume={120},
  pages={1414-1423},
  year={2025},
  doi={10.1080/01621459.2025.2495319},
  publisher={Taylor \& Francis}
}

@article{matsouaka2024overlap,
  doi={10.1080/03610918.2024.2319419},
  title={Overlap, matching, or entropy weights: what are we weighting for?},
  author={Matsouaka, Roland A and Liu, Yi and Zhou, Yunji},
  journal={Communications in Statistics-Simulation and Computation},
  pages={1--20},
  year={2024},
  publisher={Taylor \& Francis}
}

@article{liu2024average,
  doi={10.1177/09622802241269646},
  title={Average treatment effect on the treated, under lack of positivity},
  author={Liu, Yi and Li, Huiyue and Zhou, Yunji and Matsouaka, Roland},
  journal={Statistical Methods in Medical Research},
  year={2024}
}

@book{efron1994introduction,
  doi={10.1201/9780429246593},
  title={An introduction to the bootstrap},
  author={Efron, Bradley and Tibshirani, Robert J},
  year={1994},
  publisher={Chapman and Hall/CRC}
}

@article{austin2018assessing,  
  doi={10.1002/sim.7615},
  title={Assessing the performance of the generalized propensity score for estimating the effect of quantitative or continuous exposures on binary outcomes},
  author={Austin, Peter C},
  journal={Statistics in Medicine},
  volume={37},
  number={11},
  pages={1874--1894},
  year={2018},
  publisher={Wiley Online Library}
}

@article{austin2017estimating,
 doi={10.1177/0962280215601134},
  title={Estimating the effect of treatment on binary outcomes using full matching on the propensity score},
  author={Austin, Peter C and Stuart, Elizabeth A},
  journal={Statistical Methods in Medical Research},
  volume={26},
  number={6},
  pages={2505--2525},
  year={2017},
  publisher={SAGE Publications Sage UK: London, England}
}

@article{poteat2024transgender,
  doi={10.2196/64373},
  title={Transgender-Specific Differentiated HIV Service Delivery Models in the South African Public Primary Health Care System (Jabula Uzibone): Protocol for an Implementation Study},
  author={Poteat, Tonia and Bothma, Rutendo and Maposa, Innocent and Hendrickson, Cheryl and Meyer-Rath, Gesine and Hill, Naomi and Pettifor, Audrey and Imrie, John and others},
  journal={JMIR Research Protocols},
  volume={13},
  number={1},
  pages={e64373},
  year={2024},
  publisher={JMIR Publications Inc., Toronto, Canada}
}

@article{poteat2022stigma,
  doi={10.1016/S2352-3018(21)00323-4},
  title={Stigma reduction is key to improving the HIV care continuum},
  author={Poteat, Tonia C and van der Merwe, L Leigh Ann},
  journal={The Lancet HIV},
  volume={9},
  number={3},
  pages={e144--e145},
  year={2022},
  publisher={Elsevier}
}

@article{matsouaka2023variance,
  doi={10.1177/09622802221142532},
  title={Variance estimation for the average treatment effects on the treated and on the controls},
  author={Matsouaka, Roland A and Liu, Yi and Zhou, Yunji},
  journal={Statistical Methods in Medical Research},
  volume={32},
  number={2},
  pages={389--403},
  year={2023},
  publisher={SAGE Publications Sage UK: London, England}
}

@article{rosenbaum1983central,
  doi={10.1093/biomet/70.1.41},
  title={The central role of the propensity score in observational studies for causal effects},
  author={Rosenbaum, Paul R and Rubin, Donald B},
  journal={Biometrika},
  volume={70},
  number={1},
  pages={41--55},
  year={1983},
  publisher={Oxford University Press}
}

@article{yang2018asymptotic,
    doi={10.1093/biomet/asy008},
	Title ={Asymptotic inference of causal effects with observational studies trimmed by the estimated propensity scores},
	Author={Yang, S and Ding, P},
	Journal ={Biometrika},
	Year  ={2018},
	Pages ={487--493},
	Volume={105},
	Publisher ={Oxford University Press}
}

@article{gao2024does,
  doi={10.1186/s12874-024-02375-3},
  title={When does adjusting covariate under randomization help? A comparative study on current practices},
  author={Gao, Ying and Liu, Yi and Matsouaka, Roland},
  journal={BMC Medical Research Methodology},
  volume={24},
  number={1},
  pages={250},
  year={2024},
  publisher={Springer}
}

@article{robins1995analysis,
  title={Analysis of semiparametric regression models for repeated outcomes in the presence of missing data},
  author={Robins, James M and Rotnitzky, Andrea and Zhao, Lue Ping},
  journal={Journal of the American Statistical Association},
  volume={90},
  number={429},
  pages={106--121},
  year={1995},
  doi={10.1080/01621459.1995.10476493},
  publisher={Taylor \& Francis}
}

@article{zhang2024three,
  title={Three new methodologies for calculating the effective sample size when performing population adjustment},
  author={Zhang, Landan and Bujkiewicz, Sylwia and Jackson, Dan},
  journal={BMC Medical Research Methodology},
  volume={24},
  number={1},
  pages={287},
  year={2024},
doi={10.1186/s12874-024-02412-1}, 
  publisher={Springer}
}

@article{austin2009balance,
  title={Balance diagnostics for comparing the distribution of baseline covariates between treatment groups in propensity-score matched samples},
  author={Austin, Peter C},
  journal={Statistics in Medicine},
  volume={28},
  number={25},
  pages={3083--3107},
  year={2009},
  doi={10.1002/sim.3697},
  publisher={Wiley Online Library}
}

@article{hernan2022target,
  title={Target trial emulation: a framework for causal inference from observational data},
  author={Hern{\'a}n, Miguel A and Wang, Wei and Leaf, David E},
  journal={JAMA},
  volume={328},
  number={24},
  pages={2446--2447},
  year={2022},
  doi={10.1001/jama.2022.21383},
  publisher={American Medical Association}
}

@article{choi2023overlap,
  title={Overlap weight and propensity score residual for heterogeneous effects: a review with extensions},
  author={Choi, Jin-young and Lee, Myoung-jae},
  journal={Journal of Statistical Planning and Inference},
  volume={222},
  pages={22--37},
  year={2023},
doi={10.1016/j.jspi.2022.04.003},
  publisher={Elsevier}
}

@article{kahan2024estimands,
  title={The estimands framework: a primer on the ICH E9 (R1) addendum},
  author={Kahan, Brennan C and Hindley, Joanna and Edwards, Mark and Cro, Suzie and Morris, Tim P},
  journal={BMJ},
  volume={384},
  year={2024},
pages={e076316},
   doi={10.1136/bmj-2023-076316},
  publisher={British Medical Journal Publishing Group}
}

@article{hernan2025target,
  title={The Target Trial Framework for Causal Inference From Observational Data: Why and When Is It Helpful?},
  author={Hern{\'a}n, Miguel A and Dahabreh, Issa J and Dickerman, Barbra A and Swanson, Sonja A},
  journal={Annals of Internal Medicine},
  volume={178},
  number={3},
  pages={402--407},
  year={2025},
  doi={10.7326/ANNALS-24-01871},
  publisher={American College of Physicians}
}

@article{greifer2021choosing,
  title={Choosing the causal estimand for propensity score analysis of observational studies},
  author={Greifer, Noah and Stuart, Elizabeth A},
  journal={arXiv preprint arXiv:2106.10577},
  year={2021}
}

@article {Hansforde074626,
	author = {Hansford, Harrison J and Cashin, Aidan G and Jones, Matthew D and Swanson, Sonja A and Islam, Nazrul and Dahabreh, Issa J and Dickerman, Barbra A and Egger, Matthias and Garcia-Albeniz, Xavier and Golub, Robert M and Lodi, Sara and Moreno-Betancur, Margarita and Pearson, Sallie-Anne and Schneeweiss, Sebastian and Sterne, Jonathan and Sharp, Melissa K and Stuart, Elizabeth A and Hernan, Miguel A and Lee, Hopin and McAuley, James H},
	title = {Development of the TrAnsparent ReportinG of observational studies Emulating a Target trial (TARGET) guideline},
	volume = {13},
	number = {9},
	elocation-id = {e074626},
	year = {2023},
	doi = {10.1136/bmjopen-2023-074626},
	publisher = {British Medical Journal Publishing Group},
	issn = {2044-6055},
	URL = {https://bmjopen.bmj.com/content/13/9/e074626},
	eprint = {https://bmjopen.bmj.com/content/13/9/e074626.full.pdf},
	journal = {BMJ Open}
}

@article{mayo2023can,
  title={What can be achieved with the estimand framework?},
  author={Mayo, Susan and Kim, Yongman},
  journal={Statistics in Biopharmaceutical Research},
  volume={15},
  number={3},
  pages={549--553},
  year={2023},
  doi={10.1080/19466315.2023.2173645},
  publisher={Taylor \& Francis}
}

@article{boughdiri2025unified,
  title={A unified framework for the transportability of population-level causal measures},
  author={Boughdiri, Ahmed and Berenfeld, Cl{\'e}ment and Josse, Julie and Scornet, Erwan},
  journal={arXiv preprint arXiv:2505.13104},
  year={2025}
}

@article{zhuang2024assessment,
  title={Assessment of treatment effect heterogeneity for multiregional randomized clinical trials},
  author={Zhuang, Haotian and Wang, Xiaofei and George, Stephen L},
  journal={Statistics in Biopharmaceutical Research},
  volume={17},
  number={3},
  pages={315--322},
  year={2025},
   doi={10.1080/19466315.2024.2421748},
  publisher={Taylor \& Francis}
}

@article{west2022best,
  title={Best practice in statistics: The use of log transformation},
  author={West, Robert M},
  journal={Annals of Clinical Biochemistry},
  volume={59},
  number={3},
  pages={162--165},
  year={2022},
   doi={10.1177/00045632211050531},
  publisher={SAGE Publications Sage UK: London, England}
}

@article{barnard2024unified,
  title={A Unified Framework for Causal Estimand Selection},
  author={Barnard, Martha and Huling, Jared D and Wolfson, Julian},
  journal={arXiv preprint arXiv:2410.12093},
  year={2024}
}

@article{wang2025integrating,
  title={Integrating Randomized Controlled Trial and External Control Data Using Balancing Weights: A Comparison of Estimands and Estimators},
  author={Wang, Peijin and Hong, Hwanhee and Jeon, Kyungeun and Thomas, Laine Elliott},
  journal={arXiv preprint arXiv:2502.13871},
  year={2025}
}

@article{dynarski2003does,
  title={Does aid matter? Measuring the effect of student aid on college attendance and completion},
  author={Dynarski, Susan M},
  journal={American Economic Review},
  volume={93},
  number={1},
  pages={279--288},
  year={2003},
  doi={10.1257/000282803321455287},
  publisher={American Economic Association}
}

@article{heckman1997matching,
  title={Matching as an econometric evaluation estimator: Evidence from evaluating a job training programme},
  author={Heckman, James J and Ichimura, Hidehiko and Todd, Petra E},
  journal={The Review of Economic Studies},
  doi={10.2307/2971733},
  volume={64},
  number={4},
  pages={605--654},
  year={1997},
  publisher={Wiley-Blackwell}
}

@article{jin2025policy,
  title={Policy learning “without” overlap: Pessimism and generalized empirical Bernstein’s inequality},
  author={Jin, Ying and Ren, Zhimei and Yang, Zhuoran and Wang, Zhaoran},
  journal={The Annals of Statistics},
  volume={53},
  number={4},
  pages={1483--1512},
  year={2025},
   doi={10.1214/25-AOS2511},
  publisher={Institute of Mathematical Statistics}
}

@article{haldane1956estimation,
  title={The estimation and significance of the logarithm of a ratio of frequencies},
  author={Haldane, JB},
  journal={Annals of Human Genetics},
  volume={20},
  number={4},
  pages={309--311},
  doi={10.1111/j.1469-1809.1955.tb01285.x},
  year={1956}
}

@article{imai2010identification,
  title={Identification, inference and sensitivity analysis for causal mediation effects},
  journal={Statistical Science},
  author={Imai, Kosuke and Keele, Luke and Yamamoto, Teppei},
  doi={10.1214/10-STS321},
  year={2010}
}

@article{liu2025coadvise,
  title={COADVISE: Covariate Adjustment with Variable Selection and Missing Data Imputation in Randomized Controlled Trials},
  author={Liu, Yi and Zhu, Ke and Han, Larry and Yang, Shu},
  journal={arXiv preprint arXiv:2501.08945},
  year={2025}
}

@book{van2000asymptotic,
  title={Asymptotic statistics},
  author={Van der Vaart, Aad W},
  volume={3},
  year={2000},
  doi={10.1017/CBO9780511802256},
  publisher={Cambridge University Press}
}

@book{shao2012jackknife,
  title={The jackknife and bootstrap},
  author={Shao, Jun and Tu, Dongsheng},
  year={2012},
  doi={10.1007/978-1-4612-0795-5},
  publisher={Springer Science \& Business Media}
}

@article{liu2025assessing,
  title={Assessing racial disparities in healthcare expenditure using generalized propensity score weighting},
  author={Liu, Jiajun and Liu, Yi and Zhou, Yunji and Matsouaka, Roland A},
  journal={BMC Medical Research Methodology},
  volume={25},
  number={64},
  pages={1--16},
  year={2025},
  doi={10.1186/s12874-025-02508-2},
  publisher={Springer}
}

@article{liu2025targeted,
  title={Targeted Data Fusion for Causal Survival Analysis Under Distribution Shift},
  author={Liu, Yi and Levis, Alexander W and Zhu, Ke and Yang, Shu and Gilbert, Peter B and Han, Larry},
  journal={arXiv preprint arXiv:2501.18798},
  year={2025}
}

@article{simmons2011false,
  title={False-positive psychology: Undisclosed flexibility in data collection and analysis allows presenting anything as significant},
  author={Simmons, Joseph P and Nelson, Leif D and Simonsohn, Uri},
  journal={Psychological science},
  volume={22},
  number={11},
  pages={1359--1366},
  year={2011},
  doi={10.1177/0956797611417632},
  publisher={Sage Publications Sage CA: Los Angeles, CA}
}

@article{zhao2024covariate,
  title={Covariate adjustment in randomized experiments with missing outcomes and covariates},
  author={Zhao, Anqi and Ding, Peng and Li, Fan},
  journal={Biometrika},
  volume={111},
  number={4},
  pages={1413--1420},
  year={2024},
  doi={10.1093/biomet/asae017},
  publisher={Oxford University Press}
}

@article{belloni2014inference,
  title={Inference on treatment effects after selection among high-dimensional controls},
  author={Belloni, Alexandre and Chernozhukov, Victor and Hansen, Christian},
  journal={Review of Economic Studies},
  volume={81},
  number={2},
  pages={608--650},
  year={2014},
  doi={10.1093/restud/rdt044},
  publisher={Oxford University Press}
}

@article{lee2011weight,
  doi={10.1371/journal.pone.0018174},
  title={Weight trimming and propensity score weighting},
  author={Lee, Brian K and Lessler, Justin and Stuart, Elizabeth A},
  journal={PLOS One},
  volume={6},
  number={3},
  pages={e18174},
  year={2011},
  publisher={Public Library of Science San Francisco, USA}
}

@article{cole2008constructing,
  doi={doi.org/10.1093/aje/kwn164},
  title={Constructing inverse probability weights for marginal structural models},
  author={Cole, Stephen R and Hern{\'a}n, Miguel A},
  journal={American Journal of Epidemiology},
  volume={168},
  number={6},
  pages={656--664},
  year={2008},
  publisher={Oxford University Press}
}

@article{rubin1974estimating,
  doi={10.1037/h0037350},
  title={Estimating causal effects of treatments in randomized and nonrandomized studies.},
  author={Rubin, Donald B},
  journal={Journal of educational Psychology},
  volume={66},
  number={5},
  pages={688},
  year={1974},
  publisher={American Psychological Association}
}

@article{neyman1990applications,
    doi={10.1214/ss/1177012031},
	title={On the Application of Probability Theory to Agricultural Experiments. Essay on Principles. Section 9},
	author={Jerzy Splawa-Neyman and D. M. Dabrowska and T. P. Speed},
	journal={Statistical Science},
	year={1990}
}

@article{ogburn2015doubly,
  doi={10.1111/rssb.12078},
  title={Doubly robust estimation of the local average treatment effect curve},
  author={Ogburn, Elizabeth L and Rotnitzky, Andrea and Robins, James M},
  journal={Journal of the Royal Statistical Society Series B: Statistical Methodology},
  volume={77},
  number={2},
  pages={373--396},
  year={2015},
  publisher={Oxford University Press}
}

@article{li2019propensity,
  doi={10.1214/19-AOAS1282},
  title={Propensity score weighting for causal inference with multiple treatments},
  author={Li, Fan and Li, Fan},
  journal={The Annals of Applied Statistics},
  volume={13},
  number={4},
  pages={2389--2415},
  year={2019},
  publisher={JSTOR}
}

@article{li2019addressing,
  doi={10.1093/aje/kwy201},
  title={Addressing extreme propensity scores via the overlap weights},
  author={Li, Fan and Thomas, Laine E and Li, Fan},
  journal={American Journal of Epidemiology},
  volume={188},
  number={1},
  pages={250--257},
  year={2019},
  publisher={Oxford University Press}
}

@article{thomas2020overlap,
  doi={10.1001/jama.2020.7819},
  title={Overlap weighting: a propensity score method that mimics attributes of a randomized clinical trial},
  author={Thomas, Laine E and Li, Fan and Pencina, Michael J},
  journal={JAMA},
  volume={323},
  number={23},
  pages={2417--2418},
  year={2020},
  publisher={American Medical Association}
}

@article{zhou2020propensity,
  doi={10.1177/09622802209403},
  title={Propensity score weighting under limited overlap and model misspecification},
  author={Zhou, Yunji and Matsouaka, Roland A and Thomas, Laine},
  journal={Statistical Methods in Medical Research},
  volume={29},
  number={12},
  pages={3721--3756},
  year={2020},
  publisher={SAGE Publications Sage UK: London, England}
}

@article{chaudhuri2025heavy,
  title={Heavy tail robust estimation and inference for average treatment effects},
  author={Chaudhuri, Saraswata and Hill, Jonathan B},
  journal={Econometric Reviews},
  volume={44},
  number={5},
  pages={544--586},
  year={2025},
  doi={10.1080/07474938.2024.2444229},
  publisher={Taylor \& Francis}
}

@article{sasaki2022estimation,
  doi={10.1017/S0266466621000025},
  title={Estimation and inference for moments of ratios with robustness against large trimming bias},
  author={Sasaki, Yuya and Ura, Takuya},
  journal={Econometric Theory},
  volume={38},
  number={1},
  pages={66--112},
  year={2022},
  publisher={Cambridge University Press}
}

@article{ma2020robust,
  doi={10.1080/01621459.2019.1660173},
  title={Robust inference using inverse probability weighting},
  author={Ma, Xinwei and Wang, Jingshen},
  journal={Journal of the American Statistical Association},
  volume={115},
  number={532},
  pages={1851--1860},
  year={2020},
  publisher={Taylor \& Francis}
}

@article{petersen2012diagnosing,
  doi={10.1177/0962280210386207},
  title={Diagnosing and responding to violations in the positivity assumption},
  author={Petersen, Maya L and Porter, Kristin E and Gruber, Susan and Wang, Yue and Van Der Laan, Mark J},
  journal={Statistical Methods in Medical Research},
  volume={21},
  number={1},
  pages={31--54},
  year={2012},
  publisher={Sage Publications Sage UK: London, England}
}

@article{crump2009dealing,
  doi={10.1093/biomet/asn055},
  title={Dealing with limited overlap in estimation of average treatment effects},
  author={Crump, Richard K and Hotz, V Joseph and Imbens, Guido W and Mitnik, Oscar A},
  journal={Biometrika},
  volume={96},
  number={1},
  pages={187--199},
  year={2009},
  publisher={Oxford University Press}
}

@article{cheng2022addressing,
  doi={doi.org/10.1093/aje/kwac043},
  title={Addressing extreme propensity scores in estimating counterfactual survival functions via the overlap weights},
  author={Cheng, Chao and Li, Fan and Thomas, Laine E and Li, Fan},
  journal={American Journal of Epidemiology},
  volume={191},
  number={6},
  pages={1140--1151},
  year={2022},
  publisher={Oxford University Press}
}

@article{heckman1998matching,
  doi={10.1111/1467-937X.00044},
  title={Matching as an econometric evaluation estimator},
  author={Heckman, James J and Ichimura, Hidehiko and Todd, Petra},
  journal={The Review of Economic Studies},
  volume={65},
  number={2},
  pages={261--294},
  year={1998},
  publisher={Wiley-Blackwell}
}

@article{abadie2005semiparametric,
 doi={10.1111/0034-6527.00321},
  title={Semiparametric difference-in-differences estimators},
  author={Abadie, Alberto},
  journal={The review of economic studies},
  volume={72},
  number={1},
  pages={1--19},
  year={2005},
  publisher={Wiley-Blackwell}
}

@article{crump2006moving,
    title={Moving the Goalposts: Addressing Limited Overlap in the Estimation of Average Treatment Effects by Changing the Estimand},
    author={Crump, Richard K and Hotz, V. Joseph and Imbens, Guido W and Mitnik, Oscar A},
    journal={National Bureau of Economic Research (Working Paper)},
    publisher={Technical Working Paper Series},
    volume={t0330},
    pages={1--48},
    year={2006},
    doi={10.3386/t0330}
}

@article{hirano2003efficient,
  doi={10.1111/1468-0262.00442},
  title={Efficient estimation of average treatment effects using the estimated propensity score},
  author={Hirano, Keisuke and Imbens, Guido W and Ridder, Geert},
  journal={Econometrica},
  volume={71},
  number={4},
  pages={1161--1189},
  year={2003},
  publisher={Wiley Online Library}
}

@article{li2013weighting,
    doi={10.1515/ijb-2012-0030},
	author   ={Li, Liang and Greene, Tom},
	title    ={A weighting analogue to pair matching in propensity score analysis},
	pages    ={215--234},
	volume   ={9},
	journal  ={The International Journal of Biostatistics},
	publisher={De Gruyter},
	year     ={2013},
}

@article{li2018balancing,
    doi={10.1080/01621459.2016.1260466},
	author   ={Li, Fan and Morgan, Kari Lock and Zaslavsky, Alan M},
	title    ={Balancing covariates via propensity score weighting},
	pages    ={390-400},
	volume   ={113},
	journal  ={J Am Stat Assoc},
	publisher={Taylor \& Francis},
	year     ={2018},
}

@article{mao2019propensity,
    doi={10.1177/0962280218781171},
	author   ={Mao, Huzhang and Li, Liang and Greene, Tom},
	title    ={Propensity score weighting analysis and treatment effect discovery},
	pages    ={2439--2454},
	volume   ={28},
	journal  ={Statistical Methods in Medical Research},
	publisher={SAGE Publications Sage UK: London, England},
	year     ={2019},
}

@article{poteat2025social,
  title={Social determinants of HIV status and viral load suppression among transgender women in South Africa: a cross-sectional analysis},
  author={Poteat, Tonia and Liu, Yi and Adams, Darya and van der Merwe, L Leigh-Ann and Cloete, Allanise and Howard, Lauren E and McCarthy, Janice},
  journal={AIDS care},
  volume={37},
  number={9},
  pages={1507--1520},
  year={2025},
  doi={10.1080/09540121.2025.2535471},
  publisher={Taylor \& Francis}
}

@article{tao2019doubly,
    doi={10.1002/sim.7980},
	author   ={Tao, Yebin and Fu, Haoda},
	title    ={Doubly robust estimation of the weighted average treatment effect for a target population},
	pages    ={315--325},
	volume   ={38},
	journal  ={Statistics in Medicine},
	publisher={Wiley Online Library},
	year     ={2019},
}

@article{poteat2020prep,
  doi={10.1016/S2352-3018(20)30119-3},
  title={PrEP awareness and engagement among transgender women in South Africa: a cross-sectional, mixed methods study},
  author={Poteat, Tonia and Malik, Mannat and van der Merwe, L Leigh Ann and Cloete, Allanise and Adams, Dee and Nonyane, Bareng AS and Wirtz, Andrea L},
  journal={The Lancet HIV},
  volume={7},
  number={12},
  pages={e825--e834},
  year={2020},
  publisher={Elsevier}
}

@article{cao2024using,
  title={Using Overlap Weights to Address Extreme Propensity Scores in Estimating Restricted Mean Counterfactual Survival Times},
  author={Cao, Zhiqiang and Ghazi, Lama and Mastrogiacomo, Claudia and Forastiere, Laura and Wilson, F Perry and Li, Fan},
  journal={American Journal of Epidemiology},
  volume={194},
  pages={2402--2411},
  year={2024},
  doi={10.1093/aje/kwae416},
  publisher={Oxford University Press}
}

@article{zhao2019covariate,
  title={Covariate Balancing Propensity Score by Tailored Loss Functions},
  author={Zhao, Qingyuan},
  journal={The Annals of Statistics},
  volume={47},
  number={2},
  pages={965--993},
  year={2019},
  doi={10.1214/18-AOS1698},
  publisher={JSTOR}
}

\newpage

\appendix

\begin{center}\Large\bf
Online Supplemental Material 
\end{center}

\numberwithin{figure}{section}
\numberwithin{table}{section}

\section{Supplemental Figures from the Simulation Results}

\begin{figure}[ht]
    \centering
    \includegraphics[width=0.9\linewidth]{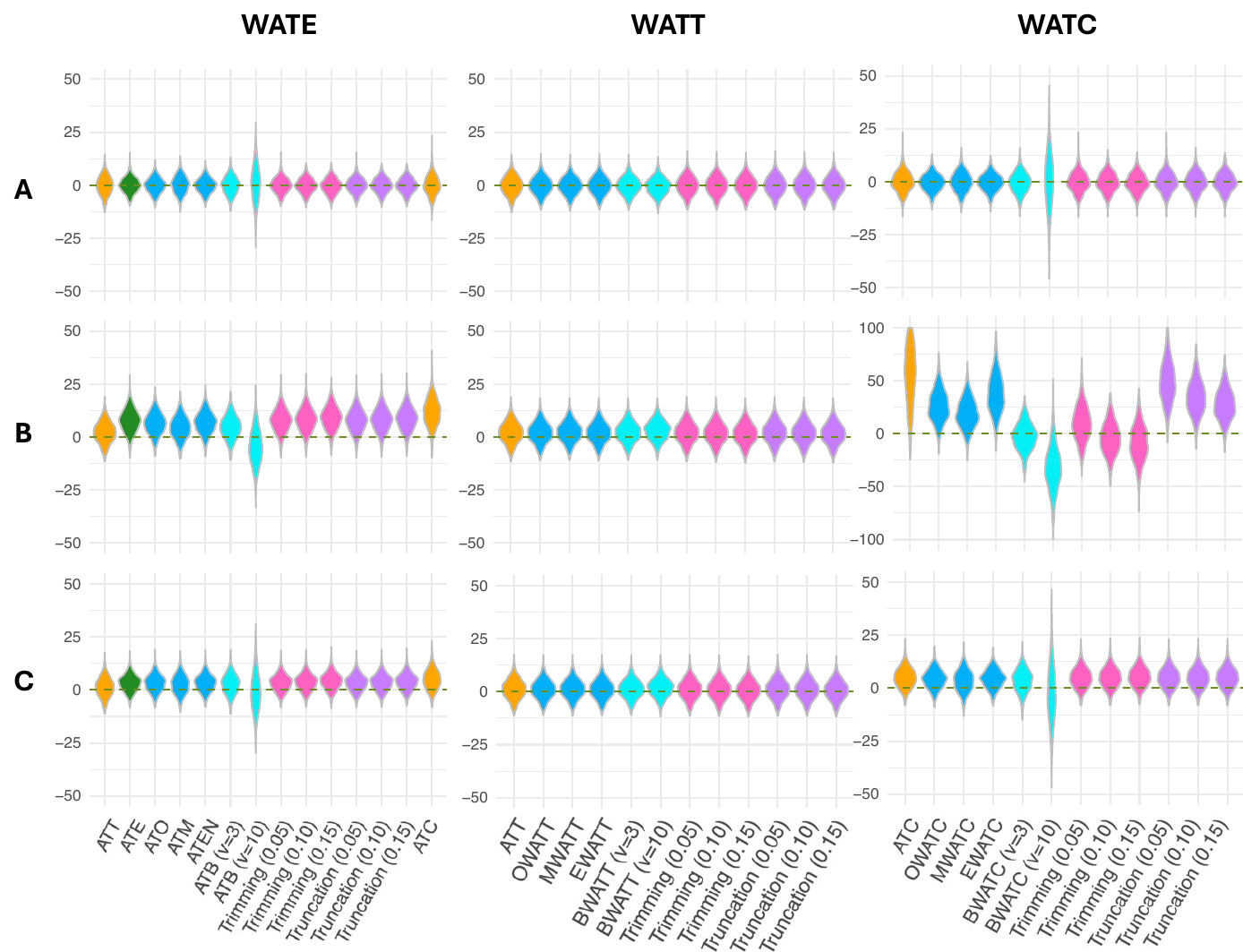}
    \caption{Estimation bias violin plots under good overlap. }
    \label{fig:simBiasgood}
\end{figure}

\begin{figure}[ht]
    \centering
    \includegraphics[width=0.75\linewidth]{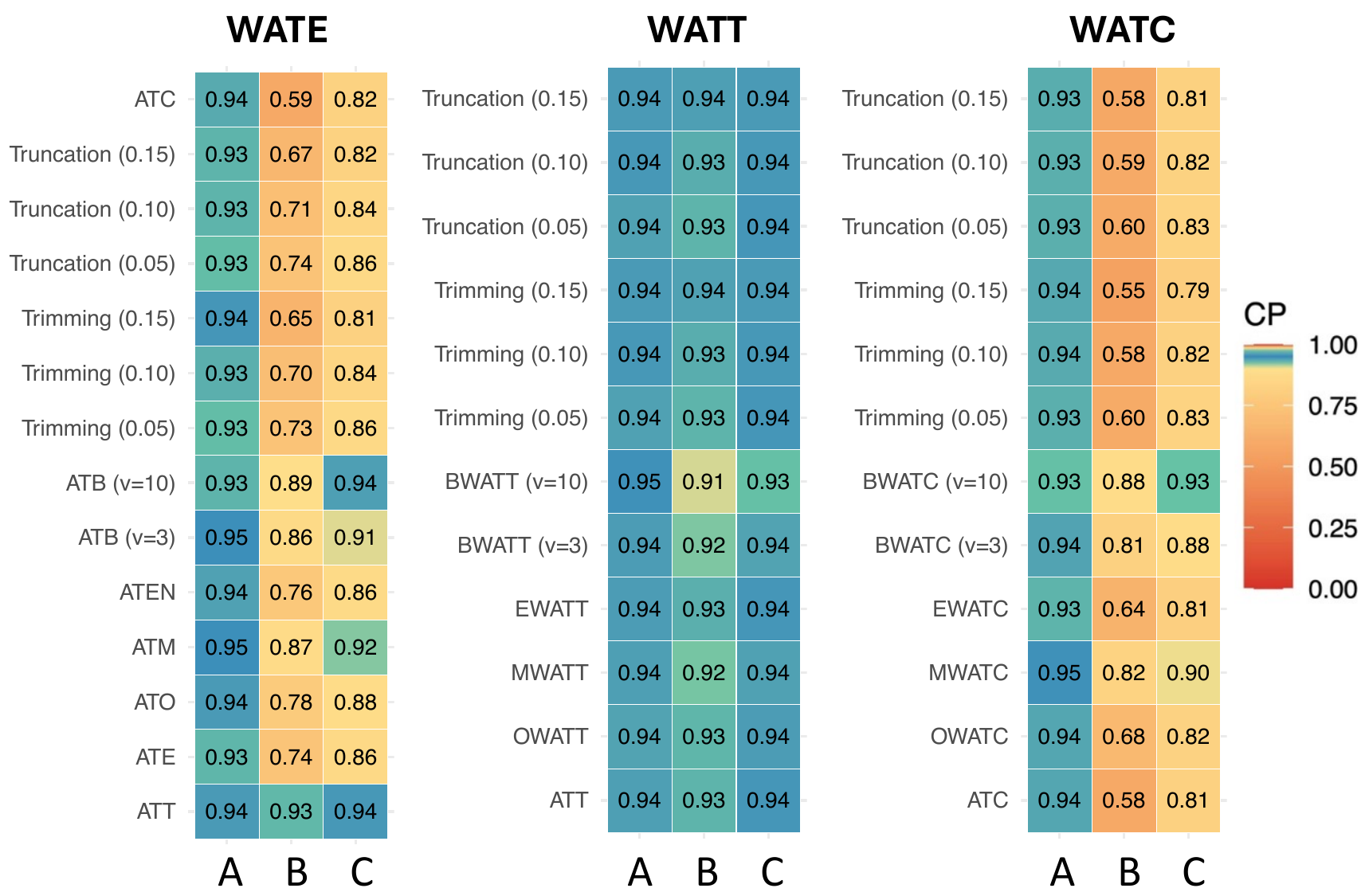}
    \caption{Coverage probability under good overlap. }
    \label{fig:simCPgood}
\end{figure}

\newpage

\section{Supplemental Tables and Figures for Data Applications}

\subsection{Study I: Effects of smoking on blood lead level}

\begin{table}[ht]
\centering
\caption{Effective sample size (ESS) of different weighting methods.}

\begin{tabular}{llrrr}
  \toprule
Estimand class & Weighting method & ESS ($A=1$) & ESS ($A=0$) & ESS (total) \\ 
& & ($N=679$) & ($N=2,661$) & ($N=3,340$)\\
  \midrule
 & IPW (ATE) & 197.9 & 2480.6 & 2678.5 \\ 
   & IPW (ATT) & 679.0 & 978.5 & 1657.5 \\ 
   & IPW (ATC) & 143.6 & 2661.0 & 2804.6 \\ 
   & OW & 622.0 & 1526.7 & 2148.7 \\ 
   & MW & 652.3 & 1243.2 & 1895.5 \\ 
   & EW & 568.4 & 1723.9 & 2292.3 \\ 
  WATE & BW ($\nu = 3$) & 622.6 & 1049.0 & 1671.6 \\ 
   & BW ($\nu = 10$) & 391.8 & 430.3 & 822.1 \\ 
   & Trimming ($\alpha=0.05$) & 395.4 & 1927.6 & 2323.0 \\ 
   & Truncation ($\alpha=0.05$) & 346.0 & 2333.6 & 2679.6 \\ 
   & Trimming ($\alpha=0.1$) & 454.5 & 1406.8 & 1861.3 \\ 
   & Truncation ($\alpha=0.1$) & 459.7 & 2085.5 & 2545.3 \\ 
   & Trimming ($\alpha=0.15$) & 469.0 & 1050.8 & 1519.8 \\ 
   & Truncation ($\alpha=0.15$) & 529.6 & 1877.0 & 2406.6 \\
  \midrule
   & IPW (ATT) & 679.0 & 978.5 & 1657.5 \\ 
   & OW & 679.0 & 726.2 & 1405.2 \\ 
   & MW & 679.0 & 635.7 & 1314.7 \\ 
   & EW & 679.0 & 772.0 & 1451.0 \\ 
   & BW ($\nu = 3$) & 679.0 & 1049.0 & 1728.0 \\ 
  WATT & BW ($\nu = 10$) & 679.0 & 430.3 & 1109.3 \\ 
   & Trimming ($\alpha=0.05$) & 679.0 & 978.5 & 1657.5 \\ 
   & Truncation ($\alpha=0.05$) & 679.0 & 978.5 & 1657.5 \\ 
   & Trimming ($\alpha=0.1$) & 679.0 & 978.5 & 1657.5 \\ 
   & Truncation ($\alpha=0.1$) & 679.0 & 978.5 & 1657.5 \\ 
   & Trimming ($\alpha=0.15$) & 679.0 & 978.5 & 1657.5 \\ 
   & Truncation ($\alpha=0.15$) & 679.0 & 978.5 & 1657.5 \\ 
    \midrule
   & IPW (ATC) & 143.6 & 2661.0 & 2804.6 \\
   & OW & 521.8 & 2661.0 & 3182.8 \\ 
   & MW & 587.8 & 2661.0 & 3248.8 \\ 
   & EW & 451.5 & 2661.0 & 3112.5 \\ 
   & BW ($\nu = 3$) & 590.0 & 2661.0 & 3251.0 \\ 
   & BW ($\nu = 10$) & 394.8 & 2661.0 & 3055.8 \\ 
  WATC & Trimming ($\alpha=0.05$) & 298.6 & 2661.0 & 2959.6 \\ 
   & Truncation ($\alpha=0.05$) & 262.7 & 2661.0 & 2923.7 \\ 
   & Trimming ($\alpha=0.1$) & 355.3 & 2661.0 & 3016.3 \\ 
   & Truncation ($\alpha=0.1$) & 365.0 & 2661.0 & 3026.0 \\ 
   & Trimming ($\alpha=0.15$) & 380.9 & 2661.0 & 3041.9 \\ 
   & Truncation ($\alpha=0.15$) & 436.6 & 2661.0 & 3097.6 \\ 
   \bottomrule
\end{tabular}
\end{table}

\begin{figure}[ht]
    \centering
    \includegraphics[width=0.7\linewidth]{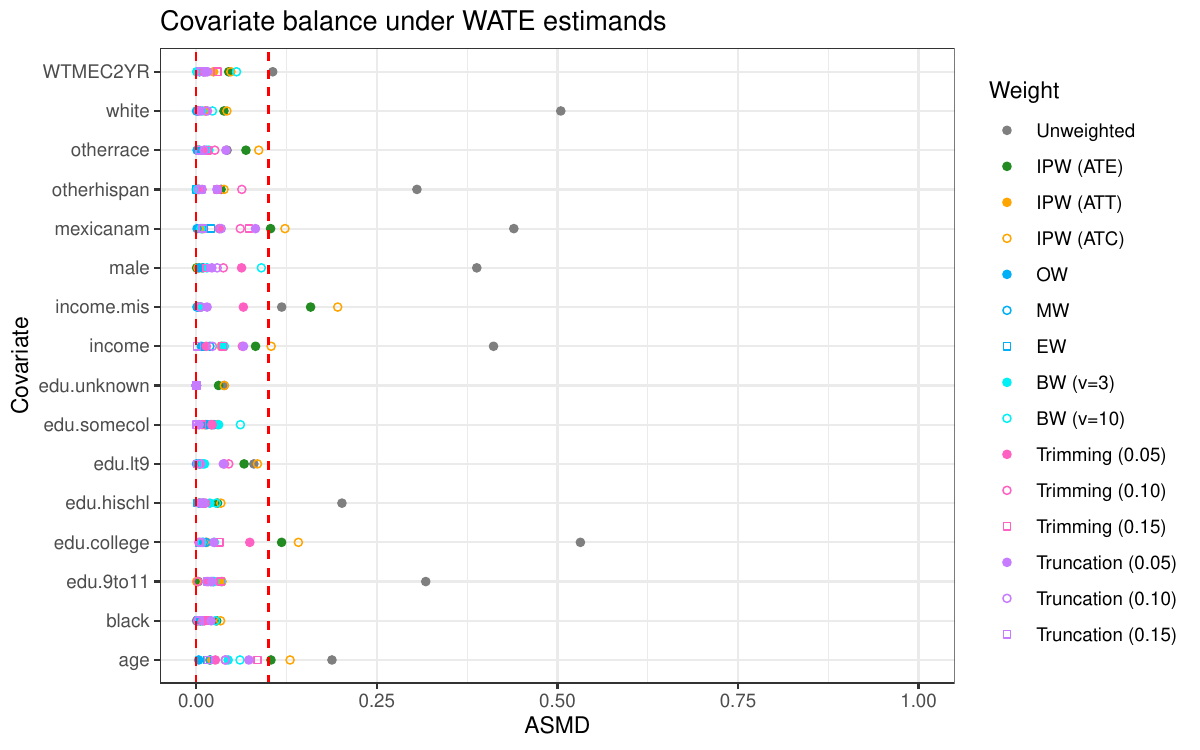}
    \caption{ASMDs using weights from WATE estimands (the red dashed lines indicate 0 and 0.1). }
    \footnotetext[0\def\thefoornote{}]{Comments on the covariate balance: Unweighted, BW ($\nu=10$) and Trimming (0.10 and 0.15) often yields large ASMDs, followed by IPW (ATE) and IPW (ATC) that occasionally have ASMDs beyond 0.1. All other methods balance covariates well, with their ASMDs below 0.1.}
    \label{fig:ASMD-blood-WATE}
\end{figure}

\begin{figure}[ht]
    \centering
    \includegraphics[width=0.7\linewidth]{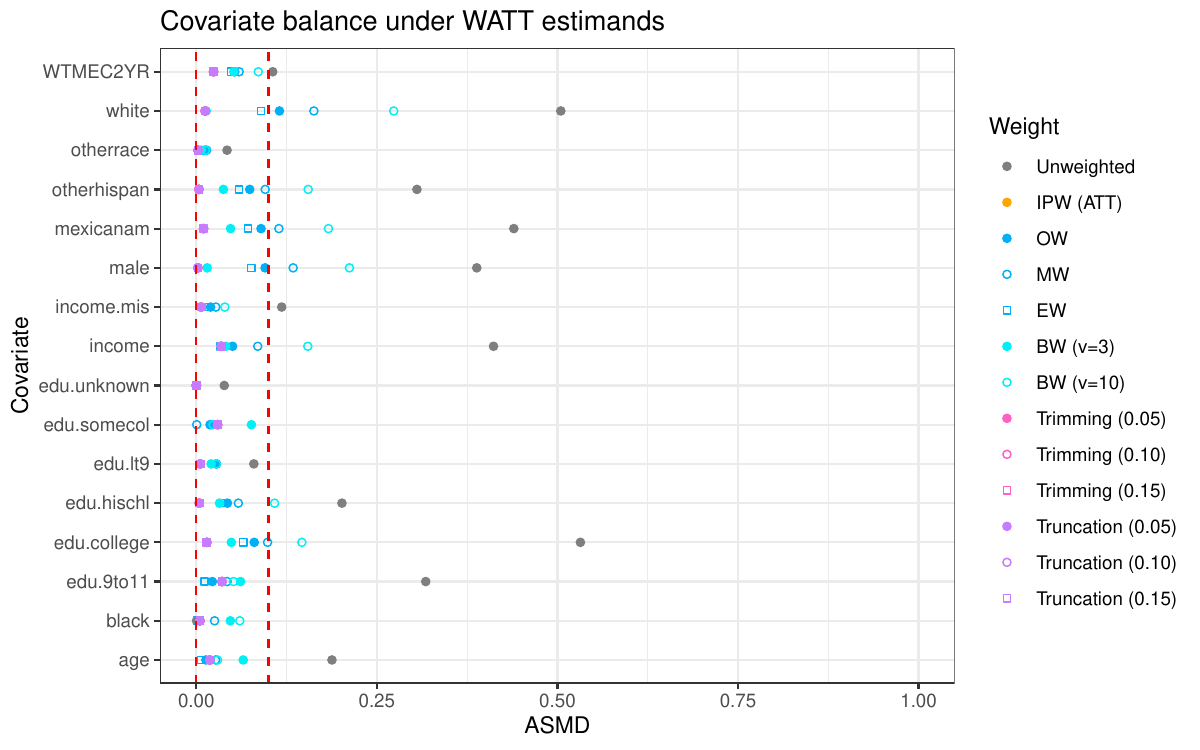}
    \caption{ASMDs using weights from WATT estimands (red dashed lines indicate 0 and 0.1).}
    \footnotetext[0\def\thefoornote{}]{Comments on the covariate balance: Except for Unweighted, BW $(\nu=10)$ and some cases of MW, all methods balance covariates well within the 0.1 ASMD threshold. }
    \label{fig:ASMD-blood-WATT}
\end{figure}

\begin{figure}[ht]
    \centering
    \includegraphics[width=0.7\linewidth]{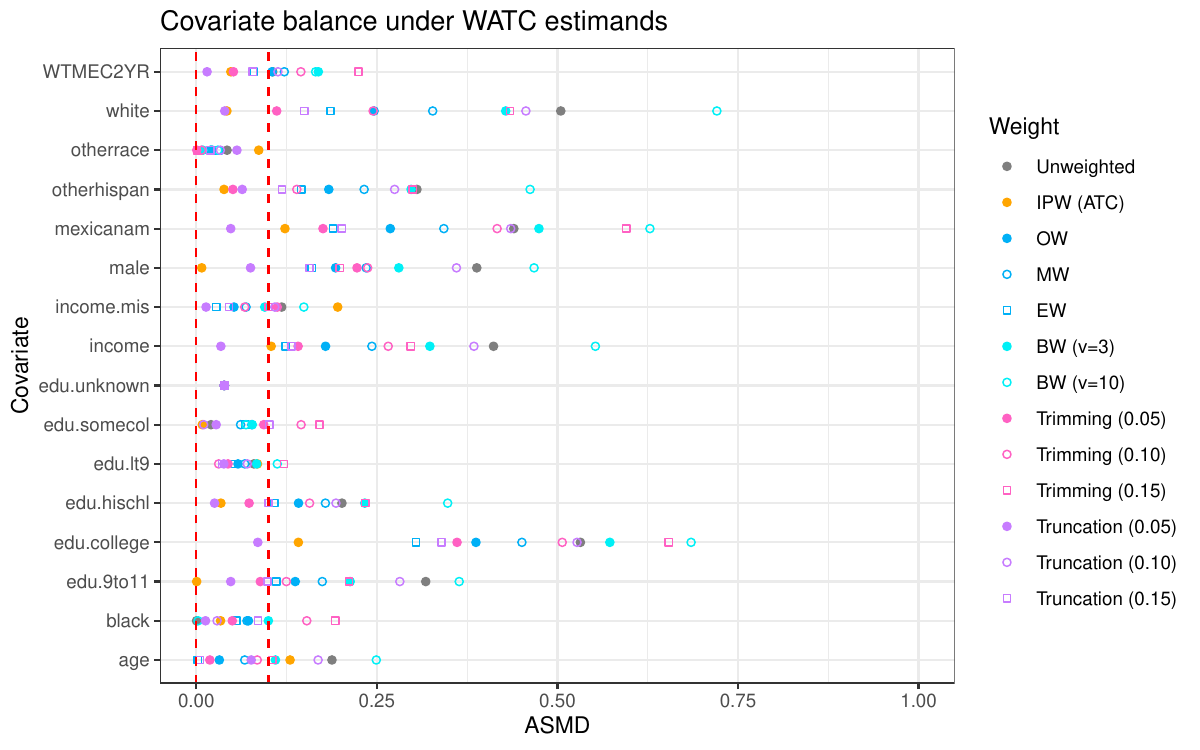}
    \caption{ASMDs using weights from WATC estimands (red dashed lines indicate 0 and 0.1).} \footnotetext[0\def\thefoornote{}]{Comments on the  covariate balance: Most methods do not balance covariates well in this example, while IPW (ATC), Truncation ($\alpha=0.05$), EW and OW perform overall better, with their ASMDs below 0.1. }
    \label{fig:ASMD-blood-WATC}
\end{figure}

\begin{table}[ht]
    \centering
    
    \caption{Covariate balance for age (overall mean age: 49.15 years). }
    \begin{tabular}{rrrrl}
  \toprule
Mean ($A=1$) & Mean ($A=0$)  & Raw difference & ASMD & Weighting method \\
\midrule
  \multicolumn{5}{c}{\bf WATE balancing} \\
  \addlinespace
46.71 & 49.78 & -3.08 & 0.19 & Unweighted \\ 
  50.91 & 49.21 & 1.69 & 0.10 & IPW (ATE) \\ 
  46.71 & 47.02 & -0.31 & 0.02 & IPW (ATT) \\ 
  51.91 & 49.78 & 2.13 & 0.13 & IPW (ATC) \\ 
  48.20 & 48.26 & -0.06 & 0.00 & OW \\ 
  47.64 & 47.95 & -0.31 & 0.02 & MW \\ 
  48.68 & 48.43 & 0.25 & 0.02 & EW \\ 
  47.05 & 47.77 & -0.72 & 0.04 & BW ($\nu=3$) \\ 
  45.23 & 46.23 & -0.99 & 0.06 & BW ($\nu=10$) \\ 
  48.64 & 48.21 & 0.43 & 0.03 & Trimming ($\alpha=0.05$) \\ 
  50.08 & 48.88 & 1.19 & 0.07 & Truncation ($\alpha=0.05$) \\ 
  47.64 & 48.30 & -0.66 & 0.04 & Trimming ($\alpha=0.1$) \\ 
  49.26 & 48.60 & 0.66 & 0.04 & Truncation ($\alpha=0.1$) \\ 
  47.16 & 48.55 & -1.39 & 0.08 & Trimming ($\alpha=0.15$) \\ 
  48.84 & 48.57 & 0.27 & 0.02 & Truncation ($\alpha=0.15$) \\ 
  \midrule
  \multicolumn{5}{c}{\bf WATT balancing} \\
  \addlinespace
46.71 & 49.78 & -3.08 & 0.19 & Unweighted \\ 
  46.71 & 47.02 & -0.31 & 0.02 & IPW (ATT) \\ 
  46.71 & 46.49 & 0.22 & 0.01 & OW \\ 
  46.71 & 46.27 & 0.44 & 0.03 & MW \\ 
  46.71 & 46.60 & 0.11 & 0.01 & EW \\ 
  46.71 & 47.77 & -1.06 & 0.07 & BW ($\nu=3$) \\ 
  46.71 & 46.23 & 0.48 & 0.03 & BW ($\nu=10$) \\ 
  46.71 & 47.02 & -0.31 & 0.02 & Trimming ($\alpha=0.05$) \\ 
  46.71 & 47.02 & -0.31 & 0.02 & Truncation ($\alpha=0.05$) \\ 
  46.71 & 47.02 & -0.31 & 0.02 & Trimming ($\alpha=0.1$) \\ 
  46.71 & 47.02 & -0.31 & 0.02 & Truncation ($\alpha=0.1$) \\ 
  46.71 & 47.02 & -0.31 & 0.02 & Trimming ($\alpha=0.15$) \\ 
  46.71 & 47.02 & -0.31 & 0.02 & Truncation ($\alpha=0.15$) \\ 
  \midrule
  \multicolumn{5}{c}{\bf WATC balancing} \\
  \addlinespace
  46.71 & 49.78 & -3.08 & 0.19 & Unweighted \\ 
  51.91 & 49.78 & 2.13 & 0.13 & IPW (ATC) \\ 
  49.26 & 49.78 & -0.52 & 0.03 & OW \\ 
  48.68 & 49.78 & -1.10 & 0.07 & MW \\ 
  49.75 & 49.78 & -0.03 & 0.00 & EW \\ 
  47.99 & 49.78 & -1.79 & 0.11 & BW ($\nu=3$) \\ 
  45.70 & 49.78 & -4.08 & 0.25 & BW ($\nu=10$) \\ 
  49.48 & 49.78 & -0.31 & 0.02 & Trimming ($\alpha=0.05$) \\ 
  51.03 & 49.78 & 1.25 & 0.08 & Truncation ($\alpha=0.05$) \\ 
  48.40 & 49.78 & -1.38 & 0.08 & Trimming ($\alpha=0.1$) \\ 
  50.14 & 49.78 & 0.35 & 0.02 & Truncation ($\alpha=0.1$) \\ 
  48.06 & 49.78 & -1.73 & 0.11 & Trimming ($\alpha=0.15$) \\ 
  49.70 & 49.78 & -0.08 & 0.01 & Truncation ($\alpha=0.15$) \\ 
  \bottomrule
\end{tabular}
\end{table}

\begin{table}[ht]
    \centering
    
    \caption{Covariate balance for income-to-poverty level (overall mean income: 2.49). }
    \begin{tabular}{rrrrl}
  \toprule
Mean ($A=1$) & Mean ($A=0$)  & Raw difference & ASMD & Weighting method \\
\midrule
  \multicolumn{5}{c}{\bf WATE balancing} \\
  \addlinespace
2.00 & 2.62 & -0.61 & 0.41 & Unweighted \\ 
  2.62 & 2.50 & 0.12 & 0.08 & IPW (ATE) \\ 
  2.00 & 2.06 & -0.05 & 0.03 & IPW (ATT) \\ 
  2.77 & 2.62 & 0.15 & 0.10 & IPW (ATC) \\ 
  2.20 & 2.21 & -0.01 & 0.01 & OW \\ 
  2.10 & 2.13 & -0.03 & 0.02 & MW \\ 
  2.28 & 2.26 & 0.01 & 0.01 & EW \\ 
  2.01 & 2.06 & -0.06 & 0.04 & BW ($\nu=3$) \\ 
  1.72 & 1.77 & -0.05 & 0.04 & BW ($\nu=10$) \\ 
  2.28 & 2.31 & -0.02 & 0.01 & Trimming ($\alpha=0.05$) \\ 
  2.52 & 2.42 & 0.10 & 0.07 & Truncation ($\alpha=0.05$) \\ 
  2.12 & 2.21 & -0.10 & 0.06 & Trimming ($\alpha=0.1$) \\ 
  2.38 & 2.34 & 0.03 & 0.02 & Truncation ($\alpha=0.1$) \\ 
  2.06 & 2.11 & -0.05 & 0.04 & Trimming ($\alpha=0.15$) \\ 
  2.30 & 2.30 & 0.00 & 0.00 & Truncation ($\alpha=0.15$) \\ 
  \midrule
  \multicolumn{5}{c}{\bf WATT balancing} \\
  \addlinespace
2.00 & 2.62 & -0.61 & 0.41 & Unweighted \\ 
  2.00 & 2.06 & -0.05 & 0.03 & IPW (ATT) \\ 
  2.00 & 1.93 & 0.07 & 0.05 & OW \\ 
  2.00 & 1.88 & 0.13 & 0.09 & MW \\ 
  2.00 & 1.95 & 0.05 & 0.03 & EW \\ 
  2.00 & 2.06 & -0.06 & 0.04 & BW ($\nu=3$) \\ 
  2.00 & 1.77 & 0.23 & 0.15 & BW ($\nu=10$) \\ 
  2.00 & 2.06 & -0.05 & 0.03 & Trimming ($\alpha=0.05$) \\ 
  2.00 & 2.06 & -0.05 & 0.03 & Truncation ($\alpha=0.05$) \\ 
  2.00 & 2.06 & -0.05 & 0.03 & Trimming ($\alpha=0.1$) \\ 
  2.00 & 2.06 & -0.05 & 0.03 & Truncation ($\alpha=0.1$) \\ 
  2.00 & 2.06 & -0.05 & 0.03 & Trimming ($\alpha=0.15$) \\ 
  2.00 & 2.06 & -0.05 & 0.03 & Truncation ($\alpha=0.15$) \\ 
  \midrule
  \multicolumn{5}{c}{\bf WATC balancing} \\
  \addlinespace
  2.00 & 2.62 & -0.61 & 0.41 & Unweighted \\ 
  2.77 & 2.62 & 0.15 & 0.10 & IPW (ATC) \\ 
  2.35 & 2.62 & -0.27 & 0.18 & OW \\ 
  2.25 & 2.62 & -0.36 & 0.24 & MW \\ 
  2.43 & 2.62 & -0.18 & 0.12 & EW \\ 
  2.14 & 2.62 & -0.48 & 0.32 & BW ($\nu=3$) \\ 
  1.79 & 2.62 & -0.82 & 0.55 & BW ($\nu=10$) \\ 
  2.41 & 2.62 & -0.21 & 0.14 & Trimming ($\alpha=0.05$) \\ 
  2.67 & 2.62 & 0.05 & 0.03 & Truncation ($\alpha=0.05$) \\ 
  2.22 & 2.62 & -0.40 & 0.27 & Trimming ($\alpha=0.1$) \\ 
  2.51 & 2.62 & -0.11 & 0.07 & Truncation ($\alpha=0.1$) \\ 
  2.17 & 2.62 & -0.44 & 0.30 & Trimming ($\alpha=0.15$) \\ 
  2.42 & 2.62 & -0.20 & 0.13 & Truncation ($\alpha=0.15$) \\ 
  \bottomrule
\end{tabular}
\end{table}

\begin{table}[ht]
\centering
\caption{Estimated weighted treatment effects (point estimate and 95\% CI) of smoking on blood lead level under WATE, WATT, and WATC frameworks.}
\begin{tabular}{lccc}
\toprule
Weighting method & WATE & WATT & WATC \\
\midrule
Overall & 0.85 (0.59, 1.10) & 0.83 (0.62, 1.03) & 0.82 (0.38, 1.26) \\
Treated & 0.83 (0.62, 1.03) & - & - \\
Control & 0.82 (0.38, 1.26) & - & - \\
Overlap (OW) & 0.84 (0.67, 1.01) & 0.81 (0.60, 1.02) & 0.97 (0.73, 1.22) \\
Matching (MW) & 0.88 (0.69, 1.07) & 0.83 (0.61, 1.05) & 1.02 (0.77, 1.27) \\
Entropy (EW) & 0.84 (0.67, 1.00) & 0.81 (0.61, 1.01) & 0.95 (0.69, 1.20) \\
Beta (BW; $\nu=3$) & 0.86 (0.66, 1.06) & 0.82 (0.59, 1.04) & 1.04 (0.78, 1.30) \\
Beta (BW; $\nu=10$) & 1.06 (0.64, 1.49) & 0.90 (0.57, 1.23) & 1.19 (0.80, 1.58) \\
Trimming ($\alpha=0.05$) & 0.80 (0.60, 1.00) & 0.83 (0.63, 1.02) & 0.98 (0.66, 1.30) \\
Trimming ($\alpha=0.1$)  & 0.77 (0.57, 0.98) & 0.83 (0.63, 1.02) & 0.96 (0.65, 1.27) \\
Trimming ($\alpha=0.15$) & 0.75 (0.52, 0.98) & 0.83 (0.63, 1.02) & 0.98 (0.68, 1.27) \\
Truncation ($\alpha=0.05$) & 0.83 (0.52, 1.14) & 0.83 (0.63, 1.02) & 0.86 (0.53, 1.19) \\
Truncation ($\alpha=0.1$)  & 0.82 (0.50, 1.14) & 0.83 (0.63, 1.02) & 0.90 (0.61, 1.18) \\
Truncation ($\alpha=0.15$) & 0.84 (0.51, 1.17) & 0.83 (0.63, 1.02) & 0.93 (0.67, 1.20) \\
\bottomrule
\end{tabular}
\begin{tablenotes}\footnotesize
    \item The ``Treated'' and ``Control'' entries are not meaningful for the WATT and WATC frameworks. Under WATE, the ``Treated'' (resp. ``Control'') results correspond exactly to the ``Overall'' results under WATT (resp. WATC), as they represent the ATT (resp. ATC) by definition. 
\end{tablenotes}
\end{table}

\begin{figure}[ht]
    \centering
    \includegraphics[width=\linewidth]{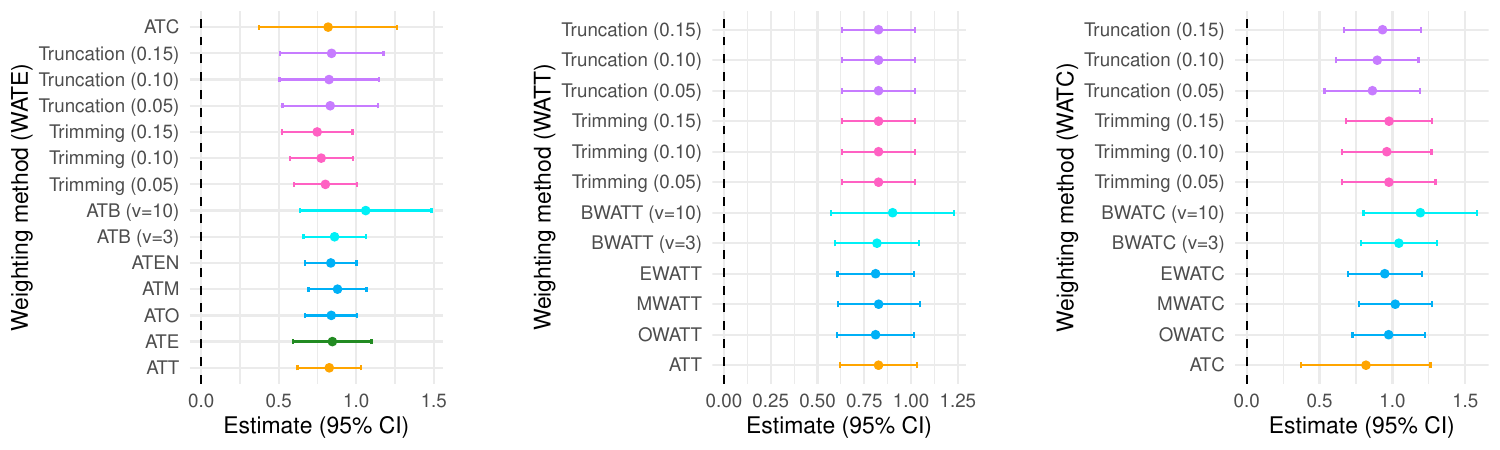}
    \caption{Forest plots for the estimated effects of smoking on blood levels. From left to right, the panels correspond to weighting methods (and estimands) in the WATE, WATT, and WATC classes, respectively. Within each class, forest plots present the estimation and 95\% CI results for the chosen weighting methods in that class (same as those considered in the simulation study).}\label{fig:forst-blood}
\end{figure}

\clearpage

\subsection{Study II: Effects of sex work history on HIV status among transgender women in South Africa}

\begin{table}[ht]
\centering
\caption{Effective sample size (ESS) of different weighting methods. }

\begin{tabular}{llrrr}
  \toprule
Estimand class & Weighting method & ESS ($A=1$) & ESS ($A=0$) & ESS (total) \\ 
& & ($N=128$) & ($N=68$) & ($N=196$)\\
  \midrule
   & IPW (ATE) & 107.2 & 34.7 & 141.9 \\ 
   & IPW (ATT) & 128.0 & 21.1 & 149.1 \\ 
   & IPW (ATC) & 48.5 & 68.0 & 116.5 \\ 
   & OW & 81.0 & 57.5 & 138.6 \\ 
   & MW & 73.3 & 59.7 & 132.9 \\ 
   & EW & 86.4 & 53.8 & 140.2 \\ 
  WATE & BW ($\nu=3$) & 67.1 & 59.7 & 126.8 \\ 
   & BW ($\nu=10$) & 38.3 & 38.9 & 77.2 \\ 
   & Trimming ($\alpha=0.05$) & 96.7 & 34.7 & 131.4 \\ 
   & Truncation ($\alpha=0.05$) & 97.2 & 34.7 & 131.9 \\ 
   & Trimming ($\alpha=0.1$) & 85.3 & 48.4 & 133.8 \\ 
   & Truncation ($\alpha=0.1$) & 87.1 & 50.8 & 137.9 \\ 
   & Trimming ($\alpha=0.15$) & 69.9 & 48.4 & 118.4 \\ 
   & Truncation ($\alpha=0.15$) & 74.6 & 50.9 & 125.5 \\ 
   \midrule
   & IPW (ATT) & 128.0 & 21.1 & 149.1 \\ 
   & OW & 128.0 & 40.6 & 168.6 \\ 
   & MW & 128.0 & 46.2 & 174.2 \\ 
   & EW & 128.0 & 36.1 & 164.1 \\ 
   & BW ($\nu=3$) & 128.0 & 59.7 & 187.7 \\ 
  WATT & BW ($\nu=10$) & 128.0 & 38.9 & 166.9 \\ 
   & Trimming ($\alpha=0.05$) & 128.0 & 21.1 & 149.1 \\ 
   & Truncation ($\alpha=0.05$) & 128.0 & 21.1 & 149.1 \\ 
   & Trimming ($\alpha=0.1$) & 128.0 & 31.7 & 159.7 \\ 
   & Truncation ($\alpha=0.1$) & 128.0 & 32.1 & 160.1 \\ 
   & Trimming ($\alpha=0.15$) & 128.0 & 31.7 & 159.7 \\ 
   & Truncation ($\alpha=0.15$) & 128.0 & 32.3 & 160.3 \\ 
   \midrule
   & IPW (ATC) & 48.5 & 68.0 & 116.5 \\ 
   & OW & 45.1 & 68.0 & 113.1 \\ 
   & MW & 44.5 & 68.0 & 112.5 \\ 
   & EW & 45.7 & 68.0 & 113.7 \\ 
   & BW ($\nu=3$) & 43.4 & 68.0 & 111.4 \\ 
  WATC & BW ($\nu=10$) & 32.6 & 68.0 & 100.6 \\ 
   & Trimming ($\alpha=0.05$) & 48.5 & 68.0 & 116.5 \\ 
   & Truncation ($\alpha=0.05$) & 48.5 & 68.0 & 116.5 \\ 
   & Trimming ($\alpha=0.1$) & 48.5 & 68.0 & 116.5 \\ 
   & Truncation ($\alpha=0.1$) & 48.5 & 68.0 & 116.5 \\ 
   & Trimming ($\alpha=0.15$) & 48.5 & 68.0 & 116.5 \\ 
   & Truncation ($\alpha=0.15$) & 48.5 & 68.0 & 116.5 \\  
   \bottomrule
\end{tabular}
\end{table}

\begin{figure}[ht]
    \centering
    \includegraphics[width=0.7\linewidth]{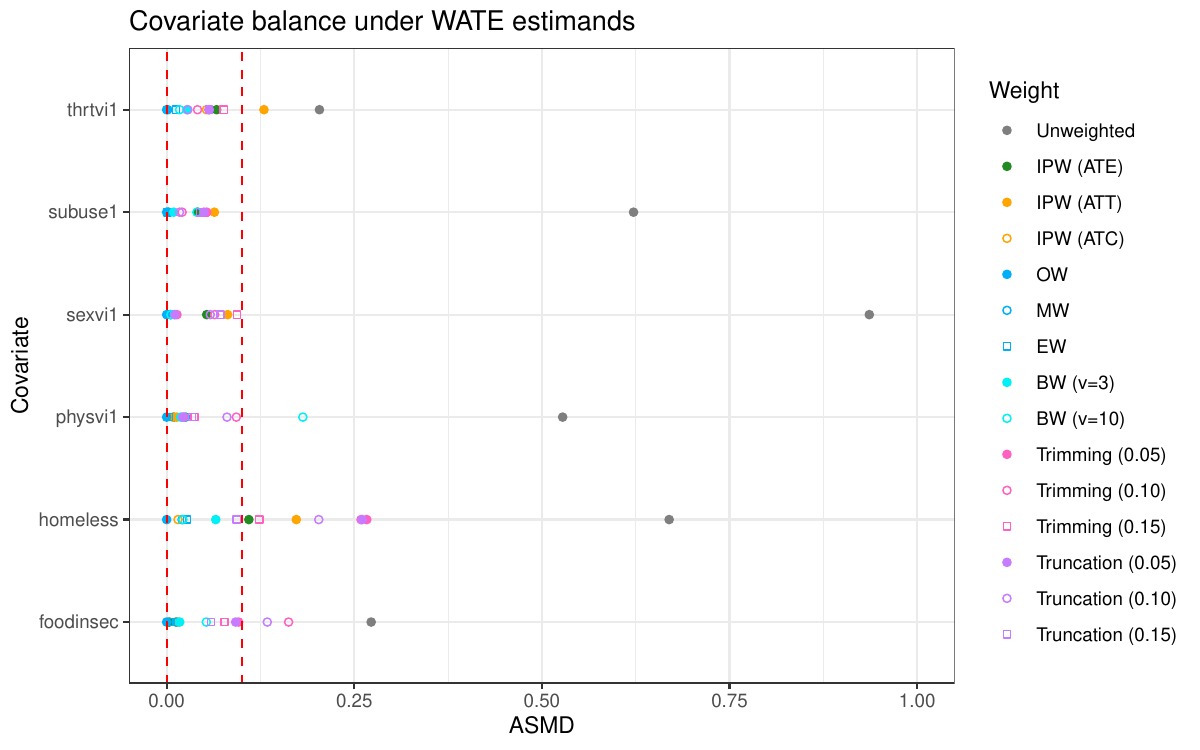}
    \caption{ASMDs using weights from WATE estimands (the red dashed lines indicate 0 and 0.1).} \footnotetext[0\def\thefoornote{}]{Comments on the covariate balance: Unweighted, BW ($\nu=10$), IPW (ATT) and Trimming (0.10 and 0.15) often yields large ASMDs, followed by IPW (ATE) and IPW (ATC) that occasionally have ASMDs beyond 0.1. All other methods balance covariates well, i.e., their ASMDs are all below the 0.1. }
\end{figure}

\begin{figure}[ht]
    \centering
    \includegraphics[width=0.7\linewidth]{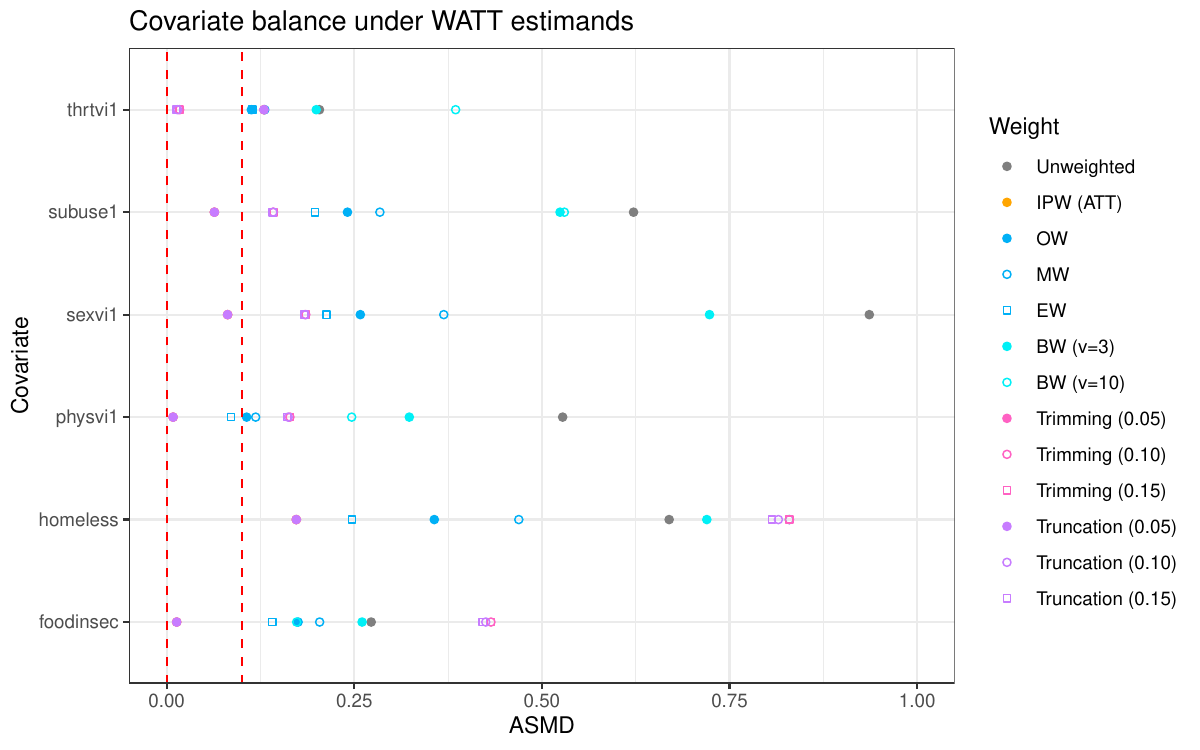}
    \caption{ASMDs using weights from WATT estimands (the red dashed lines indicate 0 and 0.1).}\footnotetext[0\def\thefoornote{}]{Comments on the covariate balance:  Many of the weights did not lead to well-balanced covariates. Overall, IPW (ATT), OW, MW, EW, Truncation (0.05) provide better balances, with some ASMDs within the interal [0.0, 0.1]. BW ($\nu=3$ and 10) as well as most trimming and truncation do not provide a good balance. }
\end{figure}

\begin{figure}[ht]
    \centering
    \includegraphics[width=0.7\linewidth]{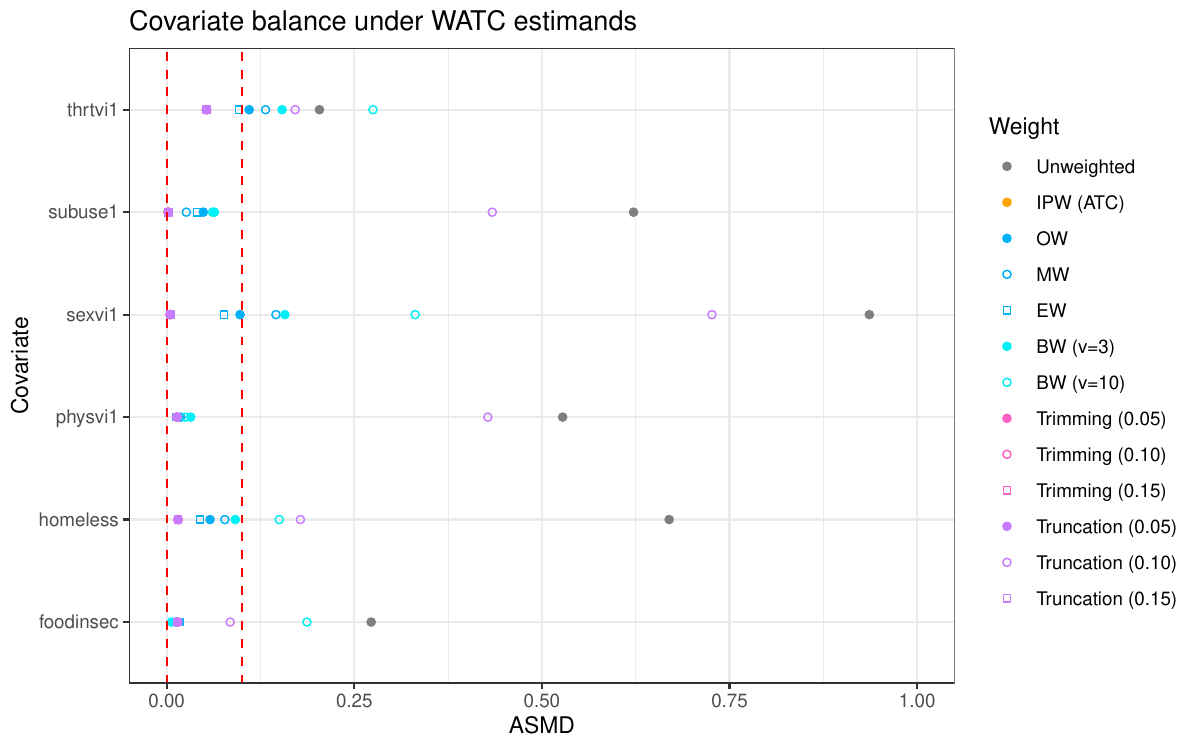}
    \caption{ASMDs using weights from WATC estimands (the red dashed lines indicate 0 and 0.1).}\footnotetext[0\def\thefoornote{}]{Comments on the covariate balance:  Unweighted, BW $(\nu=10)$, Truncation (0.10 and 0.15) and Trimming (0.10 and 0.15) provide the worst balancing results, followed by BW ($\nu=3$) and Truncation (0.05). All the other weights did balance the covariates well, with most ASMDs below the 0.1 ASMD-threshold. }
\end{figure}

\begin{table}[ht]
    \centering
    
    \caption{Covariate balance for homeless status (proportion of homeless: 22\%). }
    \begin{tabular}{rrrrl}
  \toprule
Mean ($A=1$) & Mean ($A=0$)  & Raw difference & ASMD & Weighting method \\
\midrule
  \multicolumn{5}{c}{\bf WATE balancing} \\
  \addlinespace
0.30 & 0.06 & 0.25 & 0.67 & Unweighted \\ 
  0.23 & 0.20 & 0.03 & 0.09 & IPW (ATE) \\ 
  0.30 & 0.27 & 0.04 & 0.10 & IPW (ATT) \\ 
  0.08 & 0.06 & 0.02 & 0.04 & IPW (ATC) \\ 
  0.12 & 0.12 & 0.00 & 0.00 & OW \\ 
  0.11 & 0.09 & 0.02 & 0.06 & MW \\ 
  0.13 & 0.14 & -0.01 & 0.02 & EW \\ 
  0.10 & 0.06 & 0.03 & 0.09 & BW ($\nu=3$) \\ 
  0.09 & 0.00 & 0.09 & 0.25 & BW ($\nu=10$) \\ 
  0.12 & 0.23 & -0.11 & 0.29 & Trimming ($\alpha=0.05$) \\ 
  0.12 & 0.22 & -0.10 & 0.28 & Truncation ($\alpha=0.05$) \\ 
  0.12 & 0.10 & 0.01 & 0.04 & Trimming ($\alpha=0.1$) \\ 
  0.11 & 0.11 & 0.00 & 0.00 & Truncation ($\alpha=0.1$) \\ 
  0.11 & 0.05 & 0.06 & 0.15 & Trimming ($\alpha=0.15$) \\ 
  0.11 & 0.08 & 0.04 & 0.10 & Truncation ($\alpha=0.15$) \\  
  \midrule
  \multicolumn{5}{c}{\bf WATT balancing} \\
  \addlinespace
0.30 & 0.06 & 0.25 & 0.67 & Unweighted \\ 
  0.30 & 0.27 & 0.04 & 0.10 & IPW (ATT) \\ 
  0.30 & 0.18 & 0.12 & 0.33 & OW \\ 
  0.30 & 0.14 & 0.16 & 0.44 & MW \\ 
  0.30 & 0.21 & 0.10 & 0.27 & EW \\ 
  0.30 & 0.06 & 0.24 & 0.66 & BW ($\nu=3$) \\ 
  0.30 & 0.00 & 0.30 & 0.83 & BW ($\nu=10$) \\ 
  0.30 & 0.33 & -0.02 & 0.07 & Trimming ($\alpha=0.05$) \\ 
  0.30 & 0.33 & -0.02 & 0.07 & Truncation ($\alpha=0.05$) \\ 
  0.30 & 0.15 & 0.15 & 0.41 & Trimming ($\alpha=0.1$) \\ 
  0.30 & 0.15 & 0.15 & 0.41 & Truncation ($\alpha=0.1$) \\ 
  0.30 & 0.08 & 0.22 & 0.61 & Trimming ($\alpha=0.15$) \\ 
  0.30 & 0.09 & 0.22 & 0.59 & Truncation ($\alpha=0.15$) \\ 
  \midrule
  \multicolumn{5}{c}{\bf WATC balancing} \\
  \addlinespace
  0.30 & 0.06 & 0.25 & 0.67 & Unweighted \\ 
  0.08 & 0.06 & 0.02 & 0.04 & IPW (ATC) \\ 
  0.07 & 0.06 & 0.01 & 0.03 & OW \\ 
  0.07 & 0.06 & 0.01 & 0.04 & MW \\ 
  0.07 & 0.06 & 0.01 & 0.03 & EW \\ 
  0.07 & 0.06 & 0.01 & 0.03 & BW ($\nu=3$) \\ 
  0.09 & 0.06 & 0.03 & 0.07 & BW ($\nu=10$) \\ 
  0.08 & 0.06 & 0.02 & 0.04 & Trimming ($\alpha=0.05$) \\ 
  0.08 & 0.06 & 0.02 & 0.04 & Truncation ($\alpha=0.05$) \\ 
  0.09 & 0.06 & 0.03 & 0.08 & Trimming ($\alpha=0.1$) \\ 
  0.08 & 0.06 & 0.02 & 0.05 & Truncation ($\alpha=0.1$) \\ 
  0.09 & 0.06 & 0.03 & 0.08 & Trimming ($\alpha=0.15$) \\ 
  0.08 & 0.06 & 0.02 & 0.06 & Truncation ($\alpha=0.15$) \\ 
  \bottomrule
\end{tabular}
\end{table}

\begin{table}[ht]
    \centering
    \caption{Covariate balance for food insecurity status (prop. of food-insecured participants: 34\%). }
    \begin{tabular}{rrrrl}
  \toprule
Mean ($A=1$) & Mean ($A=0$)  & Raw difference & ASMD & Weighting method \\
\midrule
  \multicolumn{5}{c}{\bf WATE balancing} \\
  \addlinespace
0.40 & 0.24 & 0.16 & 0.35 & Unweighted \\ 
  0.36 & 0.40 & -0.04 & 0.09 & IPW (ATE) \\ 
  0.40 & 0.49 & -0.09 & 0.21 & IPW (ATT) \\ 
  0.29 & 0.24 & 0.06 & 0.12 & IPW (ATC) \\ 
  0.31 & 0.31 & 0.00 & 0.00 & OW \\ 
  0.30 & 0.31 & -0.01 & 0.01 & MW \\ 
  0.32 & 0.32 & -0.01 & 0.02 & EW \\ 
  0.30 & 0.29 & 0.01 & 0.03 & BW ($\nu=3$) \\ 
  0.27 & 0.33 & -0.06 & 0.13 & BW ($\nu=10$) \\ 
  0.31 & 0.32 & -0.01 & 0.03 & Trimming ($\alpha=0.05$) \\ 
  0.31 & 0.32 & -0.01 & 0.03 & Truncation ($\alpha=0.05$) \\ 
  0.32 & 0.21 & 0.10 & 0.22 & Trimming ($\alpha=0.1$) \\ 
  0.30 & 0.23 & 0.07 & 0.15 & Truncation ($\alpha=0.1$) \\ 
  0.32 & 0.26 & 0.06 & 0.12 & Trimming ($\alpha=0.15$) \\ 
  0.30 & 0.26 & 0.04 & 0.10 & Truncation ($\alpha=0.15$) \\ 
  \midrule
  \multicolumn{5}{c}{\bf WATT balancing} \\
  \addlinespace
0.40 & 0.24 & 0.16 & 0.35 & Unweighted \\ 
  0.40 & 0.49 & -0.09 & 0.21 & IPW (ATT) \\ 
  0.40 & 0.36 & 0.04 & 0.09 & OW \\ 
  0.40 & 0.35 & 0.05 & 0.11 & MW \\ 
  0.40 & 0.38 & 0.02 & 0.04 & EW \\ 
  0.40 & 0.29 & 0.11 & 0.24 & BW ($\nu=3$) \\ 
  0.40 & 0.33 & 0.07 & 0.15 & BW ($\nu=10$) \\ 
  0.40 & 0.38 & 0.02 & 0.04 & Trimming ($\alpha=0.05$) \\ 
  0.40 & 0.38 & 0.02 & 0.04 & Truncation ($\alpha=0.05$) \\ 
  0.40 & 0.22 & 0.18 & 0.40 & Trimming ($\alpha=0.1$) \\ 
  0.40 & 0.22 & 0.18 & 0.39 & Truncation ($\alpha=0.1$) \\ 
  0.40 & 0.27 & 0.13 & 0.29 & Trimming ($\alpha=0.15$) \\ 
  0.40 & 0.27 & 0.13 & 0.28 & Truncation ($\alpha=0.15$) \\ 
  \midrule
  \multicolumn{5}{c}{\bf WATC balancing} \\
  \addlinespace
  0.40 & 0.24 & 0.16 & 0.35 & Unweighted \\ 
  0.29 & 0.24 & 0.06 & 0.12 & IPW (ATC) \\ 
  0.31 & 0.24 & 0.07 & 0.15 & OW \\ 
  0.30 & 0.24 & 0.06 & 0.14 & MW \\ 
  0.30 & 0.24 & 0.07 & 0.15 & EW \\ 
  0.31 & 0.24 & 0.07 & 0.16 & BW ($\nu=3$) \\ 
  0.28 & 0.24 & 0.04 & 0.09 & BW ($\nu=10$) \\ 
  0.29 & 0.24 & 0.06 & 0.12 & Trimming ($\alpha=0.05$) \\ 
  0.29 & 0.24 & 0.06 & 0.12 & Truncation ($\alpha=0.05$) \\ 
  0.34 & 0.24 & 0.11 & 0.24 & Trimming ($\alpha=0.1$) \\ 
  0.30 & 0.24 & 0.06 & 0.13 & Truncation ($\alpha=0.1$) \\ 
  0.34 & 0.24 & 0.11 & 0.24 & Trimming ($\alpha=0.15$) \\ 
  0.31 & 0.24 & 0.08 & 0.17 & Truncation ($\alpha=0.15$) \\ 
  \bottomrule
\end{tabular}
\end{table}

\begin{table}[ht]
\centering
\caption{Estimated {weighted risk difference} (RD; point estimate and 95\% CI) of sex work history on HIV status under WATE, WATT, and WATC frameworks. }
\begin{tabular}{lccc}
\toprule
Weighting method & WATE & WATT & WATC \\
\midrule
Overall                 & 0.22 ($-$0.14, 0.40) & 0.24 ($-$0.21, 0.47) & 0.19 (0.03, 0.42) \\
Treated & 0.24 ($-$0.21, 0.47) & $-$ & $-$ \\
Control & 0.18 (0.03, 0.34) & $-$ & $-$ \\
Overlap (OW)                 & 0.21 (0.08, 0.34)  & 0.29 (0.15, 0.46)  & 0.19 (0.04, 0.32) \\
Matching (MW)               & 0.22 (0.09, 0.34)  & 0.32 (0.17, 0.47)  & 0.19 (0.04, 0.33) \\
Entropy  (EW)               & 0.21 (0.06, 0.34)  & 0.28 (0.12, 0.46)  & 0.18 (0.04, 0.33) \\
Beta (BW; $\nu = 3$)          & 0.23 (0.09, 0.35)  & 0.34 (0.19, 0.49)  & 0.19 (0.05, 0.33) \\
Beta (BW; $\nu = 10$)         & 0.26 (0.05, 0.48)  & 0.43 (0.24, 0.52)  & 0.19 ($-$0.00, 0.43) \\
Trimming ($\alpha = 0.05$) & 0.17 (0.02, 0.35) & 0.24 (0.01, 0.47)  & 0.19 (0.04, 0.42) \\
Trimming ($\alpha = 0.10$) & 0.21 (0.02, 0.35) & 0.28 (0.07, 0.48)  & 0.19 (0.06, 0.41) \\
Trimming ($\alpha = 0.15$) & 0.18 (0.05, 0.37) & 0.23 (0.13, 0.50)  & 0.23 (0.06, 0.35) \\
Truncation ($\alpha = 0.05$) & 0.18 (0.02, 0.33) & 0.24 (0.01, 0.47)  & 0.19 (0.04, 0.42) \\
Truncation ($\alpha = 0.10$) & 0.20 (0.00, 0.33) & 0.27 (0.07, 0.48)  & 0.19 (0.05, 0.41) \\
Truncation ($\alpha = 0.15$) & 0.13 (0.00, 0.33) & 0.23 (0.13, 0.49)  & 0.19 (0.05, 0.39) \\
\bottomrule
\end{tabular}
\begin{tablenotes}\footnotesize
    \item The ``Treated'' and ``Control'' entries are not meaningful for the WATT and WATC frameworks. Under WATE, the ``Treated'' (resp. ``Control'') results correspond exactly to the ``Overall'' results under WATT (resp. WATC), as they represent the ATT (resp. ATC) by definition. 
\end{tablenotes}
\end{table}

\begin{figure}[ht]
    \centering
    \includegraphics[width=\linewidth]{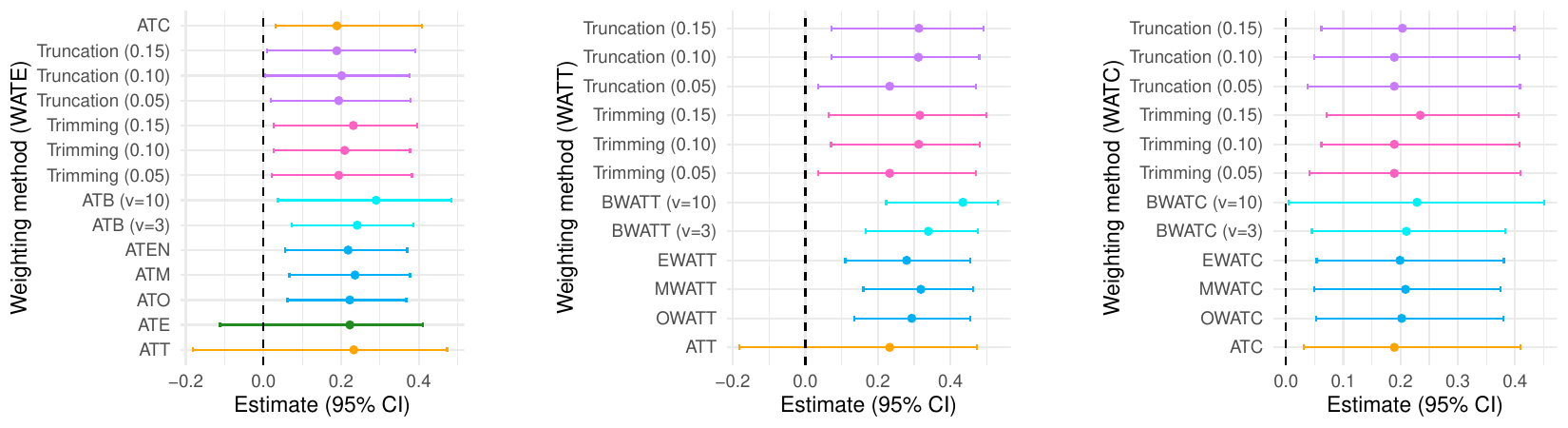}
    \caption{Forest plots for {weighted 
    risk differences (RDs)} of sex work history on HIV status among transgender women in South Africa. From left to right, the panels correspond to weighting methods (and estimands) in the WATE, WATT, and WATC classes, respectively. Within each class, forest plots present the estimation and 95\% CI results for the chosen weighting methods in that class (same as those considered in the simulation study). }
    \label{fig:hiv-RD}
\end{figure}

\begin{table}[ht]
\centering
\caption{Estimated {weighted risk ratio} (RR; point estimate and 95\% CI) of sex work history on HIV status under WATE, WATT, and WATC frameworks. }
\begin{tabular}{lccc}
\toprule
Weighting method & WATE & WATT & WATC \\
\midrule
Overall                 & 2.23 (0.73, 10.49) & 2.08 (0.67, 14.83) & 2.74 (1.11, 7.94) \\
Treated & 2.08 (0.67, 14.83) & $-$ & $-$ \\
Control & 2.74 (1.11, 7.94) & $-$ & $-$ \\
Overlap (OW)                 & 2.56 (1.38, 8.71)  & 2.60 (1.52, 8.73)  & 2.80 (1.30, 8.12) \\
Matching (MW)               & 2.94 (1.62, 8.76)  & 3.09 (1.74, 11.02) & 2.81 (1.25, 8.11) \\
Entropy (EW)                & 2.41 (1.29, 7.51)  & 2.42 (1.42, 8.98)  & 2.79 (1.26, 7.89) \\
Beta (BW; $\nu = 3$)          & 3.36 (1.64, 12.02) & 3.58 (1.77, 15.67) & 2.84 (1.29, 8.35) \\
Beta (BW; $\nu = 10$)         & 7.76 (1.39, 1209.18) & 13.40 (2.25, 1189.24) & 2.85 (0.98, 9.44) \\
Trimming ($\alpha = 0.05$) & 1.95 (1.01, 8.03) & 2.08 (1.02, 9.75) & 2.74 (1.11, 7.94) \\
Trimming ($\alpha = 0.10$) & 2.44 (1.03, 7.74) & 2.42 (1.06, 16.25) & 2.74 (1.31, 7.88) \\
Trimming ($\alpha = 0.15$) & 2.08 (0.90, 10.62)  & 1.98 (0.90, 79.01)  & 3.13 (1.48, 8.94) \\
Truncation ($\alpha = 0.05$) & 1.97 (0.97, 7.00) & 2.08 (1.02, 10.23) & 2.74 (1.17, 7.15) \\
Truncation ($\alpha = 0.10$) & 2.31 (0.96, 6.26) & 2.42 (1.16, 25.61) & 2.74 (1.30, 7.39) \\
Truncation ($\alpha = 0.15$) & 1.78 (0.90, 8.75)   & 1.99 (1.36, 52.23) & 2.85 (1.30, 7.80) \\
\bottomrule
\end{tabular}
\begin{tablenotes}\footnotesize
    \item The ``Treated'' and ``Control'' entries are not meaningful for the WATT and WATC frameworks. Under WATE, the ``Treated'' (resp. ``Control'') results correspond exactly to the ``Overall'' results under WATT (resp. WATC), as they represent the ATT (resp. ATC) by definition. 
\end{tablenotes}
\end{table}

\begin{figure}[ht]
    \centering
    \includegraphics[width=\linewidth]{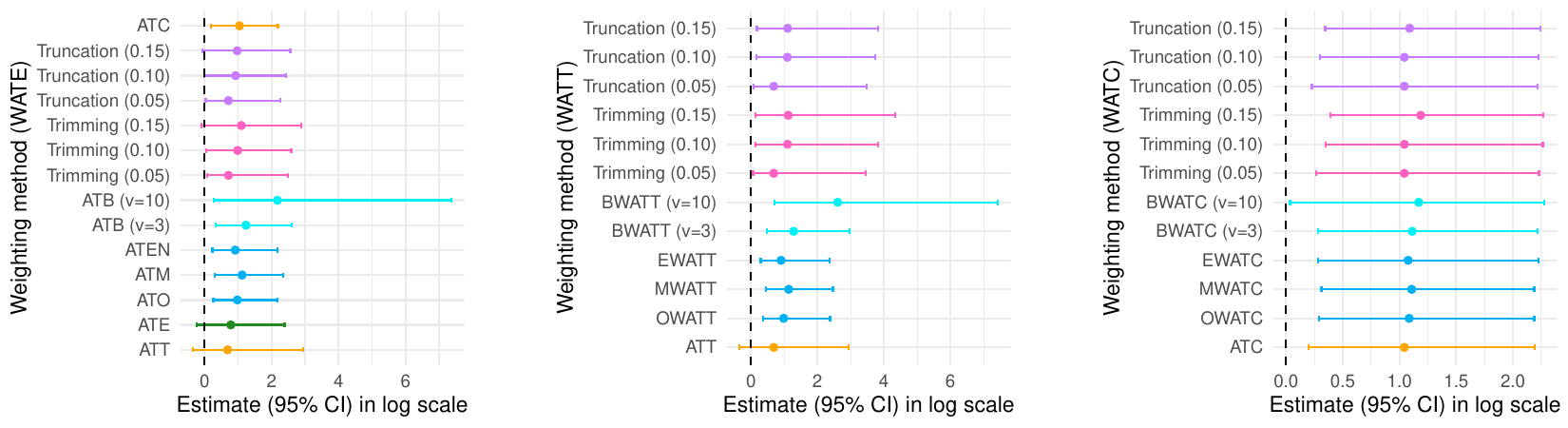}
   \caption{Forest plots for {weighted 
    risk ratios (RRs)} of sex work history on HIV status among transgender women in South Africa. From left to right, the panels correspond to weighting methods (and estimands) in the WATE, WATT, and WATC classes, respectively. Within each class, forest plots present the estimation and 95\% CI results for the chosen weighting methods in that class (same as those considered in the simulation study). }
    \label{fig:hiv-RR}
\end{figure}

\begin{table}[ht]
\centering
\caption{Estimated {weighted odds ratio} (OR; point estimate and 95\% CI) of sex work history on HIV status under WATE, WATT, and WATC frameworks. }
\begin{tabular}{lccc}
\toprule
Weighting method & WATE & WATT & WATC \\
\midrule
Overall                 & 3.08 (0.56, 16.15) & 3.03 (0.43, 28.00) & 3.42 (1.41, 12.21) \\
Treated & 3.03 (0.43, 28.00) & $-$ & $-$ \\
Control & 3.42 (1.41, 12.21)  & $-$ & $-$ \\
Overlap (OW)                 & 3.38 (1.55, 12.60) & 4.01 (1.98, 17.41) & 3.53 (1.36, 10.71) \\
Matching (MW)               & 3.92 (1.80, 12.98) & 4.93 (2.30, 21.52) & 3.55 (1.32, 11.67) \\
Entropy (EW)                 & 3.18 (1.41, 11.29) & 3.67 (1.72, 17.31) & 3.50 (1.33, 10.58) \\
Beta (BW; $\nu = 3$)          & 4.51 (1.85, 16.79) & 5.86 (2.34, 27.86) & 3.60 (1.39, 11.15) \\
Beta (BW; $\nu = 10$)         & 10.57 (1.56, 1541.59) & 24.34 (3.35, 2216.85) & 3.62 (0.97, 14.16) \\
Trimming ($\alpha = 0.05$) & 2.47 (1.02, 12.50) & 3.03 (1.03, 19.08) & 3.42 (1.41, 12.21) \\
Trimming ($\alpha = 0.10$) & 3.22 (1.05, 11.28) & 3.68 (1.12, 29.70) & 3.42 (1.39, 11.76) \\
Trimming ($\alpha = 0.15$) & 2.64 (0.90, 16.89)   & 2.85 (0.90, 130.04)  & 4.14 (1.58, 12.23) \\
Truncation ($\alpha = 0.05$) & 2.50 (0.96, 10.49) & 3.03 (1.04, 19.33) & 3.42 (1.23, 10.90) \\
Truncation ($\alpha = 0.10$) & 3.02 (0.95, 9.37)  & 3.66 (1.32, 44.32) & 3.42 (1.35, 10.52) \\
Truncation ($\alpha = 0.15$) & 2.12 (0.90, 12.07)   & 2.86 (1.70, 92.86) & 3.61 (1.40, 10.91) \\
\bottomrule
\end{tabular}
\begin{tablenotes}\footnotesize
    \item The ``Treated'' and ``Control'' entries are not meaningful for the WATT and WATC frameworks. Under WATE, the ``Treated'' (resp. ``Control'') results correspond exactly to the ``Overall'' results under WATT (resp. WATC), as they represent the ATT (resp. ATC) by definition. 
\end{tablenotes}
\end{table}

\begin{figure}[ht]
    \centering
    \includegraphics[width=\linewidth]{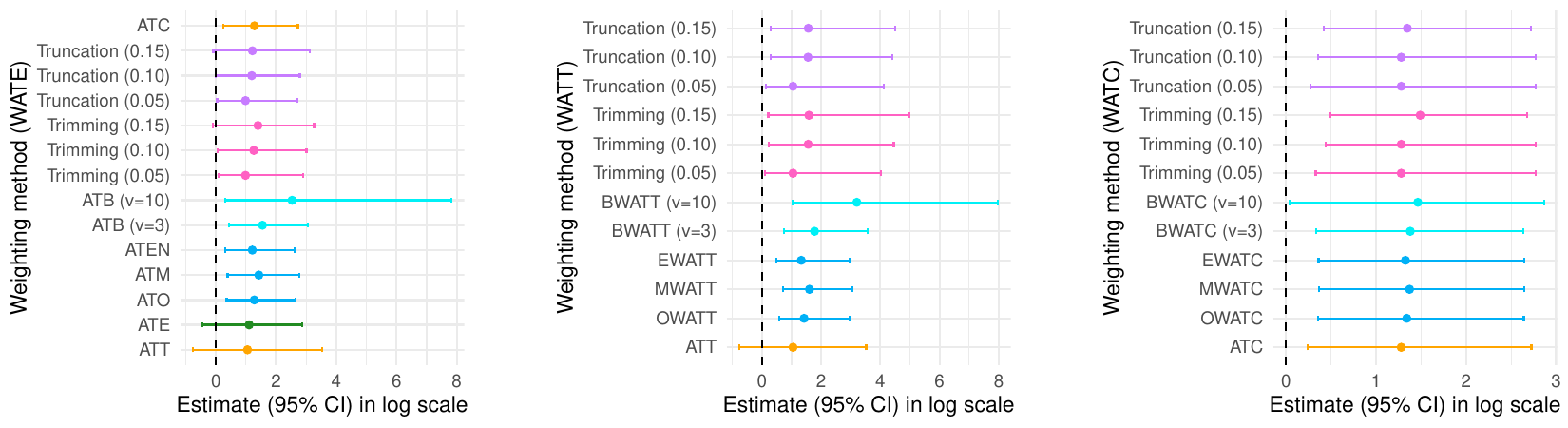}
   \caption{Forest plots for {weighted 
    odds ratios (ORs)} of sex work history on HIV status among transgender women in South Africa. From left to right, the panels correspond to weighting methods (and estimands) in the WATE, WATT, and WATC classes, respectively. Within each class, forest plots present the estimation and 95\% CI results for the chosen weighting methods in that class (same as those considered in the simulation study). }
    \label{fig:hiv-OR}
\end{figure}

\end{document}